\documentclass[%
11pt,
onecolumn,
tightenlines,
superscriptaddress,
notitlepage,
preprintnumbers,
nofootinbib,
 amsmath,amssymb,amsthm,
 aps,
eqsecnum,
]{revtex4-2}

\usepackage{isomath}
\usepackage{amsmath}
\usepackage{amsbsy}
\usepackage{amssymb}
\usepackage{amscd}
\usepackage{amsfonts}
\usepackage{stmaryrd}
\usepackage{units}
\usepackage{euscript}
\usepackage[utf8]{inputenc}
\usepackage[T1]{fontenc}
\usepackage{times}
\everymath{\displaystyle}
\usepackage{exscale}

\usepackage{graphicx}
\usepackage{boxedminipage}
\usepackage{calc}
\usepackage[usenames,dvipsnames]{xcolor}
\graphicspath{ {media/} }
\usepackage[caption=false]{subfig}

\usepackage[small,compact]{titlesec}
\usepackage{setspace}
\usepackage{enumitem}
\setitemize{noitemsep,topsep=0pt,parsep=0pt,partopsep=0pt}
\setenumerate{noitemsep,topsep=0pt,parsep=0pt,partopsep=0pt}
\setdescription{noitemsep,topsep=0pt,parsep=0pt,partopsep=0pt}

\usepackage[colorinlistoftodos, color=green!40,prependcaption]{todonotes}

\usepackage{soul}

\usepackage[,colorlinks,urlcolor=blue,citecolor=blue]{hyperref}
\usepackage[normalem]{ulem}

\makeatletter
\renewcommand{\p@subsection}{}
\renewcommand{\p@subsubsection}{}
\makeatother

\setuptodonotes{inline}

\usepackage{isomath}
\usepackage{amsmath}
\usepackage{amssymb}
\usepackage{amscd}
\usepackage{amsfonts}

\newcommand{\nhat}{\hat{\mathbold n}}

\newcommand{\half}{\frac{1}{2}}

\newcommand{\bfchi}{\mathbold {\chi}}

\DeclareMathOperator{\erf}{erf}

\DeclareMathOperator{\divergence}{div}

\newcommand{\dm}{\ \mathrm{d}}

\newcommand{\bfm}{{\mathbold m}}
\newcommand{\bfn}{{\mathbold n}}

\newcommand{\bfp}{{\mathbold p}}

\newcommand{\bfr}{{\mathbold r}}

\newcommand{\bfx}{{\mathbold x}}

\newcommand{\bfE}{{\mathbold E}}

\newcommand{\bfI}{{\mathbold I}}

\newcommand{\bfP}{{\mathbold P}}

\renewcommand\hl[1]{#1}

\usepackage[plain]{fancyref}
\usepackage{comment}
\usepackage{mathtools}
\usepackage{tabulary}
\usepackage{amsthm}
\usepackage{bm}
\graphicspath{ {figs/} }
\everymath{\displaystyle}

\numberwithin{figure}{section}

\newcommand*{\fancyrefapplabelprefix}{app}
\frefformat{plain}{\fancyrefapplabelprefix}{appendix~#1}
\Frefformat{plain}{\fancyrefapplabelprefix}{Appendix~#1}
\frefformat{main}{\fancyrefapplabelprefix}{appendix~#1}
\Frefformat{main}{\fancyrefapplabelprefix}{Appendix~#1}

\renewcommand{\ln}{\log}
\newcommand{\hessian}{\bm{\mathcal{H}}}
\newcommand{\hessianDet}{\mathcal{H}}

\newcommand{\crossModulus}{\mathcal{K}}
\newcommand{\crossModulusPara}{\crossModulus_{\parallel}}
\newcommand{\crossModulusTensor}{\bm{\crossModulus}}
\newcommand{\Enodim}{\mathcal{E}}

\newcommand{\polarAvg}{\polar_0}

\newcommand{\Std}{\sigma}

\newcommand{\nb}{\N_b} 
\newcommand{\nl}{\N_l} 
\newcommand{\lefrac}{\alpha} 
\newcommand{\chainpdf}{\mathrm{P}} 

\newcommand{\Elastic}{Y}
\newcommand{\ElasticPara}{\Elastic_{\parallel}}
\newcommand{\ElasticPerp}{\Elastic_{\perp}}

\newcommand{\ElasticParaInf}{\Elastic_{\parallel,\infty}}
\newcommand{\ElasticPerpInf}{\Elastic_{\perp,\infty}}
\newcommand{\ElasticIso}{\Elastic_{iso}}

\newcommand{\ASym}{A}
\newcommand{\BSym}{B}
\newcommand{\StdA}{\Std_{\ASym}}
\newcommand{\StdB}{\Std_{\BSym}}
\newcommand{\polarAvgA}{\polarAvg^{\ASym}}
\newcommand{\polarAvgB}{\polarAvg^{\BSym}}
\newcommand{\chainDensityRefA}{\chaindensityref^{\ASym}}
\newcommand{\chainDensityRefB}{\chaindensityref^{\BSym}}
\newcommand{\chainFracA}{\Xi^{\ASym}}
\newcommand{\chainFracB}{\Xi^{\BSym}}
\newcommand{\NA}{\N^{\ASym}}
\newcommand{\NB}{\N^{\BSym}}
\newcommand{\NWeightA}{\xi^{\ASym}}
\newcommand{\NWeightB}{\xi^{\BSym}}
\newcommand{\SusParaA}{\sus{1}^{\ASym}}
\newcommand{\SusPerpA}{\sus{2}^{\ASym}}
\newcommand{\SusParaB}{\sus{1}^{\BSym}}
\newcommand{\SusPerpB}{\sus{2}^{\BSym}}
\newcommand{\avgA}[1]{\avg{#1}^{\ASym}}
\newcommand{\avgB}[1]{\avg{#1}^{\BSym}}
\newcommand{\dsusA}{\dsus^{\ASym}}
\newcommand{\dsusB}{\dsus^{\BSym}}
\newcommand{\Nm}{\mathcal{N}^m}
\newcommand{\NmRef}{\Nm_0}

\newcommand{\polarAB}{\polar_{iso}}

\DeclareMathOperator{\diag}{diag}
\DeclareMathOperator{\vspan}{span}
\DeclareMathOperator{\ImPart}{Im}

\newcommand{\eql}[1]{#1^{\mathrm{eq}}}
\newcommand{\rstate}[1]{#1^{\star}}

\newcommand{\bcs}{\mathbb{B}}
\newcommand{\extwork}{\mathbb{W}}
\newcommand{\gibbs}{\mathcal{G}}
\newcommand{\bigo}{\mathcal{O}}
\newcommand{\bigoof}[1]{\bigo\left(#1\right)}
\newcommand{\GaussSineInt}[3]{I_{#1}\left[#2, #3\right]}
\newcommand{\GaussSineIntZero}[3]{I^0_{#1}\left[#2, #3\right]}
\newcommand{\GaussSineIntInf}[3]{I^{\infty}_{#1}\left[#2, #3\right]}

\newcommand{\Lw}{l}
\newcommand{\Le}{l_3}

\newcommand{\diracdel}{\delta} 
\newcommand{\W}{\mathcal{W}} 
\newcommand{\Wstar}{\W^{*}}
\newcommand{\F}{\mathbold{F}} 
\newcommand{\J}{J} 
\newcommand{\tmag}{t_0}
\newcommand{\tvec}{\mathbold{\tmag}}
\newcommand{\stressij}{\Sigma}

\newcommand{\pos}{\mathbold{x}} 
\newcommand{\Pos}{\mathbold{X}} 

\newcommand{\rtocmap}{\boldsymbol{\Upsilon}} 
\DeclareMathOperator{\Grad}{Grad}

\newcommand{\GradOf}[1]{\Grad #1}
\newcommand{\Rmag}{\tilde{r}}
\newcommand{\Rvec}{\mathbold{\Rmag}}
\newcommand{\chaindensity}{N}
\newcommand{\chaindensityref}{\chaindensity_0}
\newcommand{\pstch}[1]{\lambda_{#1}}
\newcommand{\pstchu}{\lambda}
\newcommand{\cbody}{\Omega}
\newcommand{\rbody}{\cbody_0}

\newcommand{\avg}[1]{\left\langle #1 \right\rangle_{\Rvec}}
\newcommand{\epot}{\varphi}
\newcommand{\epotDiff}{\Delta \epot}

\newcommand{\rspace}{\mathbb{R}^3}

\newcommand{\pstche}{\pstch{E}}
\newcommand{\takepartial}[2]{\frac{\partial #1}{\partial #2}}
\newcommand{\takepartialflat}[2]{\partial #1 / \partial #2}
\newcommand{\takepartials}[3]{\frac{\partial^{#3} #1}{\partial #2^{#3}}}

\newcommand{\takecrosspartial}[3]{\frac{\partial^2 #1}{\partial #2 \partial #3}}
\newcommand{\susceptibility}{\mathcal{X}}
\newcommand{\TemperatureZeroLimitSym}{c}
\newcommand{\TemperatureInfLimitSym}{h}
\newcommand{\susceptibilityCold}{\susceptibility^{\TemperatureZeroLimitSym}}
\newcommand{\susceptibilityHot}{\susceptibility^{\TemperatureInfLimitSym}}
\newcommand{\susceptibilityPara}{\susceptibility_{\parallel}}
\newcommand{\susceptibilityTensorCold}{\bm{\susceptibility}^{\TemperatureZeroLimitSym}}
\newcommand{\susceptibilityTensorHot}{\bm{\susceptibility}^{\TemperatureInfLimitSym}}

\newcommand{\susceptibilityHotPara}{\susceptibilityHot_{\parallel}}

\newcommand{\susceptibilityTensor}{\bm{\susceptibility}}
\newcommand{\susceptibilityTensorLab}{\susceptibilityTensor_{sec}}
\newcommand{\susceptibilityLab}{\susceptibility_{sec}}
\newcommand{\susceptibilityLabPara}{\susceptibility_{sec,\parallel}}

\newcommand{\freeSym}{\text{free}}
\newcommand{\susceptibilityFree}{\susceptibility_{\freeSym}}
\newcommand{\susceptibilityHotFree}{\susceptibilityHot_{\freeSym}}
\newcommand{\correction}[1]{\check{#1}}
\newcommand{\susceptibilityHotParaCorr}{\correction{\susceptibilityHot_{\parallel}}}
\newcommand{\susceptibilityColdParaCorr}{\correction{\susceptibilityCold_{\parallel}}}
\newcommand{\susceptibilityHotParaCorrZero}{\correction{\susceptibilityHot_{\parallel,0}}}
\newcommand{\susceptibilityHotParaCorrInf}{\correction{\susceptibilityHot_{\parallel,\infty}}}

\newcommand{\Emag}{\tilde{E}}

\newcommand{\polarizationmag}{P}
\newcommand{\polarization}{\mathbold{\polarizationmag}}
\newcommand{\Polarmag}{\tilde{P}}
\newcommand{\Polar}{\mathbold{\Polarmag}}

\newcommand{\pstchpr}{\pstchu^{\prime}}

\newcommand{\Ham}{\mathcal{H}}

\newcommand{\AHelm}{\mathcal{A}}

\newcommand{\A}{\mathcal{F}}

\newcommand{\T}{T}

\newcommand{\Sent}{\mathcal{S}}

\newcommand{\kB}{k}

\newcommand{\Gibbs}{\gibbs}

\newcommand{\shearMod}{G}
\newcommand{\shearModClassical}{\shearMod_{iso}}

\newcommand{\mlen}{b}

\newcommand{\N}{n}

\newcommand{\density}{\rho}

\newcommand{\rmag}{r}
\newcommand{\rvec}{\mathbold{\rmag}}

\newcommand{\rdir}{\hat{\rvec}}

\newcommand{\stch}{\gamma}

\newcommand{\stchr}{\stch_r}

\newcommand{\unitsphere}{\mathbb{S}^2}

\newcommand{\um}{u}

\newcommand{\nvec}{\mathbold{\hat{n}}}

\newcommand{\U}{U}

\newcommand{\sussymbol}{\chi}
\newcommand{\sus}[1]{\ifthenelse{#1 < 2}{\sussymbol_{\parallel}}{\sussymbol_{\perp}}}

\newcommand{\sustens}{\boldsymbol{\chi}_{\mu}}
\newcommand{\dsus}{\Delta \sussymbol}
\newcommand{\SusM}{\sussymbol_{M}}

\newcommand{\emag}{E}
\newcommand{\efield}{\mathbold{\emag}}
\newcommand{\edir}{\hat{\efield}}

\newcommand{\ezeromag}{\emag_0}

\newcommand{\dipolemag}{\mu}
\newcommand{\dipole}{\boldsymbol{\dipolemag}}
\newcommand{\bfmu}{\dipole}

\newcommand{\chainpolarmag}{p}
\newcommand{\chainpolar}{\mathbold{\chainpolarmag}}

\newcommand{\unodim}{\kappa}

\newcommand{\uOnodim}{\unodim_{\perp}}
\newcommand{\uslnodim}{\unodim}

\newcommand{\im}{{i\mkern1mu}}

\newcommand{\euler}{e}

\newcommand{\multmag}{\tau}
\newcommand{\mults}{\boldsymbol{\multmag}}

\newcommand{\C}{C}

\newcommand*\df{\mathop{}\!\mathrm{d}}

\newcommand{\dx}[1]{\df #1 \text{ }}

\newcommand{\azi}{\phi}
\newcommand{\polar}{\theta}

\newcommand{\iden}{\bfI}

\newcommand{\Lang}{\mathcal{L}}
\newcommand{\Langinv}{\Lang^{-1}}
\newcommand{\Langinvs}{\Langinv\left(\stch\right)}

\DeclareMathOperator{\csch}{csch}

\newcommand{\generic}{\Box}

\newcommand{\Asm}{\A_{s\multmag}}
\newcommand{\Akg}{\A_{KG}}

\newcommand{\smallparam}{\epsilon}

\newcommand{\I}[1]{I_{#1}}

\newcommand{\intoverS}[1]{\int_{0}^{\pi} \df \polar \int_{0}^{2\pi} \df \azi \mbox{ } #1 \sin \polar}
\newcommand{\intoverSns}[1]{\int_{\unitsphere} \df{A} \mbox{ } #1}
\renewcommand{\intoverSns}[1]{\int_{\unitsphere} #1 \dm A}
\newcommand{\intoverSPdf}[1]{\intoverS{\chainpdf\left(\azi, \polar\right) \left(#1\right)}}

\newcommand{\zmultzero}{\Langinv\left(\stch\right)}

\newcommand{\erfw}{\erf \left(\sqrt{\uslnodim}\right)}

\begin{document}


\preprint{To appear in Journal of the Mechanics and Physics of Solids (doi: \href{https://doi.org/10.1016/j.jmps.2020.104171}{10.1016/j.jmps.2020.104171})}

\title{\Large{Architected Elastomer Networks for Optimal Electromechanical Response}}

\author{Matthew Grasinger}
    \email{mgrasing@alumni.cmu.edu}
    \affiliation{Department of Civil and Environmental Engineering, Carnegie Mellon University}

\author{Kaushik Dayal}
    \email{Kaushik.Dayal@cmu.edu}
    \affiliation{Department of Civil and Environmental Engineering, Carnegie Mellon University}
    \affiliation{Center for Nonlinear Analysis, Department of Mathematical Sciences, Carnegie Mellon University}
    \affiliation{Department of Materials Science and Engineering, Carnegie Mellon University}
    
\date{\today}

\begin{abstract}
   Dielectric elastomers (DEs) that couple deformation and electrostatics have the potential for use in soft sensors and actuators with applications ranging from robotic, biomedical, energy, aerospace and automotive technologies.
    However, currently available DEs are limited by weak electromechanical coupling and require large electric fields for significant actuation.
    In this work, a statistical mechanics-based model of DE chains is applied to elucidate the role of a polymer network architecture in the performance of the bulk material.
    Given a polymer network composed of chains that are cross-linked, the paper examines the role of cross-link density, orientational density of chains, and other network parameters in determining the material properties of interest including elastic modulus, electrical susceptibility, and the electromechanical coupling.
    From this analysis, a practical strategy is presented to increase the deformation and usable work derived from (anisotropic) dielectric elastomer actuators by as much as $75-100\%$.  
\end{abstract}

\maketitle





\section{Introduction}

Soft functional polymeric materials that couple deformation to
electric fields, magnetic fields, or illumination, have emerged as
leading candidates for sensors and actuators with applications
across soft robotics, biomedical devices, biologically inspired
robots, advanced prosthetics, and various other technologies~\cite{bar-cohen2001electroactive,carpi2011electroactive,kim2007electroactive,huang2012giant,majidi2014soft,bartlett2016stretchable,lopez2014elastic,ware2016localized,erol2019microstructure,castaneda2011homogenization,galipeau2013finite}.
The key advantages of dielectric elastomers (DEs) are that they are generally inexpensive, lightweight, easily shaped, pliable, and can undergo significant deformations~\cite{bar-cohen2001electroactive}.
However, despite these advantages, DEs are limited by weak electromechanical coupling so that large voltages are often required to achieve meaningful actuation~\cite{bar-cohen2001electroactive}.
A better understanding of these materials -- in particular, how mesoscale polymer network characteristics affect the bulk material response -- would not only enable microstructural design with currently available DEs, but can also potentially lead to the development of advanced DEs with stronger electromechanical coupling, as discussed in this paper.

Manufacturing technologies are rapidly improving, thereby enabling the synthesis of materials with complex and hierarchical patterning, e.g. \cite{ambulo2017four,ford2019multifunctional}.
Thus, it is the goal of this work to: (1.) develop a model, based in statistical mechanics, to connect the molecular- and macromolecular-properties of a given DE to its performance as an actuator and/or energy harvester, and to (2.) use the multiscale model to develop insights and predictions into how novel DEs with enhanced electromechanical coupling may be designed and manufactured, particularly by tailoring the polymer network architecture.
Broadly, our strategy and goal is similar to work in nonlinear elasticity-based homogenization (e.g.~\cite{lopez2014elastic,siboni2014fiber,castaneda2011homogenization,galipeau2013finite}), in that we aim to tailor the microstructure to achieve desired or optimal responses, except that we focus on smaller lengthscales and use physical models that are appropriate to these scales.


The quintessential example of a DE actuator (DEA) is a thin DE film sandwiched between two compliant electrodes--a soft parallel plate capacitor.
When a voltage difference is applied across the electrodes, a positive charge density accumulates on one of the electrodes and an equal and opposite charge density accumulates on the other.
The electrodes are attracted to each through simple Coulombic forces, causing the DE film to compress in the thickness direction and, because DEs are roughly incompressible, the expands in the plane of the electrodes
\cite{pelrine2000high,wissler2007mechanical,kofod2008static,kollosche2012complex}.


Although the Coulombic attraction is an important factor, it has been pointed out (both theoretically and experimentally) that if the susceptibility of the DE film is a function of deformation, then an additional stress develops in the DE film~\cite{zhao2008electrostriction,suo2010theory,cohen2016electromechanical}.
In this work, we aim to design novel DEs that (1.) have maximal susceptibility (in their typical operating conditions) in order to increase the capacitance (and, therefore, performance) of its corresponding thin film DEA and (2.) use this additional stress in order to increase the electromechanical coupling of DEAs.


The electromechanical modeling of DEs can be grouped into two categories: (1.) continuum based approaches where the general form of the energy density (of either a polymer chain or polymer network) as a function of mechanical and electrical loads is inferred by assuming a form of the equation, which usually depends on electroelastic invariants, that leads to behavior observed in experiments (see for example~\cite{toupin1956elastic,dorfmann2006nonlinear,zhao2008electrostriction,toupin1956elastic,dorfmann2014nonlinear,henann2013modeling,zurlo2017catastrophic,liu2018emergent,krichen2019liquid,li2015geometrically,darbaniyan2019designing,shmuel2013axisymmetric,fox2008dynamic,rahmati2019nonlinear})
and
(2.) statistical mechanics based approaches that build from molecular-scale responses up to the levels of a single chain and eventually the continuum level response.

Statistical mechanics has a rich history in modeling and understanding rubber elasticity, e.g. \cite{treloar1975physics,kuhn1942beziehungen,weiner2012statistical,flory1944network,hill1986statistical}; however, its use in the modeling of DEs is much less developed.
The work by deBotton and coworkers \cite{cohen2016electroelasticity,cohen2016electromechanical} appears to be the first in this direction\footnote{There is a significant literature on so-called poly-electrolytes that are polymers in solution with ionic effects (e.g. \cite{shen2017electrostatic,wang2004self,argudo2012dependence}), but there are important differences between these systems and our interest; these differences are briefly discussed in \cite{grasinger2020statistical}.}, and an explicit approximate expression for the response of a single electro-active polymer chain was provided in \cite{grasinger2020statistical}.
This explicit approximation provides the starting point for the analysis in the current paper.

\paragraph*{Structure of the paper.}

In \Fref{sec:theory}, we provide a summary of the statistical mechanical model of a single electro-responsive polymer chain from \cite{grasinger2020statistical}, and the related issues of averaging over chains to obtain the continuum response.
In \Fref{sec:design-parameters}, we identify certain network properties as design parameters, motivated heuristically by the ease of being able to tune these parameters experimentally.
Subsequent sections (\ref{sec:rubber-elasticity},\ref{sec:susceptibility},\ref{sec:electrostrictive-stress}) examine the role of network properties on the elastic modulus, the susceptibility, and the electromechanical coupling modulus respectively.
Finally, \Fref{sec:deas} provides an application of these ideas to the setting of a simple DEA geometry.
Appendix \ref{sec:symbols} provides a summary of the key notation that is used in the paper.

\section{Modeling Framework: Multiscale Theory of Electroactive Elastomers} \label{sec:theory}

We briefly summarize the modeling framework for electro-elastic polymeric materials, based on \cite{grasinger2020statistical,grasinger2020inprep}.
This framework roughly solves the ``forward problem'', i.e. given the properties of the monomers, chain parameters and network architecture, it enables us to compute the continuum electro-elastic free energy.
This provides the starting point to optimize over candidate micro-architectures.

\subsection{Formulation of the Potential Energy of a Microstate}

\hl{This formulation was first provided in Section 2 of{~\cite{grasinger2020statistical}}.
A more concise version is reproduced here for the sake of completeness and in the interest of being self-contained.

We consider a polymer chain, subject to an electric field, and composed of $\N$ identical monomers that each carry a dipole.
We use the ensemble with specified temperature $\T$, average electric field $\efield_0$, and end-to-end vector $\bfr$.
We assume that the chain is contained in a spatial volume $\Omega$ with boundary $\partial\Omega$ with unit outward normal $\hat{\bfm}$.

The degrees of freedom describing the configuration of the polymer chain are: (1.) the spatial position of the $i$-th monomer, denoted $\bfx_i$;
(2.) the orientation of the $i$-th monomer, denoted $\nhat_i$;
(3.) the point dipole carried by the $i$-th monomer, denoted $\bfmu_i$;
and (4.) the electric field, -$\nabla\phi(\bfx)$, which, for now, is a general function of position $\bfx$.
All of these can be varied independently.

We make the following assumptions in our model at this stage.
(1.) Following standard practice, we have implicitly assumed above that only the orientation of the monomers is relevant and that there is no stretching.
That is, stretching of monomers costs energy that is much larger than $kT$, and hence the chain is assumed to be inextensible.
(2.) Again following standard practice, we assume for simplicity that the bending energy -- i.e., energy associated with the change in orientation of the monomers along the chain -- is much less than $kT$ and can hence be neglected.
(3.) While dipole effects are due to electronic and nuclear motion and hence can have interactions between monomers {\cite{babaei2017computing}}, we assume for simplicity that the dipole induced in a monomer can be modeled without regard to the environment; equivalently, we assume that the electrical energy of the monomers can be decomposed additively.

Under these assumptions, the potential energy of a microstate can be written:}
\begin{equation}
\label{eqn:energy-1}
    U 
    = 
    \sum_{i=1}^{n} \left( \tilde u(\bfmu_i,\hat\bfn_i) - \bfmu_i \cdot \bfE_i \right) 
    +
    \frac{1}{2} \int_\Omega |\nabla\phi|^2 \dm\Omega
\end{equation}
\hl{
where we have defined $\bfE_i = -\nabla\phi \Bigr|_{\bfx_i}$ as the local electrical field at the location of the monomer, and where we have chosen to work in Gaussian units (i.e. unit system in which $\epsilon_0$ is unity).
The first term in the summation is the energy $\tilde u$ required to separate charges to form a dipole $\bfmu_i$, and will be discussed further in Section {\ref{sec:monomer-E-response}}.
The second term in the summation is the energy of interaction between the local electric field at the monomer location and the induced dipole.
The volume integral is the electrical field energy.}

\subsubsection{Dipole Response of a Single Monomer to an Electric Field}
\label{sec:monomer-E-response}
\hl{
Following the classical Born-Oppenheimer approximation, we assume that electrons reach their ground state configuration -- under electric field -- rapidly compared to the timescale of the thermal motion of the atoms}~\cite{yu2016energy}\hl{.
While we do not consider quantum effects explicitly, the Born-Oppenheimer approximation justifies neglecting thermal effects in modeling the dipole response of a single monomer to an electric field.
In other words, we are assuming that the first excited state of the electrons has energy much larger than $kT$, and the system is always in the ground state with respect to the electron configuration.

An important implication of this assumption is that the dipole moment of an individual monomer, $\bfmu_i$, is uniquely determined -- through energy minimization -- given $\hat\bfn_i$, $\bfE_i$, and $\bfx_i$.
To find an expression for $\bfmu_i$ in terms of the other quantities, we differentiate} \eqref{eqn:energy-1} \hl{with respect to $\bfmu_i$ to obtain the polarization response} $\takepartialflat{\tilde u}{\bfmu_i} = \bfE_i$ \hl{at the ground state.

Here we model the monomer response through the choice:}
\begin{equation} \label{eq:dipole-response}
    \tilde u(\bfmu_i,\hat\bfn_i)
    =
    \half \bfmu_i \cdot \bfchi^{-1}(\hat\bfn_i) \bfmu_i 
    \Rightarrow
    \bfmu_i = \bfchi(\hat\bfn_i) \bfE_i
\end{equation}
\hl{where $\bfchi$ is the tensorial polarizability of the monomer.
Following {\cite{cohen2016electroelasticity}}, we model the tensor $\bfchi$ as transversely isotropic: $\bfchi(\hat\bfn) = \chi_\parallel \hat\bfn\otimes\hat\bfn + \chi_\perp (\bfI - \hat\bfn\otimes\hat\bfn)$.
The non-negative material constants $\chi_\parallel$ and $\chi_\perp$ are measures of the susceptibility along the monomer direction and transverse to the monomer direction respectively.
We refer to monomers with $\chi_\parallel > \chi_\perp$ as uniaxial, and monomers with $\chi_\parallel < \chi_\perp$ as transverse isotropic (TI).
For this choice of $\bfchi$, we have the ground state energy, $u = \tilde u - \bfmu_i \cdot \bfE_i$, as:}
\begin{equation}
\label{eq:monomer-energy}
    u(\hat\bfn_i, \bfE_i) 
    =
    -\half\bfmu_i(\hat\bfn_i,\bfE_i)\cdot\bfE_i
    =
    \half \Delta\chi \left( \bfE_i\cdot\hat\bfn \right)^2 - \half \chi_\perp |\bfE_i|^2
\end{equation}
\hl{where $\Delta\chi = \chi_\perp - \chi_\parallel$.

If $\chi_\parallel > \chi_\perp$, i.e. the monomer is uniaxial, then the monomer has minimum energy when $\hat\bfn$ is parallel or anti-parallel to $\bfE_i$, and maximum energy when $\hat\bfn$ lies in the plane orthogonal to $\bfE_i$.
If $\chi_\parallel < \chi_\perp$, i.e. the monomer is TI, the situation is reversed: the minimum energy state is when $\hat\bfn$ lies in the plane orthogonal to $\bfE_i$, and the maximum energy state is when $\hat\bfn$ is parallel or anti-parallel to $\bfE_i$.}


\subsubsection{Multiscale Structure of the Electrical Field Energy, and Consequent Nonlocal-to-Local Decoupling} \label{sec:nonlocal}

\hl{Under the assumptions in Section {\ref{sec:monomer-E-response}}, the potential energy from} \eqref{eqn:energy-1} \hl{reduces to:}
\begin{equation}
\label{eqn:energy-2}
    U 
    = 
    \sum_{i=1}^{n} \left( - \half \bfE_i \cdot \bfchi(\hat\bfn_i) \bfE_i \right) 
    +
    \frac{1}{2} \int_\Omega |\nabla\phi|^2 \dm\Omega.
\end{equation}

\hl{The energy posed in} \eqref{eqn:energy-2} \hl{has a highly nonlocal structure} \cite{marshall2014atomistic,james1990frustration,yang2011completely,liu2013energy}.
\hl{Physically, this is due to the fact that we need to solve the electrostatics equation to find the field at every monomer location in $\Omega$.
We can see this by examining the ground state with respect to the electric potential.
Taking the variation $\phi \to \phi + \psi$ and requiring this to be $0$ for all variations $\psi$, we find that:}
\begin{equation}
\label{eqn:nonlocal-electrostatics}
\begin{split}
    0 = &
    \sum_{i=1}^{n} - \bfchi \left. \bfE \right|_{\bfx_i} \cdot \left. \nabla\psi\right|_{\bfx_i} + \int_\Omega \nabla \phi \cdot \nabla \psi \dm\Omega
    \\
    & \Rightarrow 0 =
    \int_\Omega \sum_{i=1}^{n} \left(- \bfchi \bfE \cdot \nabla\psi \right) \delta_{\bfx_i} \dm\Omega + \int_\Omega \nabla \phi \cdot \nabla \psi \dm\Omega
    \\
    & \Rightarrow 0 =
    \int_\Omega \psi \divergence
    \underbrace{\left[ \sum_{i=1}^{n} \left( \bfchi \bfE \delta_{\bfx_i} \right) \right]}_{\tilde\bfp(\bfx)}
    \dm\Omega - \int_\Omega \psi \divergence \nabla \phi \dm\Omega
    \\
    & \Rightarrow \divergence \nabla \phi = \divergence \tilde\bfp, \quad \text{ subject to boundary conditions.} 
\end{split}
\end{equation}
\hl{Here, $\delta_{\bfx_i}$ is the Dirac mass located at $\bfx_i$, and $\tilde\bfp(\bfx)$ is the dipole moment of the chain, treated here as a field through the use of Dirac masses that represent the point dipoles carried by the monomers, to be interpreted in the sense of distributions.}

\hl{This final equation shows the nature of the nonlocal problem: to evaluate the energy in} \eqref{eqn:energy-2}, \hl{we need to solve a boundary value problem.}
\hl{Therefore, statistical mechanical averaging over the energy will need to average over all fields $\phi$ that are consistent with the specified average electric field ensemble.
This is challenging and would require methods of statistical field theory.}


\hl{Instead, we begin by assuming that the energy in} \eqref{eqn:energy-2} \hl{has a separation of scales such that the energy in forming a dipole by separating charges is of order $kT$, while the stored energy of the field in vacuum is of much higher order.}
\hl{This allows us to first minimize with respect to the field energy to find the electric field, and then use this electric field as a constraint in performing the statistical averaging.}

\hl{The potential energy, with the separation of scales explicitly highlighted, is:}
\begin{equation}
\label{eqn:energy-3}
    U 
    = 
    \underbrace{
        \sum_{i=1}^{n} \left( - \half \bfE_i \cdot \bfchi(\hat\bfn_i)  \bfE_i \right)
    }_{\sim kT}
    +
    \underbrace{
        \frac{1}{2} \int_\Omega |\nabla\phi|^2 \dm\Omega
    }_{\gg kT}
\end{equation}
\hl{We first minimize the field energy, of order much greater than $kT$, by setting the first variation to zero to obtain $\divergence \nabla \phi = 0$.}
\hl{We find that $-\nabla\phi = \bfE_0$, where $\bfE_0$ is a fixed (average) electric field specified by the ensemble (see} ~\cite{grasinger2020statistical} \hl{for a more detailed discussion).}

\hl{The final form of the reduced potential energy that we will use for statistical averaging reads simply:}
\begin{equation}
\label{eqn:energy-4}
    U 
    = 
   \sum_{i=1}^{n} \left( - \half \bfE_0 \cdot \bfchi(\hat\bfn_i)  \bfE_0 \right).
\end{equation}
\hl{For simplicity, moving forward, we will drop the subscript zero and let it be understood that $\bfE$ is a locally average electric field at a chain or a material point.}

\hl{We highlight two additional important simplifications that are a consequence of the assumption of separation of energy scales.
First, the degrees of freedom required to describe a microstate are vastly reduced from our general starting point.
In particular, the spatial position plays no role, as the electric field is uniform in space; the point dipole is completely specified given $\bfE$ and the orientation of the monomer; and the electric field is also completely specified $-\nabla\phi = \bfE$.
Therefore, the monomer orientations $\hat\bfn_i$ are the only remaining degrees of freedom over which to conduct statistical averaging.}
\hl{Second, the simplification from having to solve a nonlocal boundary value problem to determine $\nabla\phi$ makes it possible to perform a Legendre transform to go between the free energy written as a function of $\bfE$, which is relatively simpler to evaluate using statistical mechanics, and the free energy written as a function of $\bfp$, which is more convenient for applications since it has a minimum rather than a saddle-point structure}~\cite{grasinger2020statistical}.
\hl{We will see later on that this has implications regarding Legendre transforms of the continuum-scale free energy density as well.}

\subsection{Statistical Mechanics of a Single Electro-responsive Polymer Chain}

Given the potential energy of a microstate, we derive the free energy in \cite{grasinger2020statistical}.
To this end, we derive a mean-field theory and determine that the density of monomers oriented in the direction $\nvec$ is given by
\begin{equation} \label{eq:density}
	\density\left(\nvec\right) = \C \exp\left(-\unodim \left(\edir \cdot \nvec\right)^2 + \mults \cdot \nvec\right)
\end{equation}
where $\unodim = \emag^2 \dsus / 2 \kB \T$, and the unknowns, $\C$ and $\mults$, are determined by enforcing the constraints
\begin{equation}
	\label{eq:ccn} \N = \intoverSns{\density\left(\nvec\right)}, \quad \frac{\rvec}{\mlen} = \intoverSns{\density\left(\nvec\right) \nvec} 
\end{equation}
where $\unitsphere$ denotes the surface of the unit sphere, $\N$ is the number of monomers in the chain, $\rvec$ is the end-to-end vector, and $\mlen$ is the length of a single monomer.
In terms of the monomer density function, one can show that the free energy is approximately:
\begin{equation} \label{eq:A-approx}
	\A \approx \intoverSns{\left( \density \um + \kB \T \density \ln \density\right)} - \N \kB \T \ln \N
\end{equation}

There are still two remaining difficulties in solving for $\C$ and $\mults$: (1.) the integrals in \eqref{eq:ccn} are difficult to evaluate and (2.) the resulting systems of equations are nonlinear.
However, there are two limits in which a closed-form solution can be obtained.
Let $\stch = \rmag / \N \mlen$ denote the absolute chain stretch, where $\rmag = |\rvec|$.
In the limit of $\stch \rightarrow 0$, we have that $|\mults| \rightarrow 0$.
Thus, one can obtain a solution in the limit of small stretch by using a Taylor expansion on the $\mults \cdot \nvec$ term in the exponential of the monomer density function ($\density$ given by \eqref{eq:density})  up to linear order.
Similarly, $\density$ takes a simpler form when $\stch \rightarrow 1$.
Physically, this is because when the chain approaches the fully stretched limit, the distribution of possible monomer orientations are narrowly centered around $\rdir := \rvec/\rmag$; and in this case, all of the possible orientations are of approximately the same energy.
Hence the energy term, $-\unodim \left(\edir \cdot \nvec\right)^2$, in the exponential of $\density$ can be neglected.
Neglecting the energy term, one obtains the Kuhn and Gr\"{u}n~\cite{kuhn1942beziehungen} solution.
Let $\Asm$ denote the approximation of the free energy using the small stretch solution and similarly, let $\Akg$ denote the approximation near the fully stretched limit (i.e. the result of substituting the Kuhn and Gr\"{u}n density into \eqref{eq:A-approx}).
Then, using what is known about the limiting behavior, we construct a free energy approximation:
\begin{equation} \label{eq:asymptotic-A}
\begin{split}
	\A & = \Akg + \left( 1 - \stch^2\right)\left( \lim_{\stch \rightarrow 0} \Asm - \lim_{\stch \rightarrow 0} \Akg \right)
	\\
	& = \N \kB \T 
	\Bigg(
	-\uOnodim +  \stch \zmultzero + \log \left(\frac{\zmultzero \csch \left[\zmultzero\right]}{4 \pi}\right)
	+ \frac{\stch \unodim}{\zmultzero} 
	\\
	& \qquad \hspace{7mm} + \left(1 - \stch^2 \right) \left[-\frac{\unodim}{3} + \log \left(\frac{2 \sqrt{\unodim}}{\sqrt{\pi} \erfw}\right)\right]
	+ \unodim \left(1 - \frac{3 \stch}{\zmultzero}\right) \left(\edir \cdot \rdir\right)^2
	\Bigg)
\end{split}
\end{equation}
where $\uOnodim = \emag^2 \sus{2} / 2 \kB \T$, and $\Langinv$ is the inverse Langevin function.
Notice that \eqref{eq:asymptotic-A} recovers the exact solution when $\unodim = 0$ and is exact in the limits of zero stretch and full stretch.
This approximation has been shown to agree well with numerical experiments for a large variety of general chain conditions (e.g. stretches, orientations with respect to the electric field, $|\unodim|$, etc.)~\cite{grasinger2020statistical}.

Having obtained an approximation of the free energy, one can obtain the chain polarization, $\chainpolar$, by differentiating the free energy with respect to the electric field, which is equivalent to integrating the dipole moment over the monomer density function; i.e.
\begin{equation*} 
	\chainpolar \coloneqq \intoverSns{\dipole\left(\nvec\right) \rho\left(\nvec\right)} = -\frac{\partial \A}{\partial \efield}
\end{equation*}
The significance of the relationship between the partial derivative, $\takepartial{\A}{\efield}$ and the chain polarization is that it establishes the fact that $\A$ is the Legendre transform of $\AHelm = \AHelm\left(\rvec, \chainpolar\right)$, that is, the Helmholtz free energy of the chain.
The derivation of this relationship is given in \Fref{app:polarization}.

We note an important feature of the free energy-stretch relationship of electro-responsive chains: numerical experiments suggest that the $\A / \kB \T$ vs $\stch$ curve is convex and its minimum is at zero stretch, implying that a chain will not stretch when subject to electric field.
Therefore, the additional electrostriction that occurs in dielectric elastomers when the permittivity is deformation-dependent cannot be explained by the notion that individual chains spontaneously stretch or contract within the network due the applied field\footnote{By additional electrostriction, we mean the contribution to electromechanical coupling that is not due to the Coulomb attraction between the electrodes.}.
Physically, this feature of the $\A/\kB \T$-$\stch$ relationship can be understood as a consequence of: (1.) the electrostatic energy of the monomers being quadratic in $\efield \cdot \nvec$ (see \eqref{eq:monomer-energy}) and (2.) the assumption that monomer-monomer interactions are negligible.
This means that if a monomer's orientation is reversed (i.e. $\nvec \rightarrow -\nvec$), its energy, and hence, its contribution to the Boltzmann factor is the same.
And since there is no energy penalty associated with large or small bond angles between neighboring monomers, the chain is free to fold back on itself.
So in terms of the Boltzmann factor, a longer end-to-end vector is never any more favorable than a shorter end-to-end vector.
However, in terms of entropy, the shorter end-to-end vector is more favorable. 
For these reasons, the free energy versus stretch relationship for an electro-responsive polymer chain is expected to be convex with its minimum at zero stretch.

\subsection{Orientational Averaging Over a Polymer Chain Network}

In order to relate the continuum scale deformation to individual chains in the network, we use the affine deformation assumption~\cite{treloar1975physics} and the full network model~\cite{wu1993improved,beatty2003average}.

At each material point in the reference (stress-free) state, it is assumed that chains have their most probable length, $\mlen \sqrt{\N}$, and are randomly oriented.
For instance, consider a material that is isotropic due to the chain orientations being uniformly distributed.
That is, the probability density function of finding a chain with end-to-end vector, $\Rvec$, is $\chainpdf\left(\Rvec\right) = \frac{\diracdel\left(|\Rvec| - \mlen \sqrt{\N}\right)}{4\pi}$.

By the affine deformation assumption, each chain gets mapped from the reference configuration to the current configuration by the deformation gradient, $\F$, i.e. $\rvec = \F \Rvec$.
Finally, the free energy density is taken to be the product of the average chain free energy and the number of chains per unit volume, $\chaindensity$:
\begin{equation} \label{eq:energy-density}
  \Wstar\left(\F, \efield\right) = \chaindensity \avg{\A\left(\F \Rvec, \efield\right)} = \chaindensity \int \df \Rvec \; \chainpdf\left(\Rvec\right) \A\left(\F \Rvec, \efield\right),
\end{equation}
where $\chaindensity$ is the number of chains per unit volume and $\avg{\cdot}$ denotes an average over the distribution of chains.

Similarly, we can obtain the (continuum-scale) polarization, i.e. dipole moments per unit volume, $\bfP$ by:
\begin{equation*}
  \bfP\left(\F, \efield\right) = \chaindensity \avg{\chainpolar\left(\F \Rvec, \efield\right)} = \chaindensity \int \df \Rvec \; \chainpdf\left(\Rvec\right) \chainpolar\left(\F \Rvec, \efield\right).
\end{equation*}

Appendix \ref{app:GI} discusses an approximate closed-form expression to evaluate the integration in \eqref{eq:energy-density} for a commonly-used class of $\chainpdf$.
In general, numerical integration of \eqref{eq:energy-density} can be challenging to perform accurately~\cite{verron2015questioning}.

\subsection{Continuum Electroelasticity} \label{sec:continuum}

Since the pioneering work of~\cite{toupin1956elastic}, there has been a lot of work recently in variational methods and formulations for continuum electroelasticity~\cite{toupin1956elastic,suo2008nonlinear,suo2010theory,bustamante2009nonlinear,dorfmann2006nonlinear,dorfmann2014nonlinear,liu2013energy,liu2014energy}.
For the theory and analysis of the stability of DEAs using continuum models, see~\cite{zurlo2017catastrophic,yang2017revisiting,zhao2007method}.

Let $\rbody$ denote the body in the reference configuration and $\cbody = \rtocmap\left(\rbody\right)$ the body in the current configuration, where $\rtocmap$ is the deformation mapping.
The position of a material point in the reference configuration, $\Pos$, is mapped to a position in the current configuration by the deformation mapping, i.e. $\pos = \rtocmap\left(\Pos\right)$; and the deformation gradient is given by $\F = \GradOf{\rtocmap}$.

Then given some boundary conditions, $\bcs$, and some external work, $\extwork$,
the equilibrium configuration of the body is given by the minimization of the Gibbs free energy such that the boundary conditions are satisfied; that is:
\begin{equation} \label{eq:gibbs-min}
\begin{split}
\gibbs\left[\F, \Polar\right] =& \int_{\rbody} \W\left(\F, \Polar\right) \df V - \extwork \\
\left\{\eql{\F}, \eql{\Polar}\right\} =& \arg \min_{\F, \Polar} \text{ } \gibbs\left[\F, \Polar\right] \text{ subject to } \bcs
\end{split}
\end{equation}
Thus, the constitutive response is encoded in the form of the Helmholtz free energy density function, $\W$, which is given by
\begin{align} \label{eq:W}
\W\left(\F, \Polar; \T\right) &= \Wstar\left(\F, \efield; \T\right) + \J^{-1} \Polar \cdot \efield \\
\label{eq:Wstar}
\Wstar &= \chaindensity \avg{ \A\left( \F \Rvec, \efield; \T \right) } = \chaindensity \int_{\rspace} \df^3 \Rmag \text{  } \chainpdf\left(\Rvec\right) \A\left( \F \Rvec, \efield; \T \right)
\end{align}
where $\Polar = \J \polarization$ is the pullback of the (continuum-scale) polarization\footnote{
    Note that the definition of the pullback, $\Polar$, is not unique, see, e.g., the discussion in section 4.1.1 of~\cite{marshall2014atomistic}. We make this choice for $\Polar$ because of its simplicity and note that others have used the same definition (see, e.g.,~\cite{yang2017revisiting,liu2014energy}). In general, when deriving the free energy density from a statistical mechanics description, one derives the free energy density in the current configuration and then, once a choice of variables has been made, can formulate the free energy density in terms of the pullback quantities by substitution.
}, and $\Wstar$ is the Legendre transform (in the $\Polar$ slot) of the Helmholtz free energy density.
We will denote $\Wstar$ as the closed-dielectric free energy; more precisely, it is the Gibbs free energy when the dielectric material can be considered as a closed system in a constant electric field.
Physically, the significance of the closed-dielectric free energy is that it has a minimum principle at constant temperature and constant electric field.
See \cite{grasinger2020statistical} Section 3.3 for further discussion of this issue.

Finally, recall that because of the long-range nature of dipole-dipole interactions, the relationship between $\efield$ and $\Polar$ is nonlocal; they are related by the electrostatic equation.
Therefore, in general, the Legendre transform cannot be taken in terms of the free energy density, rather this must be done at the level of the total free energy, which is a formidable challenge as well as highly problem-specific.
However, when the monomer-monomer dipolar interactions are neglected, the nonlocal relationship becomes local \hl{(recall Section }\ref{sec:nonlocal} \hl{of this paper}, or see Section 2.2 of \cite{grasinger2020statistical}) and the transform taken in \eqref{eq:W} is indeed valid.

\subsection{Electromechanical Coupling Due to Chain Torque} \label{sec:electrostriction}

We highlight some important physical implications of the model outlined above.
We notice that an applied electric field applies a moment to individual polymer chains, due to the fact that the chain carries an effective dipole.
The importance of this is two-fold: (1.) it contributes to the stress beyond that imposed by the Coulombic attraction of the electrodes, and (2.) it provides a means to engineer the orientational distribution of chains so as to obtain a desired network architecture.

\begin{figure}
	\includegraphics[clip, trim=1.0cm 1.0cm 1.0cm 1.0cm, width=\linewidth]{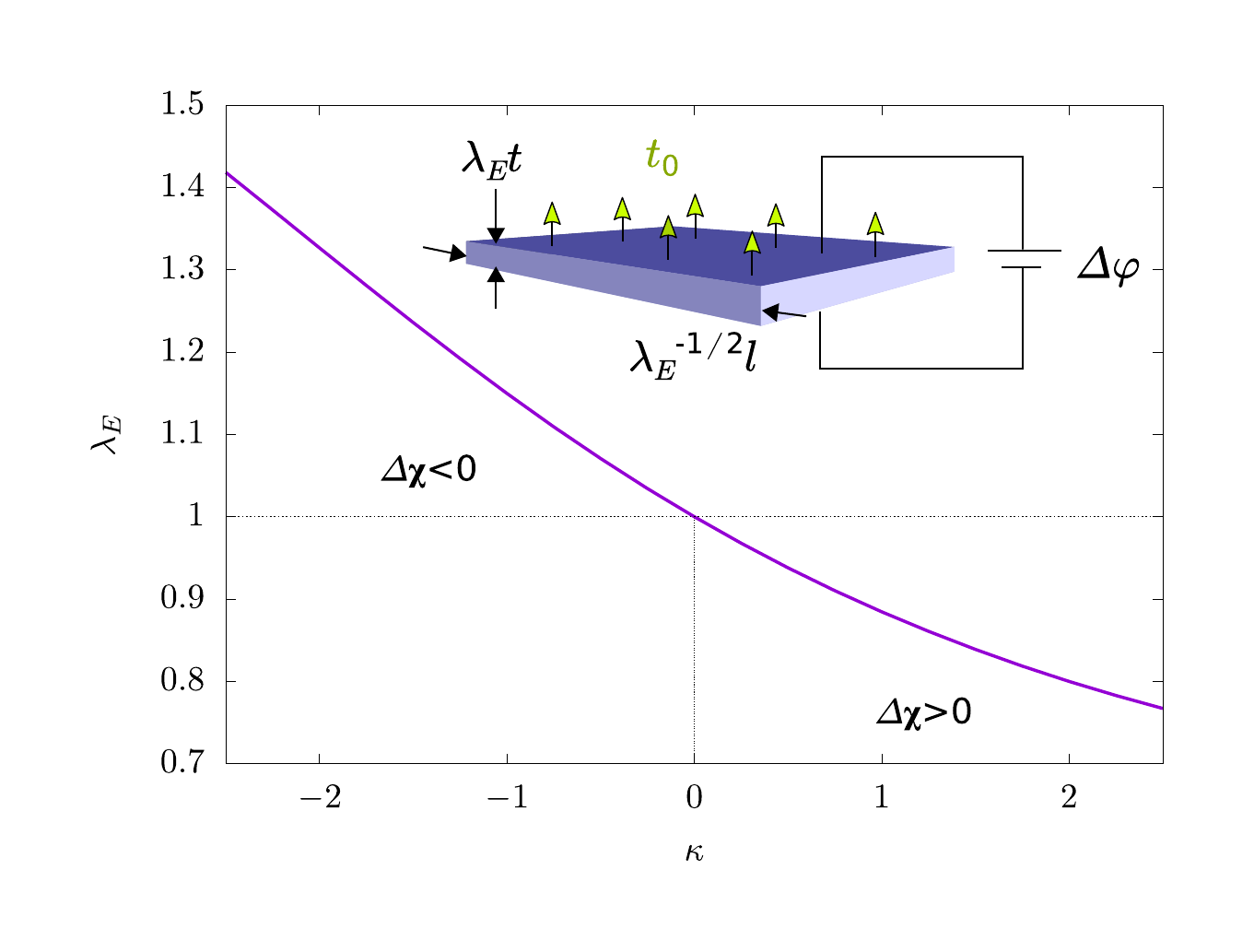}
  \caption{Electrostriction of a DEA with a fixed bottom surface and an applied traction, $\tvec$, to the top surface of the actuator that is equal and opposite to the Coulomb attraction (inset). The plot shows the stretch across the thickness, $\pstche$, as a function of $\unodim$.}
	 \label{fig:electrostriction}
\end{figure}

(1.) represents an opportunity to design for enhanced electromechanical coupling.
Specifically, one could potentially tune the network architecture such that the deformation is increased due to contributions from (1.).
To illustrate (1.), we consider a typical DEA, but with the stress caused by the electrodes counteracted.
This counteraction could be achieved, for instance, by applying a traction, $\tvec$, to the outside surfaces of the electrodes that is equal and opposite to the Coulomb attraction and by ensuring the voltage difference is adjusted to keep the electric field constant when the distance between the electrodes changes.
This is shown in \fref{fig:electrostriction} (inset).

In this case, the change in energy stored in the electric field (or equivalently, the work of the battery) and the work of the applied traction will cancel each other out; as a result, the equilibrium configuration will be the one that minimizes the free energy of the DE film.
We neglect the fringe fields and assume that the electric field is uniform in space, implying that the electric field is then aligned along the thickness direction.
Further, assume homogeneity, isotropy, and incompressibility; the stretch in the thickness direction is denoted $\pstche$, implying that the stretch in any in-plane direction is $\pstche^{-\half}$.
Evaluating the free energy for this class of deformations using \eqref{eq:W} with uniform orientational distribution $P = 1/4\pi$, we then minimize to find the energy-minimizing stretch $\pstche$.
\Fref{fig:electrostriction} shows $\pstche$ as a function of $\unodim$.
We see that the model predicts a spontaneous deformation of the film, even in the absence of pressure from the top and bottom electrodes.
When the chain is composed of TI monomers ($\unodim  > 0$), the film shrinks in the thickness direction (i.e. $\pstche < 1$).
Alternatively, when the chain is composed of uniaxial monomers ($\unodim < 0$), the film elongates in the thickness direction ($\pstche > 1$).


\section{Design Parameters} \label{sec:design-parameters}

In this section, we identify design parameters which describe the polymer network architecture.
That the design parameters constitute a description of the network is supported by the affine deformation assumption--which states that chain end-to-end vectors in the reference configuration are mapped to the current configuration by the deformation gradient--and the \emph{negligibly interacting assumption}--which allows us to decompose complex chain architectures in the network into separate linear parts.

The key variables that we aim to optimize over are related to the network connectivity and geometry, and the orientational distribution of the chains.
These aspects are relatively feasible to control using applied electromagnetic fields or stresses~\cite{ambulo2017four}.
In contrast, we treat the dielectric response, specified here by $\sus{1}, \sus{2}$ of individual monomers as a fixed and given, as modifying these characteristics require molecular-scale chemical methods that are much less developed in terms of being able to achieve target properties.

The following is a list of the design variables: mass density, i.e. $\chaindensityref \N$ where $\chaindensityref = \J \chaindensity$ is the number of chains per unit volume in the reference configuration; fraction of loose-end monomers, $\lefrac$; density of cross-links; and orientational distribution of chains, $\chainpdf$; which will be introduced shortly.


\subsection{Mass Density}

Since we take $\sus{1}$ and $\sus{2}$ to be fixed and, for many applications, we would prefer a high polarization susceptibility, we may be tempted to pack as many monomers per unit volume, $\chaindensityref \N$, into the polymer network as possible.
There are, however, some subtleties associated with this strategy.
For one, since there is a mass associated with each monomer, this would serve to increase the mass density of the material; however, it is likely that we do not want the density of the material to exceed a certain threshold for many of the applications of interest (e.g. wearable sensors, soft robotics, etc.).
Similarly, packing more monomers into the network may also affect the compliance of the material.
Therefore we will be interested in the product $\chaindensityref \N$ as a tunable parameter; and, specifically, its influence on, and interplay between, the aforementioned bulk properties.


\subsection{Negligibly Interacting Assumption} \label{sec:weakly}

Two systems $A$ and $B$ are in \emph{weak interaction} with each other if the Hamiltonian of the combined system $\Ham$ can be decomposed as $\Ham = \Ham^{A} + \Ham^{B} + \Ham^{A \leftrightarrow B}$ with $|\Ham^{A \leftrightarrow B}|^2 \ll |\Ham^{A}| ~ |\Ham^{B}|$.
Here $\Ham^{A}, \Ham^{B}$, are the Hamiltonians of systems $A, B$ respectively; and $\Ham^{A \leftrightarrow B}$, is a correction term due to interactions between system $A$ and system $B$ (see~\cite{tadmor2011modeling} section 7.3.3, for example).
In this work, we assume that the chains of the network are in weak interaction with each other such that their interactions are negligible. 
Because $\Ham^{A \leftrightarrow B}$ is negligible, we neglect it completely.
As we will see, in the negligible interacting regime, one can characterize many different chain and network architectures in a general and straightforward way.


\subsection{Individual Chain Geometry: Backbone Monomers, Loose-end Monomers, and Linear Chains}

\begin{figure}
	\centering
	\includegraphics[width=1.0\linewidth]{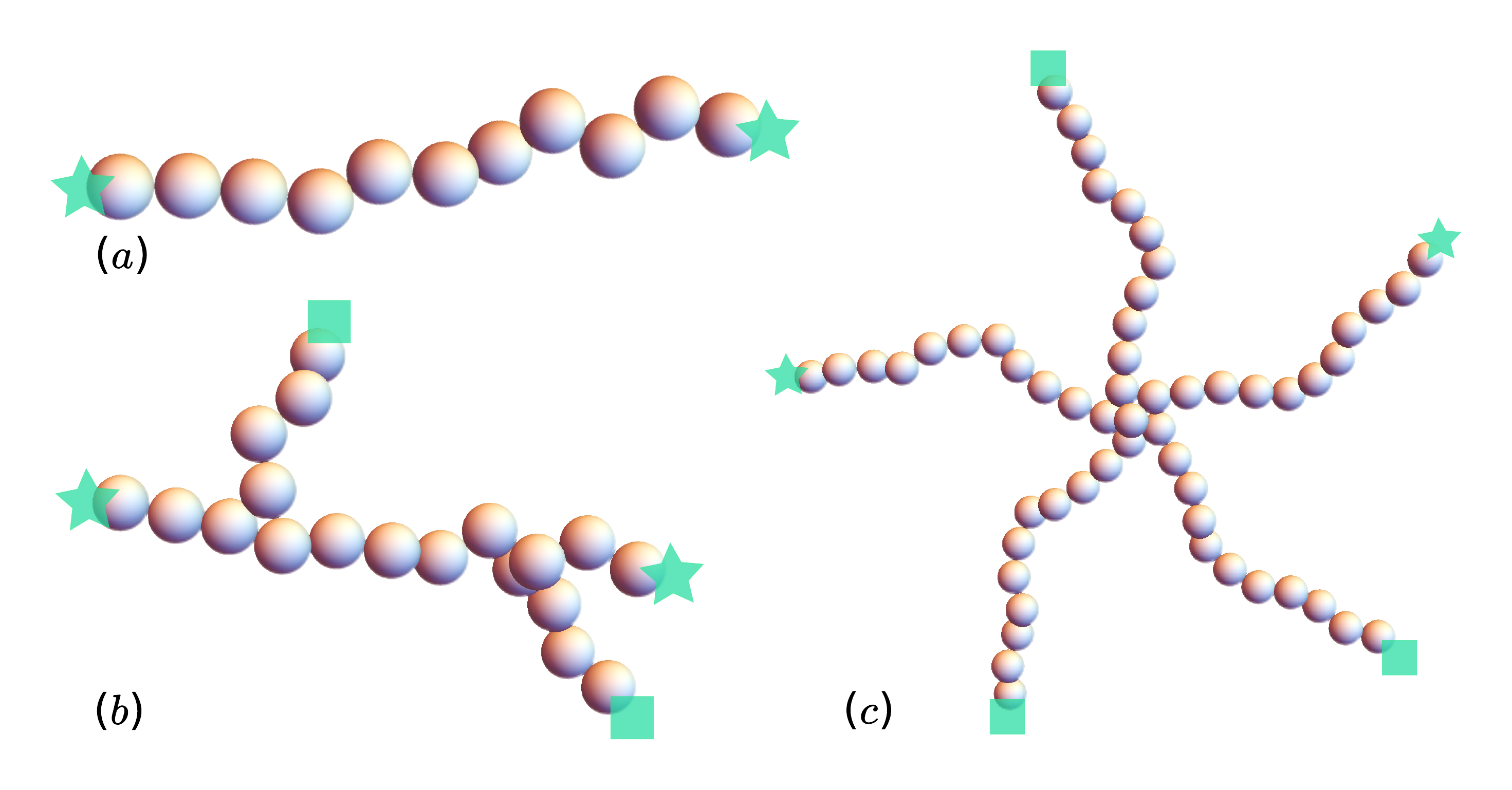}
	\caption{(a) Linear polymer chain (b) (linear) block copolymer chain (c) branched polymer (d) star polymer}
	\label{fig:polymer-chain-architectures}
\end{figure}

A key design parameter is the architecture of the chains that make up the cross-linked network.
Specifically, we mean the geometrical features of the chain, the types of monomers the chain consists of, and the various sequences that the monomer types are arranged in.
\Fref{fig:polymer-chain-architectures} shows a few examples of polymer chain architectures.
\hl{
We divide the monomers into two groups:
(1.) $n_b$ monomers that contribute to the backbone of the chain; and (2.) $\nl$ monomers that are in the loose ends of the network
}\footnote{
  In~\cite{flory1944network}, a similar decomposition into backbone monomers and loose-end monomers was used to correct for so-called ``network defects'' when relating mass density to the elastic properties of rubbers.
}.
By backbone of the chain, we mean all segments that span from cross-link to cross-link; by loose ends, we mean the non-backbone segments.
This distinction allows us to consider monomers which interact through the end-to-end vector constraint separately from those that do not need to satisfy such a constraint.
\hl{To make this clear, consider the example architectures in} \fref{fig:polymer-chain-architectures}.
\hl{If there are cross-links at each of the stars in} \fref{fig:polymer-chain-architectures}.a, \hl{then we say this is a linear chain without any loose-end monomers (i.e. $\nb = 11$, $\nl = 0$).
In contrast, if a cross-link is missing at either star, then the chain only contributes to the average number of loose-end monomers in the network, (i.e. $\nb = 0$, $\nl = 11$).
Similarly, in the branched polymer given in (b), if there are cross-links at the stars but not the squares, then our decomposition says that this is a single chain with $\nb = 11$, $\nl = 8$.
In the last example, the star polymer, (c), if there are only cross-links at the stars, this results in a single chain with $\nb = 22$ and $\nl = 33$; whereas cross-links at both the stars and squares results in} \emph{five linear chains}, \hl{each with $\nb = 11$.
}

Having decomposed complex architectures into linear parts, we make the assumption that loose end monomers are in negligible interaction with the backbone monomers and that separate chains are in negligible interaction with each other\footnote{
    The assumption of negligibly interacting chains is often used in rubber elasticity, see~\cite{treloar1975physics} section 4.2.
}.
This reduces all chain architectures to collections of negligibly interacting linear chains with some fraction of loose ends, $\lefrac = \nl / \N$.
This is justified as follows.
First, in regards to the electrical energy of the system, we have assumed that all dipole-dipole interactions will be neglected.
Second, for the entropic contributions, monomers within a chain interact through the enforcement of the end-to-end vector constraint; however, loose end monomers, do not take part in this constraint, and obviously monomers in separate chains do not interact through such a constraint.
The negligible-interaction assumption could be violated if chains are ``too short'' such that a significant portion of monomers in the chain are in the vicinity of another chain; if long-range interactions are significant; or if excluded volume effects are important.

\hl{To simplify further, we assume that the chain architectures are uniform enough to model their behaviour in terms of average numbers of backbone monomers and loose-end monomers per chain}\footnote{\hl{This is similar to the assumption, which is typically implicitly made, that the network of chains can be accurately described by a single (average) $\N$ as opposed to specifying the number of monomers in each individual chain or taking $\N$ to be a random variable.}}.
\hl{In summary, we describe the network through the number of linear chains per unit volume, $\chaindensity$, and the (average) number of backbone monomers and loose-end monomers per chain.
Thus, $\chaindensity \nb$ is the number of backbone chains per unit volume; in other words, the number of monomers per unit volume which are related to continuum scale deformation through $\F$ and chain averaging.
Whereas, $\chaindensity \nl$ is the number of loose-end chains per unit volume; in other words, the number of monomers which only contribute to the polarization of the network and do not contribute at all to the elasticity of the network.}


\subsection{Density of Cross-links}

The junction point at which chains in the network are joined together are called (effective) cross-links. 
Specifically, given our decomposition of complex chain geometries into chains with linear backbones and loose-ends, we say that there are cross-links at the beginning and end of each linear backbone.
For a given mass density then, $\chaindensityref$ is proportional to, and $\N$ is inversely proportional to, the number of cross-links per unit volume.
Therefore the density of cross-links in the network is both a parameter which is known to be controllable~\cite{treloar1975physics} and will affect the electroelasticity of the architected network (by \eqref{eq:asymptotic-A} and \eqref{eq:Wstar}, for example).



\subsection{Chain Probability Distribution Function} \label{sec:pdf-philosophy}

The orientation of polymer chains with respect to the local electric field influences the electroelastic response of the network.
As a result of both the importance of orientation and our decomposition of complex chain architectures into (multiple) linear chains, we use the orientational distribution of these linear chains within the network as a design variable.
More specifically, let $\chainpdf = \chainpdf\left(\Rvec\right)$ denote a probability density function which describes the fraction of chains with reference end-to-end vector $\Rvec$.
The design space of $\chainpdf$ is the space of all probability measures on $\rspace$; and hence, an infinite dimensional space.
However, we reduce the dimensionality of the problem as follows.

First, we assume that the length of all chains depends only on $\N$ and $\mlen$; specifically, we assume $\left|\Rvec\right| = \mlen \sqrt{\N}$, which is the expectation of the length of a random walk of $\N$ steps with step length $\mlen$.
This choice is motivated by the fact that, prior to cross-linking, chains are free to move in a random manner.
While this choice is primarily motivated by classical polymer theory, it will also be discussed below when we discuss residual stresses.

Second, having reduced the support of $\chainpdf$ to the surface of a sphere of radius $\mlen \sqrt{\N}$, we next consider the symmetries inherent in the physical problem.
For instance, in regards to chain statistical mechanics: $\A\left(\rvec, ...\right) = \A\left(-\rvec, ...\right)$; or, in other words, although chains have well defined ends (cross-linking points), either end may be identified as the start and finish.
Therefore $\chainpdf$ need only be defined on a half sphere.

Third, we further reduce the dimensionality by considering the setting of a thin film dielectric elastomer actuator with the voltage applied across the thickness of the film.
Physically, the direction across the film thickness is significant.
We therefore restrict our attention to functions $\chainpdf = \chainpdf\left(\polar\right)$ which satisfy
\begin{equation} \label{eq:pdf-symm}
  \chainpdf\left(\polar\right) = \chainpdf\left(\pi - \polar\right)
\end{equation}
where $\polar$ is the polar angle with respect to the direction of the electric field, $\edir$.
This choice is consistent with the symmetry of the DEA and with the $\rvec \rightarrow -\rvec$ symmetry.
For materials with the above symmetries and the BVP associated with a thin film DEA, we expect a homogeneous deformation of the form $\F = \diag\left(1 / \sqrt{\pstchu}, 1 / \sqrt{\pstchu}, \pstchu\right)$.

\newcommand{\GaussExpr}[3]{\exp\left[-\frac{\left(#1 - #2\right)^2}{2 #3^2}\right]}
Fourth, we obtain a drastic dimension reduction by using the following as an ansatz of $\chainpdf$:
\begin{equation} \label{eq:TI-ansatz}
    \chainpdf\left(\azi, \polar\right) 
    = \chainpdf_{\polar}\left(\polar; \polarAvg, \Std\right) 
    = \C \left(\GaussExpr{\polar}{\polarAvg}{\Std} + \GaussExpr{\pi - \polar}{\polarAvg}{\Std}\right).
\end{equation}
where $\C$ is a normalization constant such that $\int \chainpdf = 1$.
Notice that this ansatz is consistent with a transversely isotropic material.

Some examples of this ansatz are shown in \fref{fig:example-pdfs}.
The reasons for this choice of ansatz are as follows: 
(1) the properties of Gaussian distributions are well understood; 
(2) ease of evaluating various integrations, e.g. \eqref{eq:Wstar}; 
(3) the symmetry given by \eqref{eq:pdf-symm} is satisfied; 
(4) it reduces our search space from infinite-dimensional to a two-parameter space spanned by ($\polarAvg$, $\Std$); and (5) it is consistent with heuristic ideas of modeling manufacturing tolerances; for instance, tolerances could be modeled by placing lower bounds on $\Std$.
\newcommand{\auxwidth}{0.45\linewidth}
\begin{figure}[htb!]
	\begin{tabular}{c c}
		\includegraphics[width=\auxwidth]{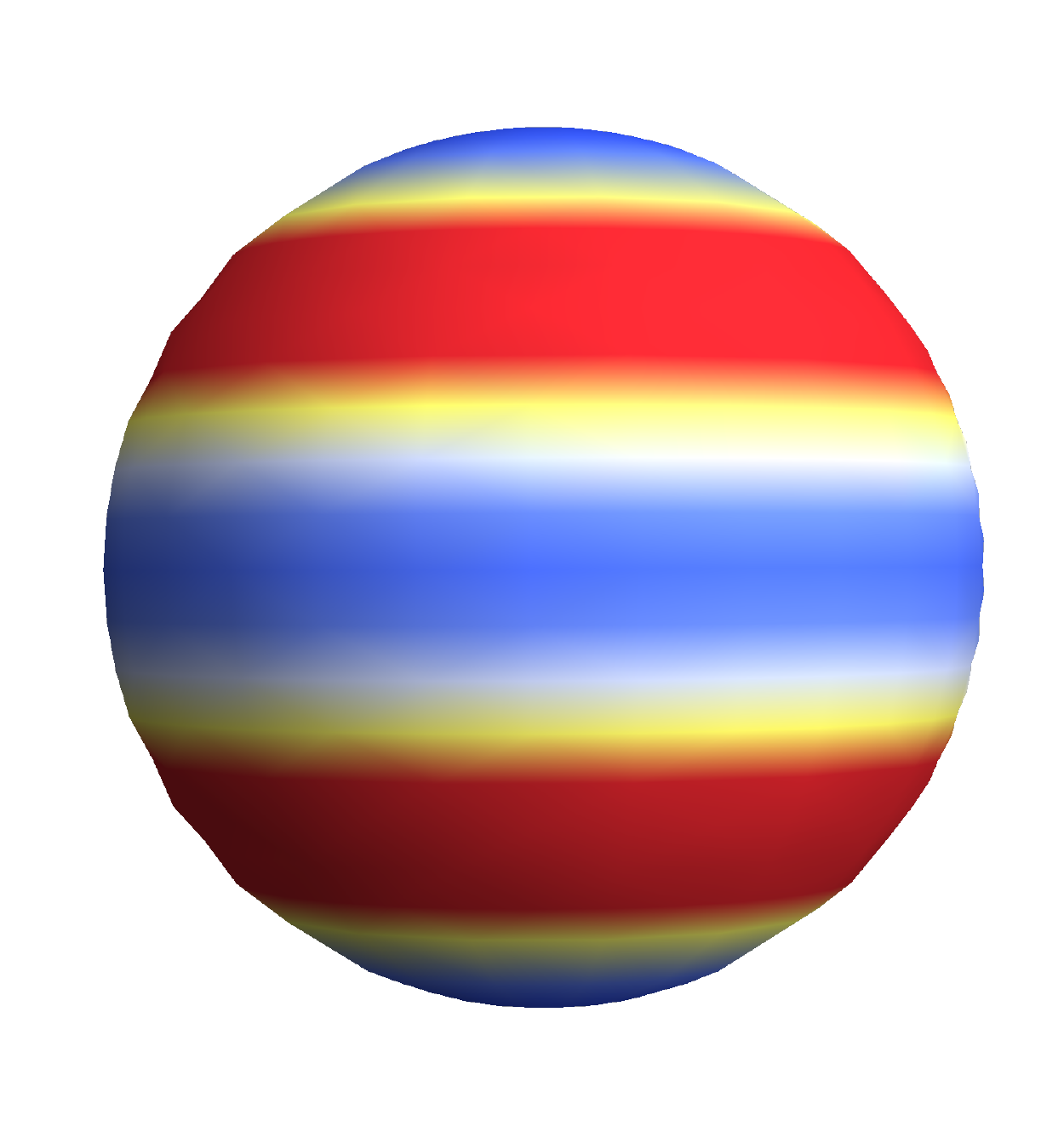} &
		\includegraphics[width=\auxwidth]{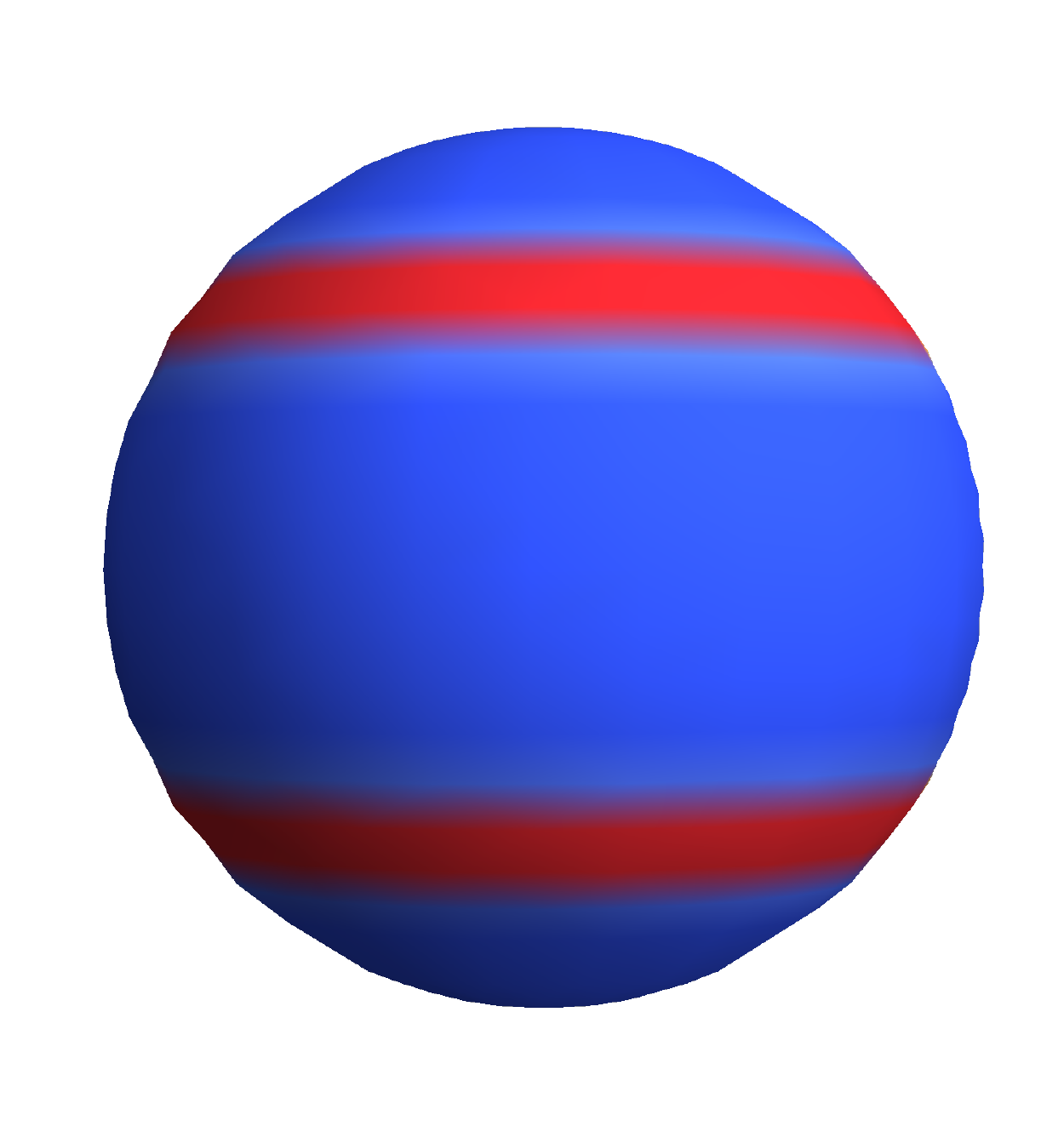} \\
		\includegraphics[width=\auxwidth]{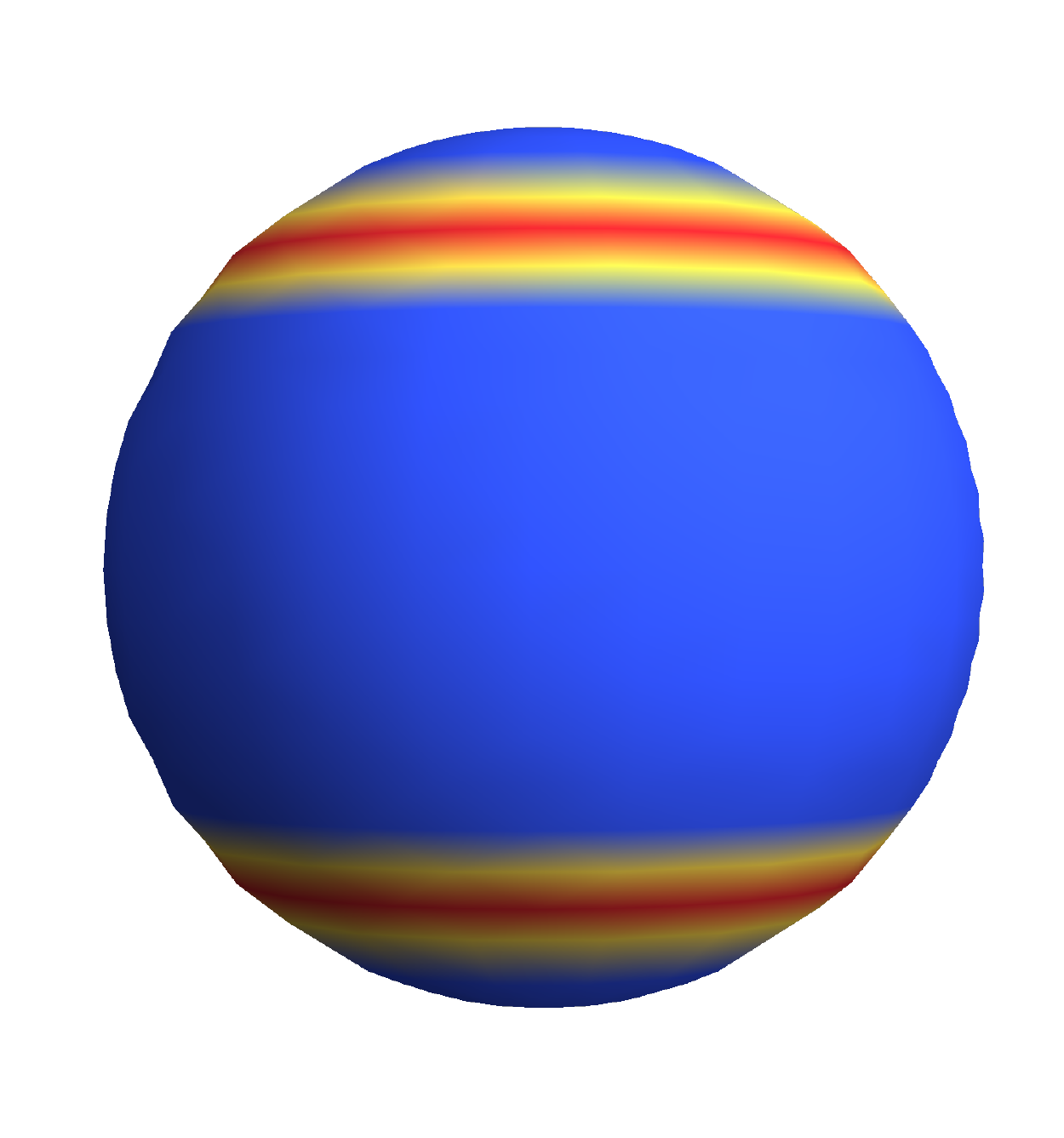} &
		\includegraphics[width=\auxwidth]{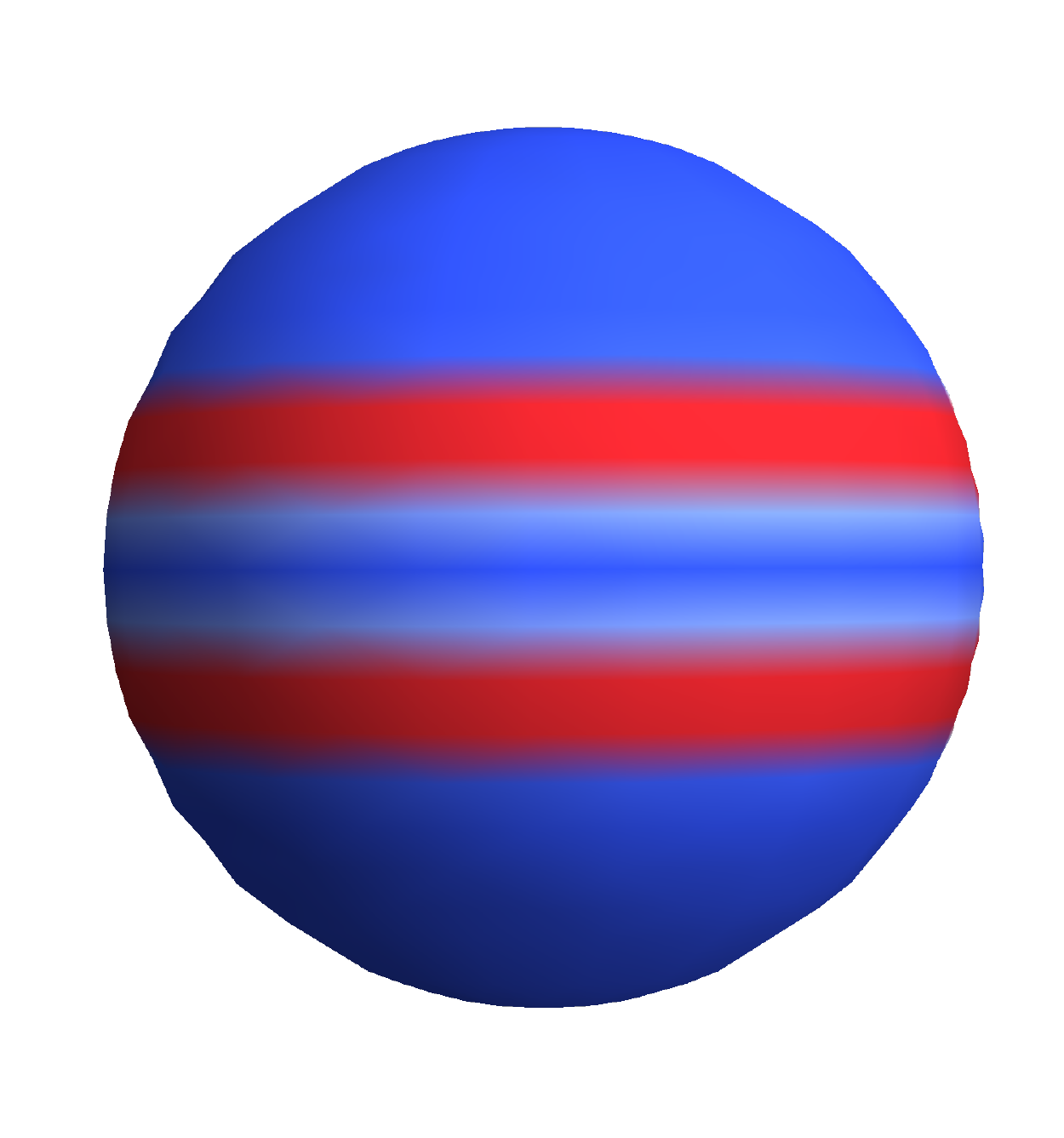}
	\end{tabular}
	\caption{Example chain pdfs of the form \eqref{eq:TI-ansatz}: $\polarAvg = \arctan \sqrt{2}$, $\Std = \pi / 8$ (top left), $\polarAvg = \arctan \sqrt{2}$, $\Std = \pi / 32$ (top right), $\polarAvg = \pi / 6$, $\Std = \pi / 32$ (bottom left), and $\polarAvg = 5 \pi / 12$, $\Std = \pi / 32$ (bottom right). Red signifies higher density and blue signifies lower density. The chain pdfs, roughly speaking, consist of a ring of width $\Std$ at angle $\polarAvg$ with respect to the polar axis (i.e. axis of symmetry) and reflected about the equator ($\polar = \pi / 2$) of the unit sphere.}
	\label{fig:example-pdfs}
\end{figure}

In regards to manufacturing for a design $\chainpdf$: we envision controlling $\chainpdf$ either by cross-linking while under an applied electric field (i.e. taking advantage of chain torque) and/or applied stresses; or by an advanced type 3D printing~\cite{ambulo2017four}.

\subsection{Reference States to Define Material Response}
\label{sec:reference-states}

Consider a uniform distribution $\chainpdf = 1 / 4\pi$.
In the absence of external field and external mechanical load, assuming incompressibility gives from symmetry that there is no deformation at equilibrium, i.e. $\eql{\F} = \iden$.

Interestingly, this is not the case for a general choice of $\chainpdf$, i.e. the material is not at equilibrium for $\F = \iden$ in the absence of external loads.
This can be interpreted as a residual stress introduced during manufacturing.
We use the relaxed Helmholtz energy minimizing deformation, denoted $\rstate{\F}$, to define linearized material response properties that we will aim to optimize.

It is for this reason that we remarked earlier that assuming fixed $|\Rvec| = \mlen \sqrt{\N}$ would be further justified.
This justification lies in the fact that it is apparent that chain lengths and orientations cannot be controlled independently of each other--they are related through the relaxation of the residual stresses.
Indeed, after $\rstate{\F}$ the chain lengths in the network will no longer be homogeneous, but will depend on $\Rvec$ (i.e. $|\rstate{\F} \Rvec| \neq \text{const.}$).

Considering the Taylor expansion of the free energy density about the relaxed configuration $\left\{\F = \rstate{\F}, \Polar = \bf0\right\}$:
\begin{equation} \label{eq:W-taylor}
\begin{split}
    \W\left(\F, \Polar\right) - \W\left(\rstate{\F}, \bf0\right) 
    = & \takepartial{\W}{F_{ij}} \left(\F - \rstate{\F}\right)_{ij} + \takepartial{\W}{\Polarmag_k} \Polarmag_k 
    \\
    & 
    + \frac{1}{2} \left.\takecrosspartial{\W}{F_{ij}}{F_{kl}}\right|_{\rstate{\F},\bf0} \left(\F - \rstate{\F}\right)_{ij} \left(\F - \rstate{\F}\right)_{kl}
    \\
    &
    + \left.\takecrosspartial{\W}{F_{ij}}{\Polarmag_{k}}\right|_{\rstate{\F},\bf0} \left(\F - \rstate{\F}\right)_{ij} \Polarmag_k 
    + \frac{1}{2} \left.\takecrosspartial{\W}{\Polarmag_i}{\Polarmag_j}\right|_{\rstate{\F},\bf0} \Polarmag_i \Polarmag_j 
    \\ 
    & + \bigoof{\smallparam_1^3} + \bigoof{\smallparam_1^2 \smallparam_2} + \bigoof{\smallparam_1 \smallparam_2^2} + \bigoof{\smallparam_2^3}
\end{split}
\end{equation}
where $\smallparam_1 = |\F - \rstate{\F}|$ and $\smallparam_2 = |\Polar|$.
We assume $\W$ is convex in $\Polar$ and its minimum is $\Polar = \mathbf{0}$, justified empirically for general DEs.
Therefore, the linear terms in \eqref{eq:W-taylor} vanish.
The constitutive response of the DE is governed by $\takecrosspartial{\W}{\F}{\F}$, $\takecrosspartial{\W}{\Polar}{\Polar}$, and $\takecrosspartial{\W}{\F}{\Polar}$.
These quantities correspond to the stiffness tensor, the inverse of the polarization susceptibility tensor (i.e. $\susceptibilityTensor^{-1}$)--which is a measure of the stiffness of the atomic bonding of charges bound to the DE--and the cross modulus tensor, respectively.
The magnitude of the cross modulus signifies the intrinsic electromechanical coupling of the material; that is, the electromechanical coupling irrespective of external loads or Coulombic attraction of external electrodes.
Thus, we will be interested in how our design variables affect these three quantities.
However, we will not, strictly speaking, seek to optimize all of these properties.
In particular, stability and positive definiteness of the energy near the equilibrium state bounds the cross modulus from above by, roughly, the geometric mean of the susceptibility and the stiffness.
Therefore, a practical strategy is to keep the stiffness from getting too high or too low, while optimizing the susceptibility and the cross modulus.

\subsection{Incompressibility}

As a final note to this section, we make a few remarks regarding the incompressibility of our hypothetical, proposed, anisotropic electroresponsive elastomer.
While most elastomers are incompressible to a very good approximation and are often modeled as such, this is empirically motivated and taken as an assumption in both statistical mechanics and continuum mechanics models.
Since we are proposing to design an anisotropic material -- which is potentially very different from existing elastomers -- it is possible that incompressibility will not be a good approximation for our designed materials.
However, for simplicity, we will assume here that the proposed anisotropic electro-responsive elastomers are incompressible.
Formally, this means that the deformation gradient must satisfy $\J = 1$ where $\J \coloneqq \det \F$.
As a consequence, $\chaindensity = \chaindensityref$ and $\polarization = \Polar$.



\section{Elastic Modulus} \label{sec:rubber-elasticity}

We first consider the elastic modulus in the purely mechanical setting.
Let $\efield\left(\pos\right) = 0$ for all $\pos \in \cbody$.
Then $\U = 0$ and consequently:
\begin{equation*}
    \A = \AHelm = -\T \Sent = \N \kB \T \left[\stch \zmultzero + \ln\left(\frac{\zmultzero}{4 \pi \sinh\left(\zmultzero\right)}\right) \right]
\end{equation*}
from~\cite{kuhn1942beziehungen,treloar1975physics,daniels1972kuhn}.
The affine deformation assumption gives that $\rvec = \F \Rvec$, and therefore the Helmholtz free energy density is given by:
\newcommand{\stchExpr}{\frac{|\F \Rvec|}{\N \mlen}}
\begin{equation} \label{eq:Wmech-full}
\begin{split}
    \W(\F) = \chaindensity \int_{0}^{\pi} \df \polar \int_{0}^{2\pi} \df \azi \mbox{ } \chainpdf \cdot \left( \N \kB \T \left[\stchExpr \Langinv\left(\stchExpr\right) + \ln\left(\frac{\Langinv\left(\stchExpr\right)}{4 \pi \sinh\left(\Langinv\left(\stchExpr\right)\right)}\right) \sin\polar \right]\right)
\end{split}
\end{equation}
where $\Rvec$ can be written as $\mlen \sqrt{\N} \left(\cos\azi \sin\polar, \sin\azi \sin\polar, \cos\polar\right)$.
This integral is difficult to evaluate in closed-form, but simplifies significantly if we use a Taylor expansion of $\Langinvs$ in powers of $\stch$:
\begin{equation} \label{eq:Wmech-linearized}
    \W = \frac{3}{2} \chaindensity \kB \T \intoverSPdf{\stchr^2 + \bigoof{\stch^4 \N}}
\end{equation}
where $\stchr \coloneqq |\rvec|/|\Rvec| = |\rvec|/\mlen\sqrt{\N}$.
For the remainder of this section, we neglect the higher-order terms in \eqref{eq:Wmech-linearized}.
For now, we consider $\F$ of the form $\diag\left(\pstch{1}, \pstch{2}, \pstch{3}\right)$; thus,
\begin{equation*}
    \stchr^2 = \pstch{1}^2 \cos^2\azi \sin^2\polar + \pstch{2}^2 \sin^2\azi \sin^2\polar + \pstch{3}^2 \cos^2\polar.
\end{equation*}
First, let $\chainpdf = 1/4 \pi$.
It is easy to show that we obtain the neo-Hookean energy density:
\begin{equation*}
    \W = \frac{\shearModClassical}{2}\left(\pstch{1}^2 + \pstch{2}^2 + \pstch{3}^2\right)
\end{equation*}
where $\shearModClassical \coloneqq \chaindensity \kB \T$ is the shear modulus that is predicted by the Gaussian chain approximation in classical rubber elasticity~\cite{treloar1975physics}.
This result is to be expected since we took a Taylor expansion of the Langevin chain statistics \eqref{eq:Wmech-full} about zero stretch and because the chosen form of $\chainpdf$ is isotropic.
Clearly then, for isotropic elastomers, the stiffness--for a constant mass density, $\chaindensityref \N$--increases with the density of cross-links; that is, increases with $\chaindensityref$ (recall: $\chaindensityref = \J \chaindensity$) (\hl{this is well known; see, for example,}~\cite{treloar1975physics}).
Moreover, the slope of the inverse Langevin function is a monotonically increasing function of its argument.
Its argument in \eqref{eq:Wmech-full} is $\bigoof{\N^{-1}}$, so increasing the density of cross-links for fixed mass density--which effectively lowers $\N$--also increases the stiffness through higher order terms in \eqref{eq:Wmech-linearized}.
Physically, this is because the higher order terms account for the finite extensibility of the chain and, as $\N$ decreases, so does the maximum length of a chain.
Similarly, the stiffness increases with the fraction of loose end monomers (for fixed $\N$) because, as $\lefrac$ increases, the maximum length of the chain, $\nb \mlen$, decreases and finite extensibility effects are more relevant for shorter stretches~\cite{treloar1975physics,kuhn1942beziehungen}\footnote{
    \hl{We remark that there is also some discussion in the literature regarding how the entanglement of loose-end chains with the rest of the network affect its elastic properties}~\cite{langley1974relation,gordon1975rubber,case1960branching}.
    \hl{Since we do not explicitly consider chain entanglements, we do not explore this further, but such considerations, as well as other structural details of the polymer network, may be worth investigating in future work.}
}.

\newcommand{\ITIts}[1]{\I{#1}\left[\polarAvg, \Std\right]}
Next we consider a transversely isotropic elastomer.
In this case, we take $\chainpdf$ to be given by the ansatz \eqref{eq:TI-ansatz}.
\hl{We make the definition:}
\begin{equation*}
    \GaussSineInt{k}{\mu}{\sigma} \coloneqq \int_{0}^{\pi} \dx{x} \left(\GaussExpr{x}{\mu}{\sigma} + \GaussExpr{\pi - x}{\mu}{\sigma}\right) \sin \left(k x\right)
\end{equation*}
\hl{where $k \in \mathbb{N}$, $\mu \in \left[0, \frac{\pi}{2}\right]$, and $\sigma \in \left[0, \infty\right)$}.
Then the average square relative stretch and the Helmholtz free energy density are:
\begin{align}
    \label{eq:avg-relative-stretch-sq}
    \avg{\stchr^2} &= \frac{1}{4} \left(\frac{\pstch{1}^2+\pstch{2}^2}{2}\left(3 - \frac{\ITIts{3}}{\ITIts{1}}\right) + \pstch{3}^2 \left(1 + \frac{\ITIts{3}}{\ITIts{1}}\right)\right) \\
    \label{eq:W-TI-mech}
    \W\left(\pstch{1}, \pstch{2}, \pstch{3}; \polarAvg, \Std\right) &= \frac{3}{8} \chaindensity \kB \T \left(\frac{\pstch{1}^2+\pstch{2}^2}{2}\left(3 - \frac{\ITIts{3}}{\ITIts{1}}\right) + \pstch{3}^2 \left(1 + \frac{\ITIts{3}}{\ITIts{1}}\right)\right)
\end{align}
We can approximate these expressions in the limits of $\Std \ll 1$ and $\Std \gg 1$ by using \eqref{eq:Izero} and \eqref{eq:Iinf}, respectively.

As mentioned previously, there is no systematic theory which connects the molecular structure of elastomers to their effective Poisson's ratio; instead, incompressibility is generally taken as an assumption, and we use this assumption here.

We consider loading the body with displacement-control along the axis of symmetry, and free on the transverse faces.
With incompressibility, the symmetry of the system gives the deformation gradient of the form $\F = \diag\left(1 / \sqrt{\pstchu}, 1 / \sqrt{\pstchu}, \pstchu\right)$, where the $3$-direction is aligned along the axis of symmetry.

The slope of the stress-strain curve gives the \emph{tangent elastic modulus} $\ElasticPara$ in the direction of the axis of symmetry.
Similarly, the tangent elastic modulus in the plane orthogonal to the axis of symmetry is denoted by $\ElasticPerp$\footnote{The property $\ElasticPerp$ corresponds to the slope of the stress-strain curve if a sample were stretched biaxially in the plane orthogonal to the axis of symmetry and traction-free on the lateral faces. 
It can be calculated from the deformation considered here by $\ElasticPerp = \frac{1}{4} \takepartials{\W}{\left(1 / \sqrt{\pstchu}\right)}{2}$, where the $1 / 4$ is a material and geometric factor related to the number of dimensions that are being stretched (i.e. 2) and the Poisson's ratio, $1 / 2$. 
}.

\begin{figure}
	\centering
	\input{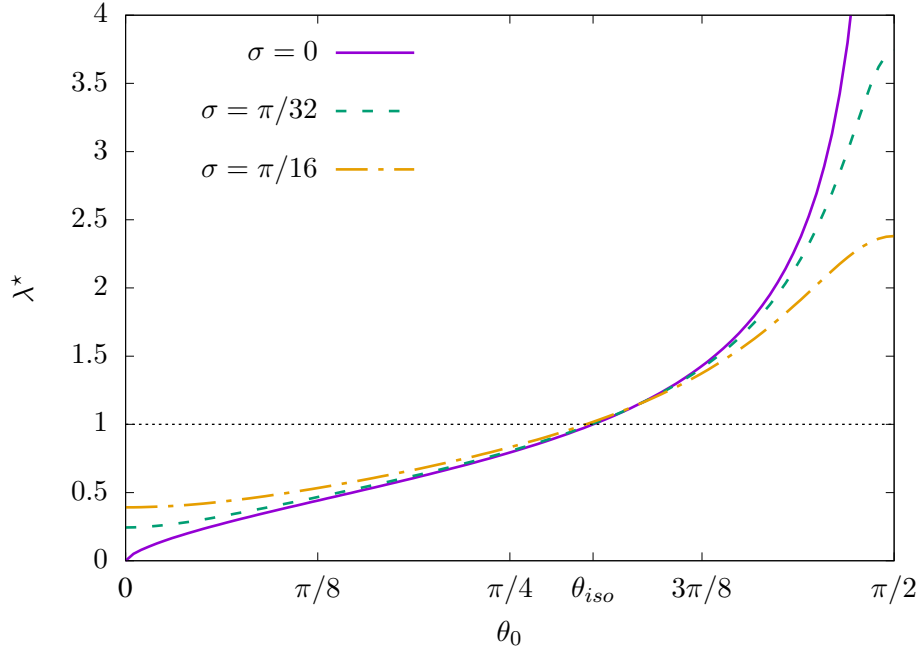}
	\caption{The load-free equilibrium stretch in the direction of the axis of symmetry for a transversely isotropic, incompressible elastomer.
		The stretch, $\rstate{\pstchu}$, is shown as a function of the angle, $\polarAvg$, that the (upper) Gaussian is centered about.}
	\label{fig:lambdastar}
\end{figure}

In the limit of $\Std \ll 1$, our model predicts:
\begin{equation} \label{eq:ElasticParaTI}
    \ElasticPara = \takepartials{\W}{\pstchu}{2} = \frac{3}{4} \chaindensityref \kB \T \left(1 + 3\pstchu^{-3} + \euler^{-4 \Std^2} \left[1 - \pstchu^{-3} + 2\left(1 - \pstchu^{-3}\right) \cos\left(2\polarAvg\right) \right] \right)
\end{equation}
which shows a strain hardening in compression and a strain softening in tension--except when $\polarAvg = 0$ and $\Std = 0$.
Interestingly, $\left(\polarAvg, \Std\right) = \left(0, 0\right)$ recovers the isotropic elastic modulus; as does any $\left(\polarAvg, \Std\right)$ at $\pstchu = 1$.
At first glance, this seems physically unreasonable, as one would expect changing the distribution of chain directions in the network to affect the stiffness in various directions.
However, recall from our discussion in \Fref{sec:pdf-philosophy} that $\pstchu = 1$ is no longer the equilibrium configuration in the absence of external loads.
Indeed, let $\rstate{\pstchu}$ be such that $\takepartial{\W}{\pstchu}\Bigr|_{\pstchu = \rstate{\pstchu}} = 0$; then,
\begin{equation} \label{eq:eql-stretch}
    \rstate{\pstchu} = \left(\frac{3 - \left[1 + 2 \cos\left(2\polarAvg\right)\right] \euler^{-4 \Std^2}}{2 + \left[2 + 4\cos\left(2\polarAvg\right)\right]\euler^{-4 \Std^2}}\right)^{1/3}.
\end{equation}
Equation \eqref{eq:eql-stretch} is shown in \Fref{fig:lambdastar} as a function of $\polarAvg$ for $\Std = 0, \pi/32$ and $\pi/16$.
Let $\polarAB$ be the polar angle for which $\rstate{\F} = \iden$ when $\Std = 0$; one can show that $\polarAB = \arctan \sqrt{2}$.
When $\polarAvg < \polarAB$, we have $\rstate{\pstchu} < 1$.
This is because there is simultaneously a higher density of chains that are oriented closer to the direction of stretch (i.e. the axis of symmetry), and a lower density of chains oriented towards the orthogonal to the direction of stretch.
The elasticity of polymer chains is due entirely to entropy in the absence of electric field in our model; consequently, polymer chains are in tension, at any finite temperature, if their end-to-end vector is finite.
Put differently, the maximum chain entropy is given by a vanishing end-to-end vector; thus, chains always want to contract if we neglecting excluded volume effects.
The additional incompressibility assumption accounts empirically for the excluded volume effects and prevents the collapse of the network to zero volume.

When incompressibility is enforced, for chains to contract in one direction requires chains to elongate in other directions.
This trade-off between stretch and elongation in different directions gives, for isotropic networks, that $\rstate{\pstchu} = 1$.
When $\polarAvg < \polarAB$, we obtain $\rstate{\pstchu} < 1$ and thus there is a contraction along the axis of symmetry, because there are more chains oriented toward this direction and hence a net increase in entropy can be gained from some $\rstate{\pstchu} < 1$ compared to the decrease in entropy of fewer chains that must elongate in other directions.
Conversely, when $\polarAvg > \polarAB$, there are fewer chains along the axis of symmetry, and a net increase in entropy can be gained by contracting orthogonal to the axis despite the decrease in entropy to the fewer chains along the symmetry axis that must elongate.
In the limit of $\sigma=0$ when all the chains are aligned precisely along a single direction, this leads to singularities at $\left(\polarAvg = 0, \Std = 0\right)$ and $\left(\polarAvg = \pi / 2, \Std = 0\right)$ as can be seen in \Fref{fig:lambdastar}.

\newcommand{\ExpTermHelp}{\euler^{-4 \Std^2}}
We return our attention to the elastic modulus.
In particular, we are interested in $\rstate{\ElasticPara} \coloneqq \ElasticPara\left(\pstchu = \rstate{\pstchu}\right)$; that is, the stiffness of the material at its equilibrium state when no loads are applied.
Using \eqref{eq:eql-stretch} in \eqref{eq:ElasticParaTI}, we obtain:
\begin{equation} \label{eq:ElasticParaStarTI}
    \rstate{\ElasticPara}\left(\polarAvg, \Std\right) = \frac{9}{4} \chaindensity \kB \T \left(1 + \left[1 + 2\cos\left(2\polarAvg\right)\right]\ExpTermHelp\right)
\end{equation}
Similarly,
\begin{equation} \label{eq:ElasticPerpStarTI}
    \rstate{\ElasticPerp}\left(\polarAvg, \Std\right) = \frac{9}{8} \chaindensity \kB \T \left(3 - \left[1 + 2\cos\left(2\polarAvg\right)\right]\ExpTermHelp\right).
\end{equation}
As expected, the maximum $\rstate{\ElasticPara}$ occurs at $\left(\polarAvg, \Std\right) = \left(0, 0\right)$--as this is the case which has the maximum amount of chains oriented in the direction of the axis of symmetry for a given $\chaindensityref$.
Interestingly, although, as previously mentioned, there is a singularity at $\left(\polarAvg, \Std\right) = \left(0, 0\right)$ such that $\rstate{\pstchu} = 0$, the elastic modulus is finite.
The maximum $\rstate{\ElasticPara}$ is given by $9 \chaindensity \kB \T$, which is \emph{the number of spatial dimensions times the elastic modulus for an isotropic network} ($\ElasticIso = \ElasticPara\left(\polarAB, 0\right) = \ElasticPerp\left(\polarAB, 0\right) = 3 \shearModClassical = 3 \chaindensity \kB \T$).
Once again, as expected, $\rstate{\ElasticPara}$ is a minimum and, more specifically, vanishes at $\left(\polarAvg, \Std\right) = \left(\pi / 2, 0\right)$. Similarly, $\rstate{\ElasticPerp}$ is a maximum at $\left(\polarAvg, \Std\right) = \left(\pi / 2, 0\right)$ and vanishes at its minimum $\left(\polarAvg, \Std\right) = \left(0, 0\right)$.
\begin{figure}
	\centering
	\input{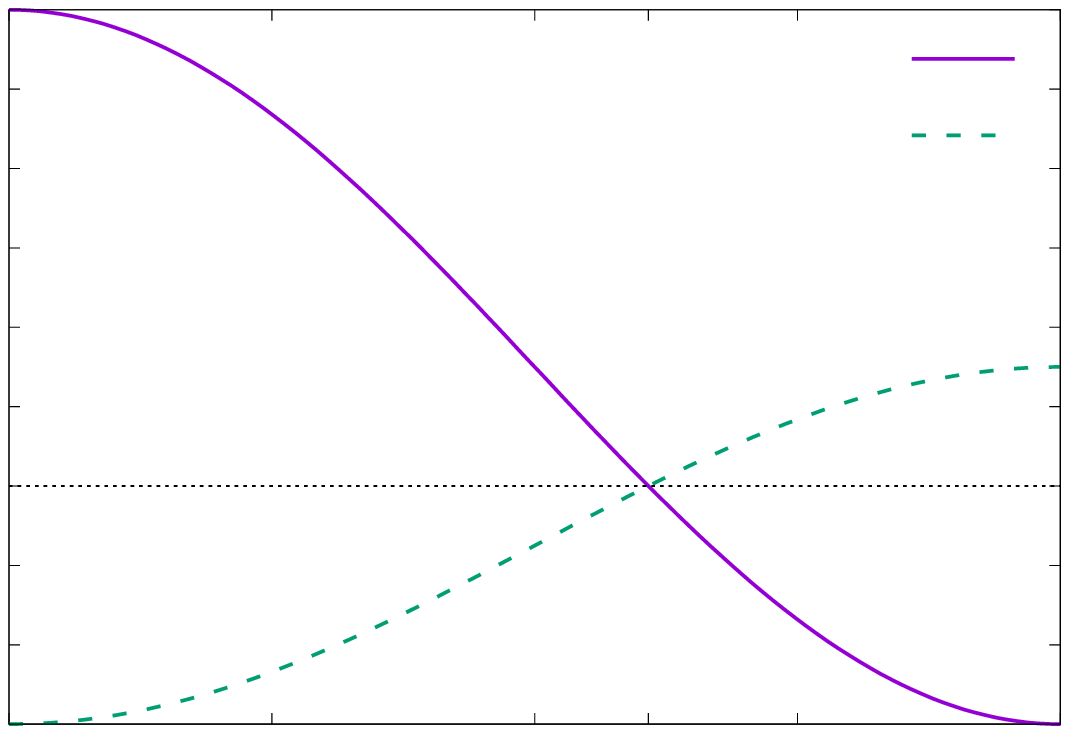}
	\caption{The dimensionless elastic moduli in the directions of the axis of symmetry, $\rstate{\ElasticPara}$, and orthogonal to the axis of symmetry, $\rstate{\ElasticPerp}$, as a function of the upper Gaussian center, $\polarAvg$, when $\Std = 0$.
		The dashed, black line represents the elastic modulus for an isotropic network, $\ElasticIso = 3 \shearModClassical = 3 \chaindensity \kB \T$}
	\label{fig:elasticTI}
\end{figure}

This analysis establishes upper bounds
\footnote{We emphasize that these bounds are obtained in the limit $\sigma\ll 1$.}
for $\rstate{\ElasticPara}$ and $\rstate{\ElasticPerp}$ as $9 \chaindensity \kB \T$ and $9/4 \chaindensity \kB \T$, respectively; and shows that theoretically, zero stiffness can be achieved.
However, as previously mentioned, there are many applications, such as soft robotics, when it will be desirable to control the DE stiffness within some range while optimizing over other properties.
In this case, \eqref{eq:ElasticParaStarTI} and \eqref{eq:ElasticPerpStarTI} can be used as design tools in the limit when $\Std \ll 1$; that is, when directionality in the network is highly controlled.

\begin{figure}[htb!]
	\centering
	\input{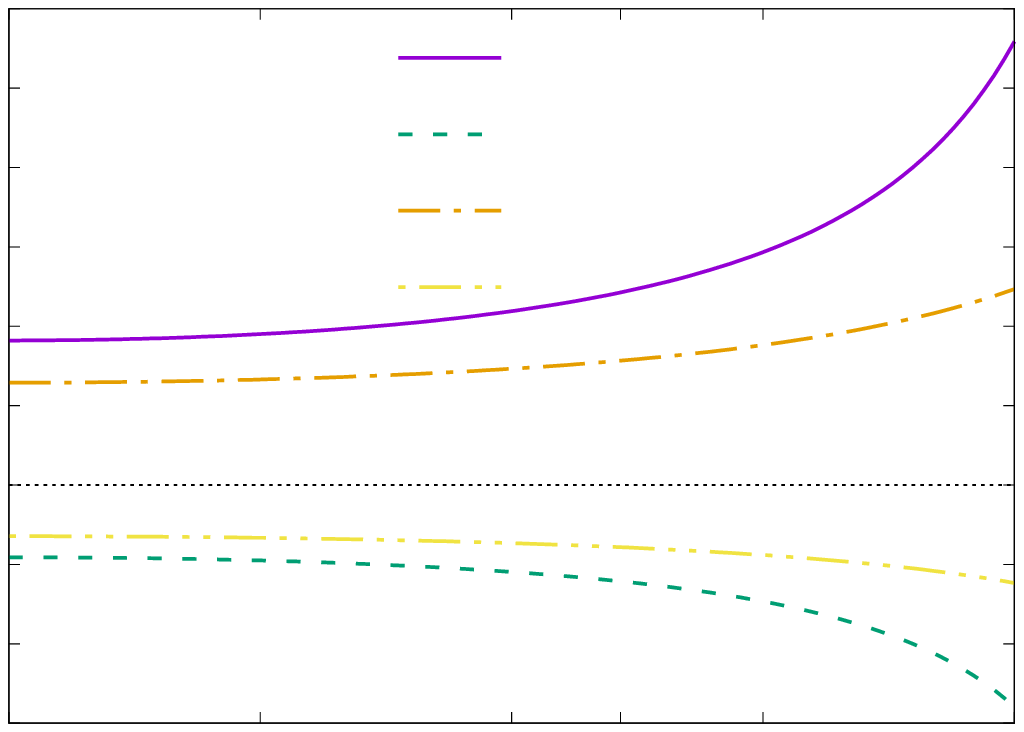}
	\caption{The dimensionless elastic moduli in the directions of the axis of symmetry, $\rstate{\ElasticPara}$, and orthogonal to the axis of symmetry, $\rstate{\ElasticPerp}$, as a function of the upper Gaussian center, $\polarAvg$, for finite $\Std$: $4 \pi / 3$ and $3 \pi / 2$.
		The dashed, black line represents the elastic modulus for an isotropic network, $\ElasticIso = 3 \shearModClassical = 3 \chaindensity \kB \T$.}
	\label{fig:elasticTIInf}
\end{figure}

We now carry out a similar analysis in the limit of weak directional control (i.e.$\Std \gg 1$).
In this case, $\GaussSineInt{k}{\polarAvg}{\Std} \approx \GaussSineIntInf{k}{\polarAvg}{\Std}$, where $\GaussSineIntInf{k}{\polarAvg}{\Std}$ is given in \eqref{eq:Iinf}.
Explicit evaluation gives:
\begin{equation} \label{eq:ElasticApproximationsInf}
\begin{split}
    \ElasticParaInf &= \chaindensity \kB \T \left[\frac{\left(2+\pstchu^3\right)f\left(\polarAvg, \Std\right) - 28\pstchu^3 - 80}{\pstchu^3\left(f\left(\polarAvg, \Std\right) - 36\right)}\right] 
    \\
    \rstate{\pstchu} &= \left[\frac{f\left(\polarAvg, \Std\right) - 40}{f\left(\polarAvg, \Std\right) - 28}\right]^{1/3} 
    \\
    \rstate{\ElasticParaInf} &= 3 \chaindensity \kB \T \left[\frac{f\left(\polarAvg, \Std\right) - 28}{f\left(\polarAvg, \Std\right) - 36}\right] 
    \\
    \rstate{\ElasticPerpInf} &= 3 \chaindensity \kB \T \left[\frac{f\left(\polarAvg, \Std\right) - 40}{f\left(\polarAvg, \Std\right) - 36}\right] 
    \\
    \text{ where } f\left(\polarAvg, \Std\right) &= 18\Std^2 - 18\pi \Std - 36\polarAvg^2 + 9\pi^2
\end{split}
\end{equation}
These approximations for the nondimensional elastic moduli, $\ElasticPara / \chaindensity \kB \T$ and $\ElasticPerp / \chaindensity \kB \T$, are shown in \Fref{fig:elasticTIInf} for $\Std = 4 \pi / 3$ and $3 \pi / 2$.
Clearly, \eqref{eq:ElasticApproximationsInf} recovers $\rstate{\ElasticParaInf} = \rstate{\ElasticPerpInf} = \ElasticIso$ in the limit of $\Std \rightarrow \infty$, as expected.
The approximations in \eqref{eq:ElasticApproximationsInf} share the same physical character as the $\Std \ll 1$ approximations in the sense that, when $\Std$ is large enough, $\rstate{\ElasticParaInf}$ is at its maximum (on the interval $\left[0, \pi / 2\right]$) at $\polarAvg = 0$ and minimum at $\polarAvg = \pi / 2$ (this can also be seen in \Fref{fig:elasticTIInf}).
The situation is reversed for $\rstate{\ElasticPerpInf}$; that is, its maximum is at $\polarAvg = \pi / 2$ and its minimum is at $\polarAvg = 0$.
Again, this is due to the fact that there is a greater stiffness in the directions of higher chain densities.
Lastly, note that the approximations in \eqref{eq:ElasticApproximationsInf} predict a nonphysical singularity on the interval $\polarAvg \in \left[0, \pi / 2\right]$ when $\Std$ is not large enough ($\Std \lessapprox \pi$).
This can result in, incorrectly, predicting negative and/or diverging $\ElasticPara$ or $\ElasticPerp$.
Thus, one should take care that $\Std$ is large enough when using \eqref{eq:ElasticApproximationsInf} for design of transversely isotropic elastomers.

\section{Susceptibility} \label{sec:susceptibility}

Using the negligibly interacting assumption, we have that $\Wstar = \chaindensity \avg{\A}$.
However, \eqref{eq:A-approx} is challenging to evaluate in closed-form.
Proceeding as in the case of the elastic modulus, we use a Taylor expansion of the inverse Langevin function about zero stretch to find $\stch / \Langinvs = \frac{1}{3} - \frac{1}{5} \frac{\stchr^2}{\N} + \bigoof{\stch^4}$.
This provides an approximation for the free energy:
\newcommand{\auxf}{w_f^{*}}
\begin{equation} \label{eq:Wstar-linearized}
    \Wstar\left(\F, \efield\right) 
    = 
    \chaindensity \kB \T \left( 
        \N \left[ \auxf\left(\unodim\right) - \uOnodim\right] 
        + \left[\frac{3}{2} + \frac{2 \unodim}{15} - \auxf\left(\unodim\right)\right]\avg{\stchr^2} 
        + \frac{3 \unodim}{5} \avg{\stchr^2 \left(\edir \cdot \rdir\right)^2} 
        + \avg{\bigoof{\stch^4 \N}}
    \right)
\end{equation}
where
\begin{equation} \label{eq:free-polarization-Astar}
    \auxf\left(\unodim\right) = \ln\left(\frac{2 \sqrt{\unodim}}{\sqrt{\pi}\erfw}\right).
\end{equation}
The four terms on the right side of \eqref{eq:Wstar-linearized} can be understood as follows.
\begin{enumerate}
    \item the first term is the closed-dielectric free energy density of a collection of $\left(\chaindensity \N\right)$ monomers that are kinematically free; that is, monomers that are not constrained to satisfy a given end-to-end vector; 
    \item a correction to the first term that is related to the average magnitude of chain stretch; 
    \item a further correction that takes into account the average amount that the chain is stretched parallel to the direction of the electric field; 
    \item higher order terms related to the finite extensibility of the chain.
Note that the first term is invariant when changing the cross-linking density at fixed mass density.
\end{enumerate}
The second and third terms contribute to electromechanical coupling, while only the third term captures the effect of chain torque.
Interestingly, if one neglects $\bigoof{\stch^4 \N}$ terms in \eqref{eq:Wstar-linearized}, then it is invariant under $\lefrac$.
Therefore changing the fraction of loose end monomers can only have an effect on higher order terms, terms which are related to the finite extensibility of the chain.

\subsection{Definition of the Susceptibility} \label{sec:sus-prelude}

In \Fref{sec:reference-states} we defined the inverse susceptibility tensor: $\susceptibilityTensor^{-1} \coloneqq \takecrosspartial{\W}{\Polar}{\Polar}$, where the derivative is evaluated at the reference state.
However, \eqref{eq:Wstar-linearized} gives us $\Wstar$; that is, the Legendre transform of $\W$ in the polarization slot.
Further, although it is clear that 
\begin{equation*}
  \W\left(\F, \Polar\right) = \Wstar\left(\F, \efield\left(\Polar\right)\right) + \J^{-1} \Polar \cdot \efield\left(\Polar\right)
\end{equation*}
(where by $\efield\left(\Polar\right)$ we mean the electric field as a function of polarization), it is not straightforward to write out $\W$ explicitly.
This is because, while we have derived the polarization as a function of the electric field, we are unable to invert this function, for general $\F$ and $\Polar$, to obtain $\efield = \efield\left(\Polar\right)$.
Therefore, we instead look for a correspondence between derivatives of $\W$ and of $\Wstar$.
The desired identity is as follows\footnote{
  As discussed in \fref{sec:continuum}, the local nature of the relationship between $\efield$ and $\Polar$ depends upon monomer-monomer interactions being negligible.
}:
\begin{equation*}
    -\takecrosspartial{\Wstar}{\efield}{\efield} = \left(\takecrosspartial{\W}{\Polar}{\Polar}\right)^{-1} = \susceptibilityTensor
\end{equation*}
This is shown in \Fref{app:legendre-transform-stuff}.

\subsection{Tangent and Secant Susceptibility}

\begin{figure}
	\centering
	\input{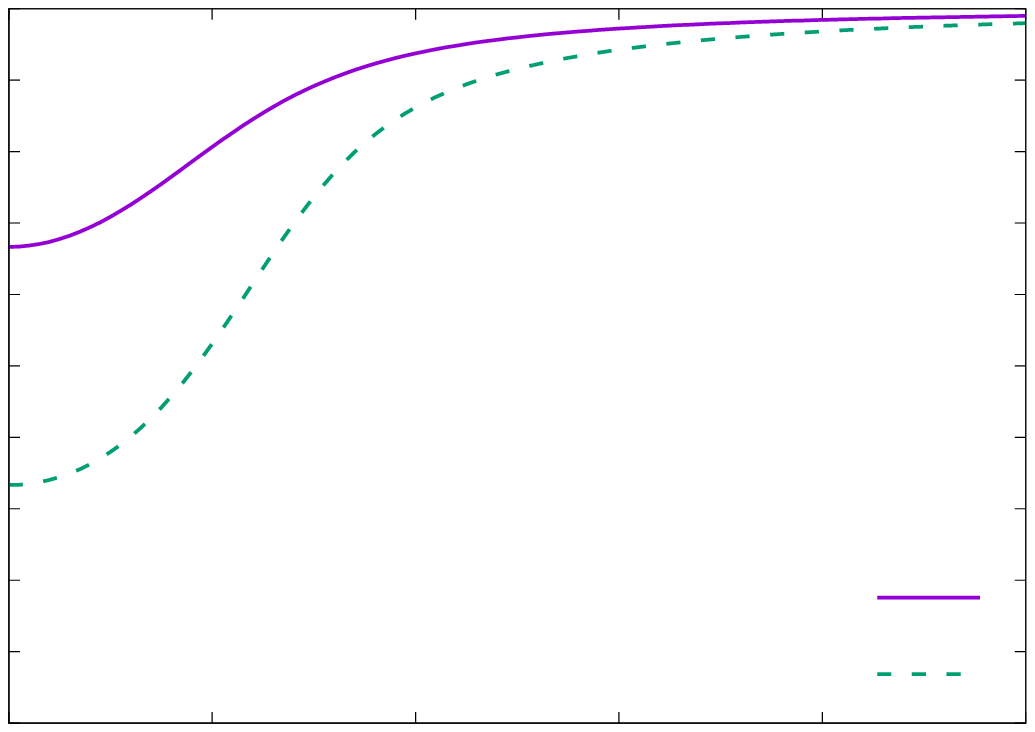}
	\caption{The nondimensional secant susceptibility per monomer for a collection of kinematically free monomers--or, in other words, the ratio of chain polarization to the characteristic scale, $\N \SusM \emag$, where $\SusM \coloneqq \max\left(\sussymbol_{\parallel}, \sussymbol_{\perp}\right)$--shown as a function of $\sqrt{|\unodim|}$.
		The chains consist of TI ($|\dsus| = \sussymbol_{\perp}$) and uniaxial ($|\dsus| = \sussymbol_{\parallel}$) monomers.
		In the limit of small electric field or large temperature (i.e. $\sqrt{|\unodim|} \rightarrow 0$), the nondimensional (lab) susceptibility per monomer is $2/3$ for TI monomers and $1/3$ for uniaxial monomers.
		In the limit of large electric field or small temperature, the chain polarization approaches its theoretical maximum: $\N \SusM \emag$.}
	\label{fig:free-polarization}
\end{figure}

We return our attention to the first term in \eqref{eq:Wstar-linearized}--the term that corresponds to a dielectric that consists of monomers that are unconstrained.
We call the result of taking $-\partial / \partial \efield$ of this term the \emph{free polarization}.
The ratio of free chain polarization to its maximum theoretical value is shown in \Fref{fig:free-polarization} as a function of the applied electric field.
We distinguish between the \emph{secant susceptibility}, $\susceptibilityTensorLab$, and the \emph{tangent susceptibility} $\susceptibilityTensor$.
The secant susceptibility is defined such that $\polarization = \susceptibilityTensorLab \efield$.
Thus, \fref{fig:free-polarization} shows a measure of the secant susceptibility of free monomers.
Clearly $\susceptibilityTensorLab \neq \susceptibilityTensor = -\takecrosspartial{\Wstar}{\efield}{\efield}$ in general; in fact, $\susceptibilityTensor = \susceptibilityTensorLab + \takepartial{\susceptibilityTensorLab}{\efield} \efield$.
They are equivalent when the dielectric is linear.

All of the dielectric nonlinearity in our DE materials is encoded in the function $\auxf$, which is a per monomer contribution to the closed-dielectric free energy free energy for an unconstrained polymer chain (see \eqref{eq:free-polarization-Astar}).
In fact, the closed-dielectric free energy of an unconstrained polymer chain is $\N \kB \T \left(\auxf - \uOnodim\right)$.
Taking $-\partial / \partial \efield$ and then dividing by $\emag$, we arrive at the result shown in \Fref{fig:free-polarization}.
The vertical axis corresponds to the secant susceptibility with a single free chain per unit volume, where the secant susceptibility is measured in units of $\N \SusM$ and $\SusM \coloneqq \max\left(\sus{1}, \sus{2}\right)$.
In regards to the physical character of the secant susceptibility for the free chain, there are three regimes that can be seen in \Fref{fig:free-polarization}.

The first regime corresponds to $\sqrt{|\unodim|} \rightarrow 0$; then $\susceptibilityLab / \N \rightarrow \left(\sus{1} + 2 \sus{2}\right) / 3$.
In this limit (i.e. $\emag \rightarrow 0$ and/or $\kB \T \rightarrow \infty$), the pdf of monomer orientations is uniformly distributed over the unit sphere because the electrostatic energy is vanishingly small as compared to the thermal energy.
The factors of $1 / 3$ for $\sus{1}$ and $2 / 3$ for $\sus{2}$ correspond to the dimensionality of each dipole susceptibility: the monomers are in three dimension space while $\sus{1}$ is the dipole susceptibility along a line (i.e. $\vspan \nvec$) and $\sus{2}$ is the dipole susceptibility in the plane orthogonal to $\nvec$.
In addition, note that $\susceptibilityLab$ does not vanish as $\kB \T \rightarrow \infty$.
This is because $\sustens$ is quadratic in $\nvec$.
Even though $\nvec$ is uniformly distributed, the polarization does not cancel.

In the second regime $\sqrt{|\unodim|} \rightarrow \infty$; then $\susceptibilityLab / \N \rightarrow \max\left(\sus{1}, \sus{2}\right)$.
In this limit (i.e. $\emag \rightarrow \infty$ and/or $\kB \T \rightarrow 0$), the monomers are frozen in their  minimum potential energy orientation; that is, $\nvec = \pm \edir$ when $\sus{1} > \sus{2}$ and $\nvec$ such that $\nvec \cdot \edir = 0$ when $\sus{2} > \sus{1}$.
In this case, the chain has reached its maximum theoretical secant susceptibility.

Finally, there is a transition regime between the limits of $\sqrt{|\unodim|} \rightarrow 0$ and $\sqrt{|\unodim|} \rightarrow \infty$.

There are two main takeaways that are relevant to this discussion: (1.) the dielectric nonlinearity of our DE materials arises because the average alignment of the monomers is determined by a balancing of electrostatic potential energy and the entropy, and because the dipole susceptibility of a monomer depends on its alignment with respect to the applied electric field; and (2.) despite this nonlinearity, we have $\susceptibility = \susceptibilityLab$ in the limits of $\sqrt{|\unodim|} \rightarrow 0$ and $\sqrt{|\unodim|} \rightarrow \infty$.
Moving forward, we will consider both $\susceptibility$ and $\susceptibilityLab$; as they will both prove useful.
Specifically, $\susceptibility$ is particularly relevant to material stability while $\susceptibilityLab$ is relevant to the capacitance--and hence, electromechanical coupling--for a given thin film DEA geometry.
Moreover, for convenience, we make the definitions: 
\begin{align*}
    \susceptibilityTensorHot &\coloneqq \lim_{\sqrt{|\unodim|} \rightarrow 0} \susceptibilityTensor = \lim_{\sqrt{|\unodim|} \rightarrow 0} \susceptibilityTensorLab \\
    \susceptibilityTensorCold &\coloneqq \lim_{\sqrt{|\unodim|} \rightarrow \infty} \susceptibilityTensor = \lim_{\sqrt{|\unodim|} \rightarrow \infty} \susceptibilityTensorLab
\end{align*}

We now consider what we have learned about free polarization in the context of the design of anisotropic dielectric elastomers.
When used for a DEA, it is desirable that the secant susceptibility of the DE be as large is possible.
This is because the susceptibility serves to increase the capacitance of the DEA.
The increased capacitance means a greater accumulation of charge on the electrodes for a given voltage difference--which leads to a greater Coulomb attraction between the electrodes and hence a greater electromechanical coupling of the DEA. 
Now, for illustrative purposes, consider a single chain.
If the chain is allowed to contract to zero stretch, then all the terms in \eqref{eq:Wstar-linearized} vanish except the first one--which is consistent with our previous discussion.
In this case, the susceptibility is isotropic and is such that $\susceptibilityTensorLab = \susceptibilityLab \iden, \susceptibilityLab \in \chaindensity \N \left[\left(\sus{1} + 2 \sus{2}\right) / 3, \max\left(\sus{1}, \sus{2}\right)\right]$.
Recall (\Fref{fig:free-polarization}) that the secant susceptibility is maximized at low temperature or large electric field (i.e. $\sqrt{|\unodim|} \gg 1$).
This is an issue, from a practical standpoint, because we would prefer not to need to operate our DEA under these conditions.
Both temperature control as well as large applied field are challenging and cumbersome to impose.

Now consider this same chain, but in the reference, load-free state of an isotropic network such that $\stchr = 1$.
Then the second and third terms--the electromechanical terms--are $\bigoof{1}$ while the first is of $\bigoof{\N}$.
This means, for a dielectric elastomer that consists of ``long chains'' (i.e. $\N \gg 1, \N \gtrsim 1000$), that the electromechanical terms are negligible when the elastomer is in the load-free state (and in many cases, they are even negligible at large macroscopic deformations).
Thus, if we are to significantly improve on the small $\sqrt{|\unodim|}$ secant susceptibility in the load-free state, then one may first want to increase the density of cross-links of the elastomer (i.e. increase $\chaindensityref / \N$)--while keeping in mind that the stiffness also scales with the density of cross-links.
Before moving on, it is also worth recalling the discussion on the ``residual stresses'' of our hypothetical anisotropic elastomers.
Since the load-free state has some initial deformation to relieve the ``residual stresses'', $\stchr \neq 1$ in general.
At first glance, one may consider this as an opportunity to increase $\rstate{\left(\susceptibilityHot\right)}$.
However, it is easy to show that $\rstate{\avg{\stchr^2}} \leq 1$ and that equality only holds for an isotropic network.
The key is that, in the absence of electrical loads, the relaxed state should maximize the entropy of the elastomer and, by our approximation, the chain entropy as proportional to $-\stchr^2$; thus, for entropy to not decrease with respect to the manufactured state (i.e. $\avg{-\stchr^2} = 1$), we require $\rstate{\avg{\stchr^2}} \leq 1$.
Similarly, the lower bound on $\rstate{\avg{\stchr^2 \left(\edir \cdot \rdir\right)^2}}$ is clearly zero since it is strictly nonnegative and it vanishes when the elastomer is manufactured such that all of the chains are orthogonal to the eventual direction of the applied electric field.
The upper bound, however, is much less clear and warrants further investigation.
For the purposes of this investigation, we split the susceptibility into two contributions: one associated with free monomers, $\susceptibilityFree$, and a correction term due to the electromechanical coupling of the material, $\correction{\susceptibilityTensor}$; that is:
\begin{equation*}
\susceptibilityTensor = \susceptibilityFree \iden + \correction{\susceptibilityTensor}.
\end{equation*}
Taking derivatives of \eqref{eq:Wstar-linearized} with respect to $\emag$ and taking the appropriate limits, we obtain:
\begin{equation} \label{eq:sus-correction-hot}
\susceptibilityHotParaCorr = \frac{\chaindensity \left(\dsus\right)}{5}\left(\avg{\stchr^2} - 3\avg{\stchr^2\left(\edir \cdot \rdir\right)^2}\right),
\end{equation}
and
\begin{equation} \label{eq:sus-correction-cold}
\susceptibilityColdParaCorr = -\frac{\chaindensity \left(\dsus\right)}{15}\left(2\avg{\stchr^2} + 9\avg{\stchr^2\left(\edir \cdot \rdir\right)^2}\right).
\end{equation}

\subsection{Design for susceptibility}

We again consider transversely isotropic materials and use \eqref{eq:TI-ansatz} as our ansatz for $\chainpdf$.
This gives us:\newcommand{\Irat}[2]{\frac{\GaussSineInt{#1}{\polarAvg}{\Std}}{\GaussSineInt{#2}{\polarAvg}{\Std}}}
\begin{equation} \label{eq:avg-stretch-Edotr-sq}
    \avg{\stchr^2 \left(\edir \cdot \rdir\right)^2} = \frac{\pstchu^2}{4} \left(1 + \Irat{3}{1}\right).
\end{equation}
\newcommand{\auxe}[1]{\euler^{#1 \Std^2}}
\newcommand{\auxc}[1]{\cos\left(#1 \polarAvg\right)}
In the limit of $\Std \ll 1$, we have the approximation:
\begin{equation*}
    \avg{\stchr^2 \left(\edir \cdot \rdir\right)^2} \approx \frac{\pstchu^2}{4} \left[1 + \auxe{-4}\left(1 + 2\auxc{2}\right)\right]
\end{equation*}

We could then derive, using \eqref{eq:avg-stretch-Edotr-sq} and \eqref{eq:avg-relative-stretch-sq}, $\Wstar$ for transversely isotropic DEs.
However, we turn our attention directly to $\susceptibilityHotParaCorr$.
In the limit of $\Std \ll 1$, we have the approximation:
\begin{equation*} 
\frac{\rstate{\left(\susceptibilityHotParaCorrZero\right)}}{\chaindensity \left(\dsus\right)} = 0
\end{equation*}
where again, $\rstate{\generic}$ denotes a quantity evaluated at $\pstchu = \rstate{\pstchu}$.
Interestingly, the electromechanical correction term vanishes at $\rstate{\pstchu}$ for all $\polarAvg$ and $\Std$ when $\Std \ll 1$; and hence, at least in the limit of the load-free state, the DE effectively behaves as a collection of free monomers.

\begin{figure} 
	\centering
	\input{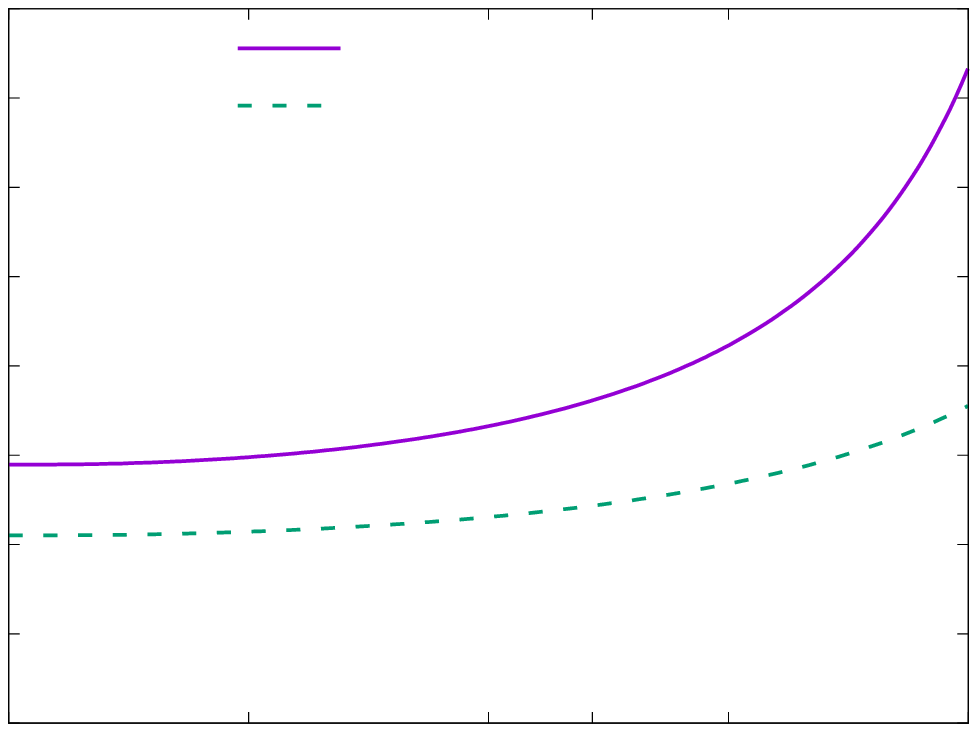}
	\caption{Correction to the free monomer susceptibility in the load-free state in the limit of $\Std \gg 1$.
		The correction to the free monomer susceptibility is much less in this limit than the $\Std \ll 1$ limit, and it vanishes as $\Std \rightarrow \infty$.}
	\label{fig:free-sus-correction-inf}
\end{figure}

Similarly, using \eqref{eq:Iinf} in \eqref{eq:avg-relative-stretch-sq} and \eqref{eq:avg-stretch-Edotr-sq}, then subsequently \eqref{eq:sus-correction-hot} and \eqref{eq:ElasticApproximationsInf}, we arrive at an approximation of $\rstate{\left(\susceptibilityHotParaCorr\right)}$ in the limit of weak directional control (i.e. $\Std \gg 1$).
This approximation is:
\begin{equation*}
-\frac{16 \dsus \left(\polarAvg - \Std\right)\left[\pi - 3\left(\polarAvg+\Std\right)\right]}{5\left(2\polarAvg^2-2\pi\polarAvg-4\Std^2+\pi^2-4\right)\left[f\left(\polarAvg,\Std\right)-40\right]^{1/3}\left[f\left(\polarAvg,\Std\right)-28\right]^{2/3}}
\end{equation*}
where $f$ was first defined in \eqref{eq:ElasticApproximationsInf}.
The approximation is shown in \Fref{fig:free-sus-correction-inf}.
Notice that, while the electromechanical correction (at $\pstchu = \rstate{\pstchu}$) does not vanish, $\rstate{\left(\susceptibilityHotParaCorrInf\right)} / \chaindensity \dsus$ is small compared to $1$ in the limit of $\Std \gg 1$.
Since the free monomer susceptibility is $\bigoof{\chaindensity \N}$, this correction term is negligible; and, what is more, is that it vanishes as $\Std \rightarrow \infty$.

\begin{figure}
	\centering
	\input{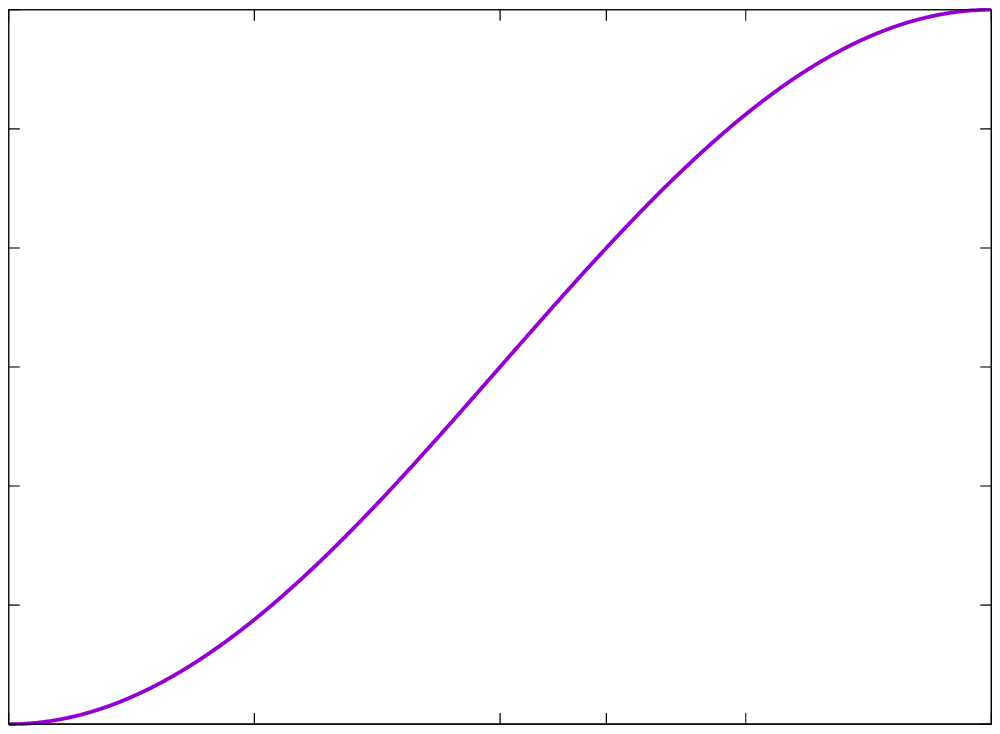}
	\caption{Approximate $\susceptibilityHotParaCorrZero$ for $\Std = 0$ and $\pstchu = 1$.
		In this case, $\susceptibilityHotParaCorrZero$ does not vanish and is $\bigoof{\chaindensity}$.}
	\label{fig:free-sus-correction-0-l1}
\end{figure}

The electromechanical correction to the susceptibility in the load-free state, $\rstate{\left(\susceptibilityHotParaCorr\right)}$, vanishes in the limit of $\Std \ll 1$ and is negligible in the limit of $\Std \gg 1$.
It would seem then that there is little hope for increasing the initial susceptibility of our design DEs.
However, it is clear from \eqref{eq:sus-correction-hot} that $\susceptibilityHotParaCorr$ is deformation dependent.
It is worth considering whether or not $\susceptibilityHotParaCorrZero$, for instance, vanishes when $\pstchu \neq \rstate{\pstchu}$.
To this end, we visualize $\susceptibilityHotParaCorrZero$ for $\Std = 0$ and $\pstchu = 1$ in \Fref{fig:free-sus-correction-0-l1}.
Notice that, in this case, $\susceptibilityHotParaCorrZero$ does not vanish and is, in fact, has a positive contribution of $\bigoof{\chaindensity}$ when $\dsus > 0$ and $\polarAvg \rightarrow \pi / 2$, and when $\dsus < 0$ and $\polarAvg \rightarrow 0$.
Thus, to maximize $\rstate{\left(\susceptibilityHotParaCorrZero\right)}$, one must find a way to alter the polarization properties while simultaneously maintaining a (nearly) mechanically isotropic network.

\subsection{Hybrid networks: maximizing the operating susceptibility}

It was clear, particularly in \Fref{fig:free-sus-correction-0-l1}, that $\rstate{\left(\susceptibilityHotParaCorr\right)}$ could be made greater if somehow the initial, load-free deformation, $\rstate{\pstchu}$, could be as close to unity as possible.
For this reason, we propose the following (rough) manufacturing process and design algorithm for optimizing $\rstate{\left(\susceptibilityHotParaCorr\right)}$.
First, imagine that we have two types of polymer chains that are compatible with each other in the sense that they can be cross-linked together in a network.
Further, imagine that the monomers for the one type of chain are such that $\dsus > 0$ and the other are such that $\dsus < 0$.
Call these type $A$ and type $B$, respectively.
Now, if an electric field is applied in a constant direction just prior to and during cross-linking, then the density of $A$ chains oriented orthogonal to the electric field direction will increase, as will the density of $B$ chains oriented parallel to the electric field.
Let $\StdA$ and $\StdB$ be the design standard deviations for chains of type $A$ and type $B$, respectively.
Then, clearly, $\StdA$ and $\StdB$ would depend on the magnitude of the applied electric field; and, $\StdA$ and $\StdB$ would likely not be able to be controlled independently of each other--or rather, the envisioned manufacturing process would need to be specialized further in order to control the two independently of each other.
In this scenario, the monomer susceptibilities $\SusPerpA$, $\SusParaA$, $\SusPerpB$, and $\SusParaB$ are given; or at best, selected from a catalog of possible monomer types.
Then the design variables consist of $\chainDensityRefA$, $\NA$, $\polarAvgA$, $\StdA$, $\chainDensityRefB$, $\NB$, $\polarAvgB$, $\StdB$, where $\chainDensityRefA$ is the number of chains of type $A$ per unit volume, $\NA$ is the number of monomers per chain in chains of type $A$, $\polarAvg$ is the center of the Gaussian in the upper half of the unit sphere for chains of type $A$, $\StdA$ is the standard deviations of the pdf for chains of type $A$, and the remaining quantities are the same but for chains of type $B$.
The closed-dielectric free energy density for this hybrid DE is:
\begin{equation*}
    \Wstar = \J^{-1} \left(\chainDensityRefA \avgA{\A} + \chainDensityRefB \avgB{\A}\right)
\end{equation*}
where
\begin{equation*}
    \avg{\A}^{\generic} = \C^{\generic} \int_{0}^{\pi} \df \polar \int_{0}^{2\pi} \df \azi \mbox{ } \Bigg(\left\{\GaussExpr{\polar}{\polarAvg^{\generic}}{\Std_{\generic}} + \GaussExpr{\pi - \polar}{\polarAvg^{\generic}}{\Std_{\generic}}\right\} 
    \cdot \A\left(\F \Rvec, \efield; \N^{\generic}, \sus{2}^{\generic}, \sus{1}^{\generic}\right) \sin \polar\Bigg).
\end{equation*}

The design space for this problem is of a higher dimension than those that we have considered thus far.
Therefore, we proceed by using some of the intuition that we have gained thus far.
By \Fref{fig:free-sus-correction-inf} and \Fref{fig:free-sus-correction-0-l1}, we reason that an optimal $\rstate{\left(\susceptibilityHotPara\right)}$ will result from taking $\polarAvgA = \pi / 2$, $\polarAvgB = 0$, and $\StdA = \StdB = 0$.
Now let $\NmRef \coloneqq \chainDensityRefA \NA + \chainDensityRefB \NB$ be the number of monomers per unit volume in the reference configuration and $\Nm \coloneqq \J^{-1} \NmRef$ be the number of monomers per unit volume in the current configuration.
The free monomer susceptibility scales with the number of monomers per unit volume; thus, assume that $\NmRef$ is already taken as large as possible or desired given some range of acceptable DE mass density.
Similarly, $\susceptibilityHotParaCorr$ scales with $\chaindensityref$, but so does the stiffness.
Thus, one should take $\chaindensityref$ as large as possible while still keeping the stiffness within some desired range.
Let $\chainFracA \coloneqq \chainDensityRefA / \chaindensityref$ so that $\chainDensityRefB = \chainFracB \chaindensityref = \left(1 - \chainFracA\right) \chaindensityref$.
Again, considering \Fref{fig:free-sus-correction-0-l1}, we want to pick $\chainFracA \in \left[0, 1\right]$ such that $\rstate{\pstchu} = 1$.
In other words, we require that $\chainFracA$ satisfies:
\begin{equation*}
\takepartial{}{\pstchu}\left[\left(\chainFracA \avgA{\A} + \left(1 - \chainFracA\right)\avgB{\A}\right)\right]\Bigr|_{\pstchu=1} = 0,
\end{equation*}
which sets $\chainFracA = 2/3$.

Next, we need to pick $\NA$ and $\NB$.
We define the ratios $\NWeightA \coloneqq \chaindensityref \NA / \NmRef$ and $\NWeightB \coloneqq \chaindensityref \NB / \NmRef$.
Since, by assumption, the total number of monomers is already given, by definition, we require:
\begin{equation*}
\NmRef = \chainDensityRefA \NA + \chainDensityRefB \NB.
\end{equation*}
Consequently: $\NWeightA \in \left[0, 1/\chainFracA \right]$ and $\NWeightB = \left(1 - \chainFracA \NWeightA\right) / \left(1 - \chainFracA\right)$.
Because the only contribution to $\susceptibilityHotPara$ that depends on $\Nm$ is the free monomer susceptibility, if $\left(\SusParaA + 2\SusPerpA\right) > \left(\SusParaB + 2\SusPerpB\right)$ then we try to reach the limit $\NWeightA \rightarrow 1 / \chainFracA$.
Where as, if $\left(\SusParaA + 2\SusPerpA\right) < \left(\SusParaB + 2\SusPerpB\right)$ then we try to reach the limit $\NWeightA \rightarrow 0$.
Lastly, if $\left(\SusParaA + 2\SusPerpA\right) = \left(\SusParaB + 2\SusPerpB\right)$ then, in our truncated theory, $\susceptibilityHotFree$ is invariant with respect to $\NWeightA$ (in this case, it is likely best to let $\NA = \NB = \Nm / \chaindensityref$).
Let $M = \arg^{\generic} \max \left[\SusParaA + 2\SusPerpA, \SusParaB + 2\SusPerpB\right]$.
If the limiting process is successfully carried out such that $\susceptibilityHotFree$ is maximized, then:
\begin{equation} \label{eq:sus-hot-max}
\begin{split}
\rstate{\left(\susceptibilityHotPara\right)} &= \left(\susceptibilityHotFree\right)^M + \frac{\chaindensity}{5} \left[\chainFracA |\dsusA| + 2 \left(1 - \chainFracA\right) |\dsusB|\right], \\
&= \frac{\Nm}{3} \left(\sus{1}^M + 2\sus{2}^M\right) + \frac{\chaindensity}{5} \left[\chainFracA |\dsusA| + 2 \left(1 - \chainFracA\right) |\dsusB|\right],\\
&= \chaindensity \left(\frac{\NA+\NB}{3} \left(\sus{1}^M + 2\sus{2}^M\right) + \frac{2}{15} \left[|\dsusA| + |\dsusB|\right]\right).
\end{split} 
\end{equation}

In summary, the proposed design process for maximizing $\susceptibilityHotPara$ is as follows:
\begin{enumerate}
	\item Choose a preferred mass density, $\NmRef$.
	\item Let $\polarAvgA = \pi / 2$ and $\polarAvgB = 0$.
	Try to approach $\Std \rightarrow 0$.
	\item Let $\chainFracA = 2/3$ and, consequently, $\chainFracB = 1/3$.
	\item Maximize $\chaindensityref / \NmRef$ without exceeding the desired stiffness threshold(s).
	Note, when $\chainFracA = 2/3, \chainFracB = 1/3$:
	\begin{equation*}
	\begin{split}
	\ElasticPara &= \chaindensity \kB \T \left(1 + 2\pstchu^{-3}\right) \\
	\ElasticPerp &= \chaindensity \kB \T\left(\frac{1}{2} + \frac{5}{2}\pstchu^{3}\right)
	\end{split}
	\end{equation*}
	\item Choose $\NWeightA$; let $\NWeightB = \left(1 - \chainFracA \NWeightA\right) / \left(1 - \chainFracA\right)$:
	\begin{enumerate}
		\item The target $\NWeightA$ should be determined by:
		\begin{equation*}
		\NWeightA = \begin{cases}
		1/\chainFracA & \SusParaA + 2 \SusPerpA > \SusParaB + 2 \SusPerpB \\
		0 & \SusParaA + 2 \SusPerpA < \SusParaB + 2 \SusPerpB \\
		\NmRef / \chaindensityref & \SusParaA + \SusPerpA \approxeq \SusParaB + \SusPerpB
		\end{cases}
		\end{equation*}
		\item However, approaching either the upper limit of $\NWeightA$, which would result in chains of type $B$ to be ``short'' (i.e. $\NB$ small) or the lower limit of $\NWeightA$, which would result in chains of type $A$ to be ``short'', may affect the electromechanical response of the DE.
		This is because, as chains become shorter, higher order terms in \eqref{eq:Wstar-linearized} become more relevant at even moderate deformations.
		Specifically, these higher order terms would cause monomers to be constrained toward the direction of chain stretch more quickly as the DE deforms.
		This will lead to an increase in both strain hardening and electromechanical coupling.
		An additional consideration is the effect of chain length on the validity of the negligibly interacting assumption; it could be that, as chain lengths become shorter, interactions between chains become more important.
		Thus, either limit should be approached iteratively until it is determined how closely the limit can be approached without affecting the desired stiffness properties.
	\end{enumerate}
\end{enumerate}

We have shown that $\rstate{\left(\susceptibilityHotFree\right)}$ can theoretically be improved upon by deliberately designing and manufacturing the network architecture of a dielectric elastomer.
However, practically speaking, it can be seen from \eqref{eq:sus-hot-max} that, unless the network has a high density of cross-links and therefore consists of ``short chains'' (i.e. $\NmRef / \chaindensityref \leq 10$) the increase is modest at best (e.g. $\lesssim 1\%$ for $\NmRef / \chaindensityref \gtrsim 100$); and a high density of cross-links may not be desirable because the stiffness per mass density scales linearly with the density of cross-links.
In addition, the theoretical improvement may be lost entirely once manufacturing error inevitably occurs.

While this analysis shows that the susceptibility can only be marginally improved at best, it does provide useful insights:
(1.) it provides theoretical limitations of what can be achieved through the design of dielectric elastomer network architectures, and 
(2.) although the gains are modest for $\rstate{\left(\susceptibilityHotFree\right)}$, we will see below that there can be significant increases in the electromechanical coupling.
Indeed, recall that $\susceptibilityHotFree$ is also a function of deformation; so that larger increases in susceptibility may be realized at deformations such that $\pstchu \neq \rstate{\pstchu}$.
This effect will be investigated in the next subsection.

\subsection{Mechanically induced susceptibility} \label{sec:mechanically-induced-susceptibility}

The susceptibility of dielectric elastomers is deformation-dependent, because deformation can cause chains in the network to rotate (thereby changing the average monomer orientation in each of the chains) and cause chains to stretch (thereby increasing the concentration of monomers oriented toward the direction of stretch for a given chain).
Using the results developed thus far, we consider a few examples.

First, as a baseline, we consider an isotropic dielectric elastomer such that $\chainpdf = 1/4\pi$.
Using \eqref{eq:avg-relative-stretch-sq}, \eqref{eq:avg-stretch-Edotr-sq}, and \eqref{eq:sus-correction-hot}:
\begin{equation} \label{eq:def-induced-sus-iso}
\susceptibilityHotParaCorr^{uni} = \frac{2\chaindensity}{15}\left(\dsus\right)\left(\pstchu^{-1} - \pstchu^2\right)
\end{equation}
Notice that when $\dsus > 0$, $\susceptibilityHotParaCorr^{uni}$ increases (relative to the reference configuration) when the DE is compressed in the direction of the axis of symmetry ($\pstchu < 1$) and decreases when stretched; when $\dsus < 0$, vice versa.

For the hybrid network, where $\polarAvgA = \pi / 2, \polarAvgB = 0, \StdA = \StdB = 0, \chainFracA = 2/3, \chainFracB = 1/3$, we have:
\begin{equation} \label{eq:def-induced-sus-hybrid}
\susceptibilityHotParaCorr^{hybrid} = \frac{2 \chaindensity}{15}\left(|\dsusA|\pstchu^{-1} + |\dsusB|\pstchu^{2}\right)
\end{equation}
Interestingly, since both coefficients are strictly nonnegative and physically we require $\pstchu > 0$, it is the case that $\susceptibilityHotParaCorr^{hybrid}$ is semi-convex in $\pstchu$ for the domain of admissible $\pstchu$.
Further, it is convex when $|\dsusA| \neq 0$ and $|\dsusB| \neq 0$.
This is significant because it means that when either compressing (i.e. $\pstchu < 1$) or stretching (i.e. $\pstchu > 1$) -- even though initially there may be a drop in $\susceptibilityHotParaCorr$ -- eventually $\susceptibilityHotParaCorr$ will begin increasing again and do so monotonically.
\emph{The electromechanical increase of $\susceptibilityHotPara$ is bidirectional for the hybrid network}.
If $|\dsusA| = 2|\dsusB|$, then $\susceptibilityHotParaCorr$ has its minimum at $\pstchu = 1$ so that any deformation increases $\susceptibilityHotParaCorr$.

Also, importantly, \eqref{eq:def-induced-sus-iso} and \eqref{eq:def-induced-sus-hybrid} show that the mechanically induced susceptibility of the hybrid network is greater than that of the uniform network for all admissible deformations, $\pstchu$.

Next, in contrast to either a uniformly distributed network or the hybrid network, we consider a network that consists of a single type of monomer and has been manufactured in the limit of high control, i.e. $\Std \ll 1$.
In this case: using \eqref{eq:Izero}, \eqref{eq:avg-relative-stretch-sq}, \eqref{eq:avg-stretch-Edotr-sq}, and \eqref{eq:sus-correction-hot}:
\begin{equation} \label{eq:def-induced-sus-0}
    \susceptibilityHotParaCorrZero = -\frac{\chaindensity }{20} \left(\dsus\right) \left( \left[\left(1+2\auxc{2}\right)\auxe{-4}-3\right]\pstchu^{-1} + \left[2\left(1+2\auxc{2}\right)\auxe{-4}+2\right]\pstchu^{2} \right)
\end{equation}
When $\polarAvg = 0$ the coefficient of $\pstchu^{-1}$ vanishes and when $\polarAvg = \pi / 2$ the coefficient of $\pstchu^2$ vanishes.
This is expected because the sign and coefficient of $\pstchu^{-1}$ and $\pstchu^{2}$ determine the effect of compression and stretching, respectively, on $\susceptibilityHotParaCorr$.
The above equation could be used as a design tool for anisotropic elastomers that consist of a single monomer type.
For further physical insight, we let $\Std = 0$.
Then \eqref{eq:def-induced-sus-0} simplifies to:
\begin{equation*}
    \susceptibilityHotParaCorrZero\Bigr|_{\Std=0} = -\frac{\chaindensity\left(\dsus\right)}{10}\left\{\left[\auxc{2}-1\right]\pstchu^{-1} + 2\left[\auxc{2}+1\right]\pstchu^2\right\}.
\end{equation*}
Importantly, the theoretical factor for mechanically induced susceptibility can be larger in this case than it is for the uniform or hybrid networks.
That is because, in this case, we can orient all of the chains in their preferred electromechanical susceptibility direction instead of a fraction.
The maximum factors occur at $\polarAvg = 0$ and $\polarAvg = \pi/2$, as expected.
At $\polarAvg = 0$, the $\pstchu^{-1}$ factor (i.e. compression factor) vanishes and the $\pstchu^{2}$ factor (i.e. expansion factor) is maximized--its value being $2 \chaindensity / 5$.
And at $\polarAvg = \pi / 2$, the $\pstchu^{2}$ factor vanishes and the $\pstchu^{-1}$ factor is maximized--its value being $\chaindensity / 5$.
These factors are $3 / 2$ and $2$ times larger than the hybrid network factors, respectively.
However, recall that $\ElasticPerp$ vanishes as $\polarAvg \rightarrow 0$ and $\ElasticPara$ vanishes as $\polarAvg \rightarrow \pi/2$.
Thus, while a greater mechanically-induced susceptibility can be achieved (over the uniform and hybrid networks), there is a trade-off in terms of stiffness and mechanical stability.
The implications of this in terms of the electromechanical coupling, operation, and failure of DEAs will be explored in the next section.

\section{Cross modulus, Electrostriction, and Material Stability}

\label{sec:electrostrictive-stress}

We now turn our attention to the electromechanical or cross modulus, $\takecrosspartial{\W}{\Polar}{\F}$, which provides a measure of the electromechanical coupling of the material.
Among other things, we will show here that it is related to both the stress that develops in the DE when loaded by a constant electric field and as well as the stability of the material.

To derive the cross modulus of our DEs, we begin with the relation:
\begin{equation*}
	\takepartial{\W}{\Polar} = \efield = \susceptibilityTensorLab^{-1} \polarization.
\end{equation*}
Let $\crossModulusTensor$ denote the cross modulus tensor.
Then, taking derivatives of both sides of the above equation with respect to the deformation gradient, we obtain:
\begin{equation} \label{eq:cross-modulus}
	\crossModulusTensor \coloneqq \takecrosspartial{\W}{\Polar}{\F} = \takepartial{\left(\susceptibilityTensorLab^{-1}\right)}{\F} \Polar
\end{equation}
which can be directly related to derivatives of $\Wstar$, as outlined in \Fref{sec:susceptibility}.
In order to highlight some important properties of the cross modulus, we will again consider the setting of a thin film as described in \Fref{sec:electrostriction}, but now using transversely isotropic materials.
By symmetry, we expect $\F = \diag\left(1 / \sqrt{\pstchu}, 1 / \sqrt{\pstchu}, \pstchu\right)$ in a Cartesian basis with the $3$ direction aligned along the thickness direction.
Further, the symmetries of the material, deformation, and applied electric field provide a simplified form of \eqref{eq:cross-modulus} along the axis of symmetry
\begin{equation}
    \crossModulusPara = -\frac{\emag}{\susceptibilityLabPara} \takepartial{\susceptibilityLabPara}{\pstchu}, \quad \text{ where } \susceptibilityLabPara\left(\pstchu, \emag\right) = \polarizationmag / \emag.
\end{equation}
\hl{Note that since $\crossModulusPara \propto \emag$, we do not see a frozen in polarization (i.e. the material does not spontaneously become an electret) for any relaxation stretch, $\rstate{\pstchu}$, as expected.}

\begin{figure} 
	\centering
	\input{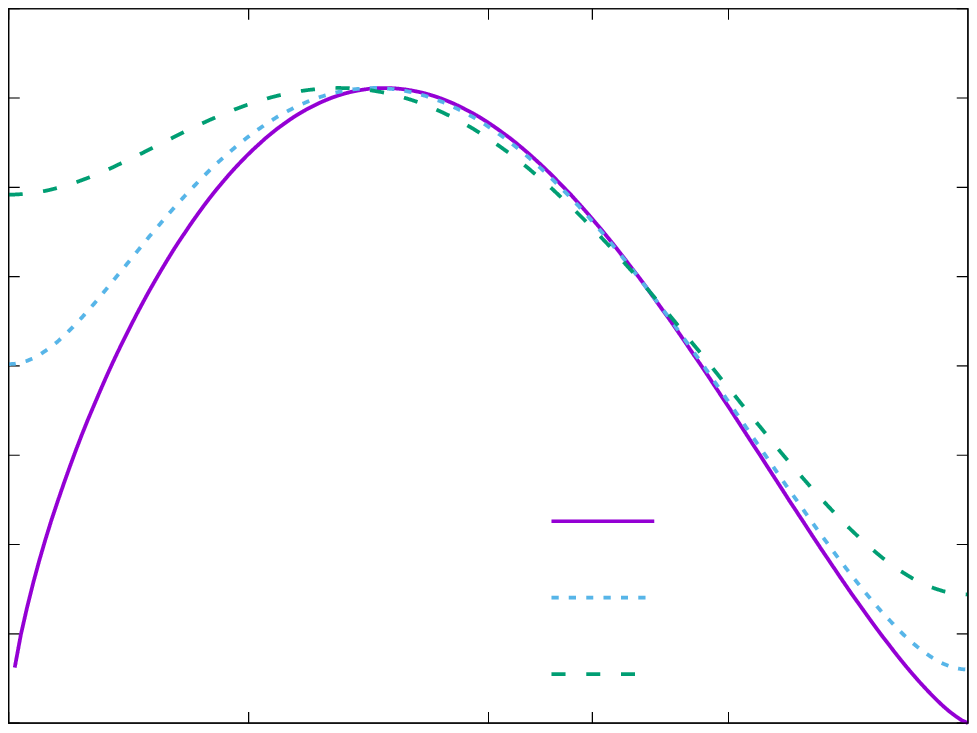}
	\caption{Cross modulus as a function of $\polarAvg$ for $\unodim = 1$. 
  The absolute value of the cross modulus does not obtain its maximum on the bounds of $\polarAvg$, nor does it decay with $\Std$.}
	\label{fig:CrossModulus-vs-theta0_TI}
\end{figure}

\begin{figure} 
	\centering
	\input{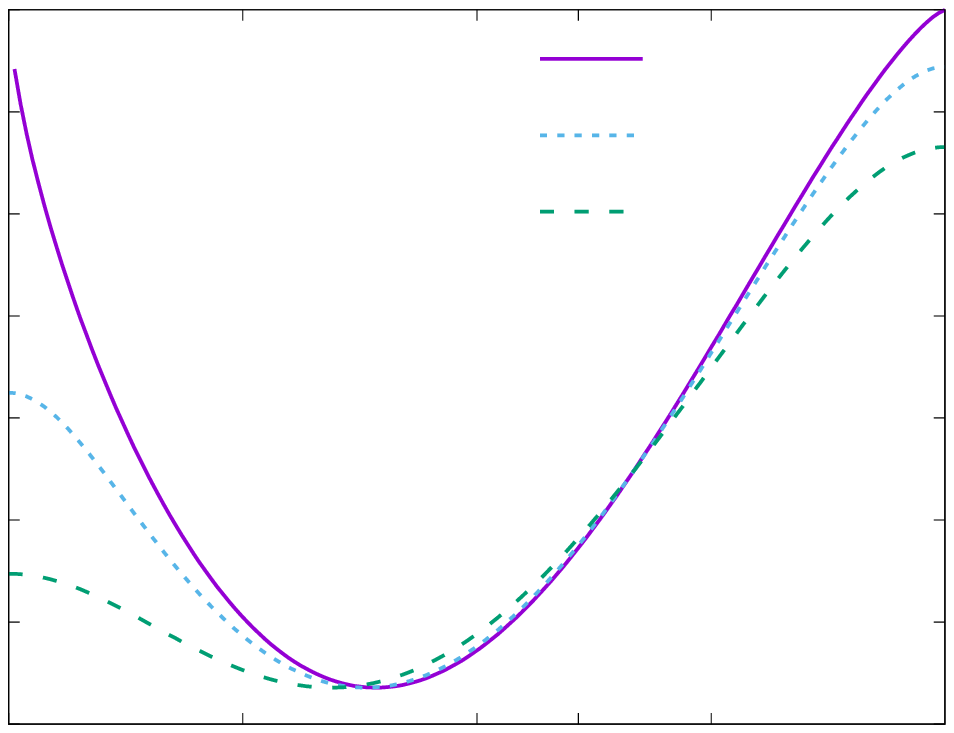}
	\caption{Cross modulus as a function of $\polarAvg$ for $\unodim = -1$. 
  The absolute value of the cross modulus does not obtain its maximum on the bounds of $\polarAvg$, nor does it decay with $\Std$.}
	\label{fig:CrossModulus-vs-theta0_Uni}
\end{figure}

For brevity, we restrict our attention to the limit of $|\Std| \ll 1$.
The cross modulus as a function of $\polarAvg$ is shown in \Fref{fig:CrossModulus-vs-theta0_TI} and \Fref{fig:CrossModulus-vs-theta0_Uni} $\unodim = 1$ and $\unodim = -1$, respectively.
For both figures, $\N = 100$.
Interestingly, the absolute value of the cross modulus does not have its maximum, in either case, at $\polarAvg = 0, \pi / 2$, or $\polarAB$; and it does not decay as the distribution spreads out ($\Std > 0$).
Instead, the cross modulus vanishes for $\Std = 0$ at $\polarAvg = 0$ and $\pi / 2$.

The significance of the cross modulus is subtle and is perhaps best understood through the lens of the experiment shown in \Fref{fig:electrostriction}: a constant electric field, $\ezeromag$, is applied to a dielectric elastomer with (effectively) traction free boundary conditions.
Since the deformation gradient and electric field are both homogeneous, the Gibbs free energy minimization (\eqref{eq:gibbs-min}) can be carried out pointwise.
Furthermore, the polarization is work-conjugate to the applied electric field; thus, the Gibbs free energy density is given by:
\begin{equation*}
	\W\left(\pstchu, \Polarmag\right) - \Polarmag \ezeromag
\end{equation*}
or, equivalently, $\Wstar$.
Now consider an anisotropic DE that has been manufactured and allowed to relax to $\pstchu = \rstate{\pstchu}$.
Then, after relaxation, the electric field is applied.
The stress that develops in the DE as a result of the electric field (denote by $\stressij$) is given by:
\begin{equation*}
	\stressij \coloneqq \takepartial{\Wstar}{\pstchu}\Bigr|_{\pstchu = \rstate{\pstchu}, \emag = \ezeromag} \propto \rstate{\left(-\takepartial{\susceptibilityLabPara}{\pstchu}\right)} \propto \rstate{\crossModulusPara}
\end{equation*}
There is a clear connection between the cross modulus and the stress as they are both proportional to change in the secant susceptibility with respect to deformation.
After our discussion in \Fref{sec:mechanically-induced-susceptibility}, there should be no surprise that this contribution to stress exists.
The existence of a mechanically induced susceptibility implies a stress induced by changing susceptibility, and vice versa.

For comparison to the cross modulus (depicted in \Fref{fig:CrossModulus-vs-theta0_Uni} and \Fref{fig:CrossModulus-vs-theta0_Uni}), we also present the formula for the electrically induced stress (which we will refer to as the \emph{electrostrictive stress}):
\begin{equation*}
	\frac{\stressij}{\chaindensity \kB \T} = \frac{3 \unodim \auxe{-4}}{10\times 2^{1/3}}  \left[3\auxe{4} - 1 - 2\auxc{2}\right]^{1/3} \left[\auxe{4} + 1 + 2\auxc{2}\right]^{2/3}
\end{equation*}
Notice that the stress is also proportional to $\unodim$ and its sign depends on whether the chains consist of TI monomers or uniaxial monomers.
The stress attains its maximum absolute value at
\begin{equation*}
	\polarAvg = \frac{1}{2} \arccos\left(\frac{1}{3}\right)
\end{equation*}
which, to a good approximation, is also where $\rstate{\crossModulusPara}$ attains its maximum absolute value.

There are a few important takeaways here that will be relevant to the design of DEs for dielectric elastomer actuators in \Fref{sec:deas}.
First, there are two apparent contributions to the electromechanical coupling in DEAs: (1.) the Coulomb attraction between the top and bottom electrodes of the DEA and (2.) the contribution to the stress in the DE due to the change in the secant susceptibility with respect to deformation.
As discussed previously, the former can be increased by increasing the susceptibility of the DE.
The latter, as we have shown in this section, can be increased by increasing the cross modulus of the material.
Since the susceptibility and cross modulus are optimal for different network architectures, there will be a competition between the two when optimizing the network architecture for a DEA.

The second takeaway is this: \emph{the amount of electrostriction of the anisotropic DEs under consideration, for a fixed $\ezeromag$, is independent of the chosen network architecture}; that is, if we make the decomposition $\pstchu = \pstchpr \rstate{\pstchu}$ where $\rstate{\pstchu}$ is the initial relaxation of the DE and $\pstchpr$ is the deformation that occurs as a result of the electrically induced stress, then $\pstchpr$ is invariant with respect to $\chainpdf$, $\chaindensity$, $\N$, and $\lefrac$.
This is indeed surprising, but we can begin to explain it as follows: 
\begin{enumerate}
	\item There are two consequences of the linearized theory that result in invariance with respect to $\lefrac$ and $\N$:
	\begin{enumerate}
		\item As before, $\lefrac$ is not accounted for in the linearized free energy density because it is has a higher order effect related to the finite extensibility of the chains in the network.
		\item Although the change in susceptibility with respect to deformation does depend on $\N$ (since it depends on the absolute stretch of the chain), this is also a higher order effect related to finite extensibility.
	\end{enumerate}
	\item The invariance with respect to $\N$ occurs because both $\ElasticPara$ and $\stressij$ are linearly proportional to $\chaindensity \kB \T$.
	\item The invariance with respect to $\chainpdf$ is surprising since there is clearly an influence of the parameters $\left(\polarAvg, \Std\right)$ on the cross modulus and the stress.
	However, note that $\ElasticPara$ also depends on $\left(\polarAvg, \Std\right)$; and, importantly, $\crossModulusPara$ and $\ElasticPara$ both depend on the deformation (see \eqref{eq:cross-modulus} and \eqref{eq:ElasticParaTI}, respectively).
	This interplay between $\polarAvg, \Std, \pstchu, \ElasticPara$ and $\crossModulusPara$ results in $\pstchpr$ being invariant with respect to $\polarAvg$ and $\Std$.
\end{enumerate}
For completeness, we also provide the formula for $\pstchpr$:
\begin{equation} \label{eq:ECtrl-stretch}
\pstchpr = \left(\frac{45 + 4\unodim - 30 \auxf}{45 + 22\unodim - 30 \auxf}\right)^{1/3}
\end{equation}
where $\auxf$ was defined earlier in \eqref{eq:free-polarization-Astar}.

\subsection{Material stability} \label{sec:material-stability}

\newcommand{\eq}[1]{#1^{\text{eq}}}
\newcommand{\pert}[1]{\Delta #1}

In addition to the electromechanical coupling of our anisotropic, design DEs, we will also be interested in designing against failure; here, we define failure as loss of stability of the free energy, not considering fracture, breakdown \&c.
For this purpose, consider again the Taylor expansion of the Helmholtz free energy density given in \eqref{eq:W-taylor}.
However, here we take the expansion about a general equilibrium configuration, $\left(\eq{\pstchu}, \eq{\Polarmag}\right)$.
Then, for perturbations about the equilibrium configuration, $\left\{\pert{\pstchu}, \pert{\Polarmag}\right\}$, the change in the free energy density is:
\begin{equation*}
	\W\left(\pstchu, \Polarmag\right) - \W\left(\eq{\pstchu}, \eq{\Polarmag}\right) = \frac{1}{2} \begin{Bmatrix}
		\pert{\pstchu} \\ \pert{\Polarmag}
	\end{Bmatrix}^T \left[\begin{matrix}
		\eq{\ElasticPara} & \eq{\crossModulusPara} \\
		\eq{\crossModulusPara} & \left(\eq{\susceptibilityPara}\right)^{-1}	\end{matrix}\right] \begin{Bmatrix}
	\pert{\pstchu} \\ \pert{\Polarmag}
	\end{Bmatrix} + \bigoof{\smallparam^3}
\end{equation*}
where $|\smallparam| = \max\left(|\pert{\pstchu}|, |\pert{\Polarmag}|\right) \ll 1$ is a small parameter and the superscript, $\eq{\generic}$, on the material properties are there to emphasize that each of the second partial derivatives are evaluated at $\left(\pert{\pstchu}, \pert{\Polarmag}\right)$.
Moving forward, we drop the superscript as it will be clear what is meant from the context. 
Next, define the Hessian of the free energy density:
\begin{equation*}
	\hessian \coloneqq \left[\begin{matrix}
	\ElasticPara & \crossModulusPara \\
	\crossModulusPara & \susceptibilityPara^{-1} \end{matrix}\right], \quad \hessianDet \coloneqq \det\hessian,
\end{equation*}
For a material to be stable (at a given equilibrium configuration), we require that the energy increase for all \emph{allowable} perturbations\footnote{
	As pointed out in~\cite{ericksen1998introduction}, although we are actually interested in changes in the free energy (as opposed to the free energy \emph{density}), it often suffices to consider stability in the context of the free energy density.
    \hl{Here we do not consider global instabilities, such as wrinkling and buckling, which are often geometry and/or boundary condition dependent, and instead focus on local failure modes.
    For dielectric elastomers, the principal local failure mode is characterized by a dramatic thinning}~\cite{zhao2007method,zurlo2017catastrophic}.
    \hl{Typical global failure modes include wrinkling, buckling, and dielectric breakdown.
    See}~\cite{zurlo2017catastrophic,yang2017revisiting} \hl{for a more comprehensive analysis and discussion.
    Note that, when failure is studied and observed in dielectric elastomers, the material is generally subjected to a Maxwell stress.
    Importantly, we show later in this section that, for both standard isotropic and the proposed anisotropic materials alike, a local failure mode will not occur in the absence of a Maxwell stress.}
}.
Here, we neglect the role of constraints, thus, a sufficient condition for stability is that $\hessian$ be positive definite.

As a first step toward understanding the properties of $\hessian$, let us consider its diagonal: $\ElasticPara$ and $\susceptibilityPara^{-1}$.
Directly, we obtain
\begin{equation*}
\begin{split}
    \ElasticPara = \frac{\auxe{-4}}{60} \chaindensity \kB \T \Bigg( & 
        18\unodim\left(1 + \auxe{4} + 2\auxc{2}\right) \\
        & + \left[\auxe{4}\left(1+3\pstchu^{-3}\right)+\left(1-\pstchu^{-3}\right)\left(1+2\auxc{2}\right)\right]\left[4\unodim+15\left(3+\ln\pi\right)-30\auxf\right] \Bigg)
\end{split}
\end{equation*}
and, further, using $\pstchu = \pstchpr \rstate{\pstchu}$ (with \eqref{eq:eql-stretch} and \eqref{eq:ECtrl-stretch})
\begin{equation*}
	\ElasticPara = \frac{1}{20} \chaindensity \kB \T \left[1 + \auxe{-4}\left(1 + 2\auxc{2}\right)\right]\left(45 + 22\unodim - 30\auxf\right).
\end{equation*}
It is not difficult to see that the quantity enclosed in the braces is nonnegative and $45 + 22\unodim - 30\auxf$ is strictly positive.
Thus, the stiffness is never negative in this case.
Similarly, although the formula for $\susceptibilityPara$ is tediously long, we can easily reason that it too is nonnegative.
The quantity $\susceptibilityPara$ is the result of averaging across scales, starting at the monomer dipole susceptibility.
Since the monomer dipole susceptibility is nonnegative, the subsequent averages that we obtain must also be nonnegative.
Further, $\susceptibilityPara^{-1} > 0$, since by similar reasoning there must be an upper bound on $\susceptibilityPara$ (this upper bound is, of course, $\chaindensity \N \sussymbol_M$ where, as before, $\sussymbol_M = \max\left(\sus{1}, \sus{2}\right)$).
Now, ordinarily
\begin{equation*}
	\hessianDet \geq 0
\end{equation*}
would only be a necessary, and not a sufficient, condition for the positive semi-definiteness (positive definiteness when $\hessianDet > 0$) of $\hessian$.
However, since the trace is also nonnegative, the above condition is also sufficient.
Thus, if $\hessianDet > 0$ then stability is guaranteed; otherwise, the material \emph{may be} unstable.
We forego the nuances of exactly when the design DE is unstable, for simplicity, and focus on how $\hessianDet$ depends on different network architectures and loading conditions.

\begin{figure}[htb!]
	\centering
	\input{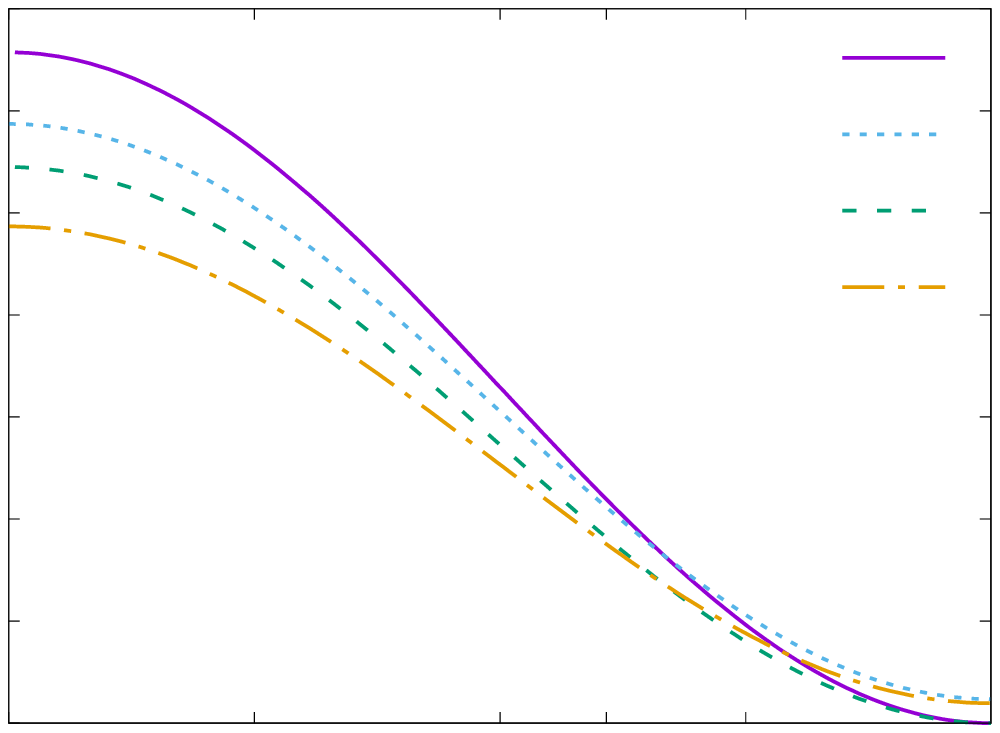}
	\caption{$\hessianDet$ as a function of $\polarAvg$ for different network architectures: $\left(\Std = 0, \unodim = 1\right)$, $\left(\Std = \pi / 16, \unodim = 1\right)$, $\left(\Std = 0, \unodim = -1\right)$, and $\left(\Std = \pi / 16, \unodim = 1\right)$.
	For all architectures considered, $\N = 100$.
	Similar to the $\unodim = 0$ case, $\hessianDet$ only vanishes when the stiffness vanishes.}
	\label{fig:Determinant-vs-theta0}
\end{figure}

Since $\crossModulusPara$ is linear in $\Polarmag$, when $\unodim = 0$, $\hessianDet$ is trivially positive for all network architectures, $\left(\polarAvg, \Std\right)$--except when $\left(\polarAvg = \pi / 2, \Std = 0\right)$ and $\ElasticPara$ theoretically vanishes (in which case, $\hessianDet = 0$).
\Fref{fig:Determinant-vs-theta0} shows $\hessianDet$ as a function of $\polarAvg$ for $\left(\Std = 0, \unodim = 1\right)$, $\left(\Std = \pi / 16, \unodim = 1\right)$, $\left(\Std = 0, \unodim = -1\right)$, and $\left(\Std = \pi / 16, \unodim = 1\right)$ ($\N = 100$ for all cases).
Similar to the $\unodim = 0$ case, $\hessianDet$ only vanishes when the stiffness vanishes.

\begin{figure}[htb!] 
	\centering
	\input{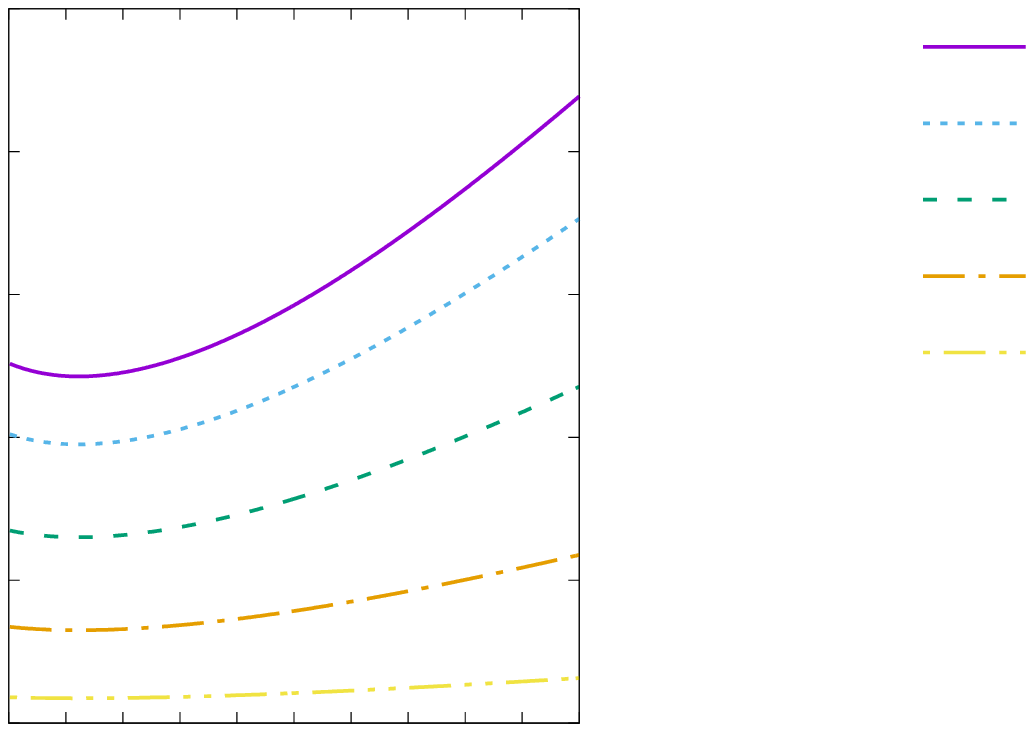}
	\caption{$\hessianDet$ as a function of $|\unodim|$ for networks consisting of TI chains and with $\N = 100$ and $\Std = 0$. 
	The $\polarAvg$ considered are $\pi / 12$, $\pi / 6$, $\pi / 4$, $\pi / 3$, and $5\pi/12$.
		$\hessianDet$ decreases slightly before increasing with $|\unodim|$; and importantly, it does not approach zero.
	Therefore, material instability does not occur.}
	\label{fig:Determinant-vs-kappa_TI}
\end{figure}

\begin{figure}[htb!]
	\centering
	\input{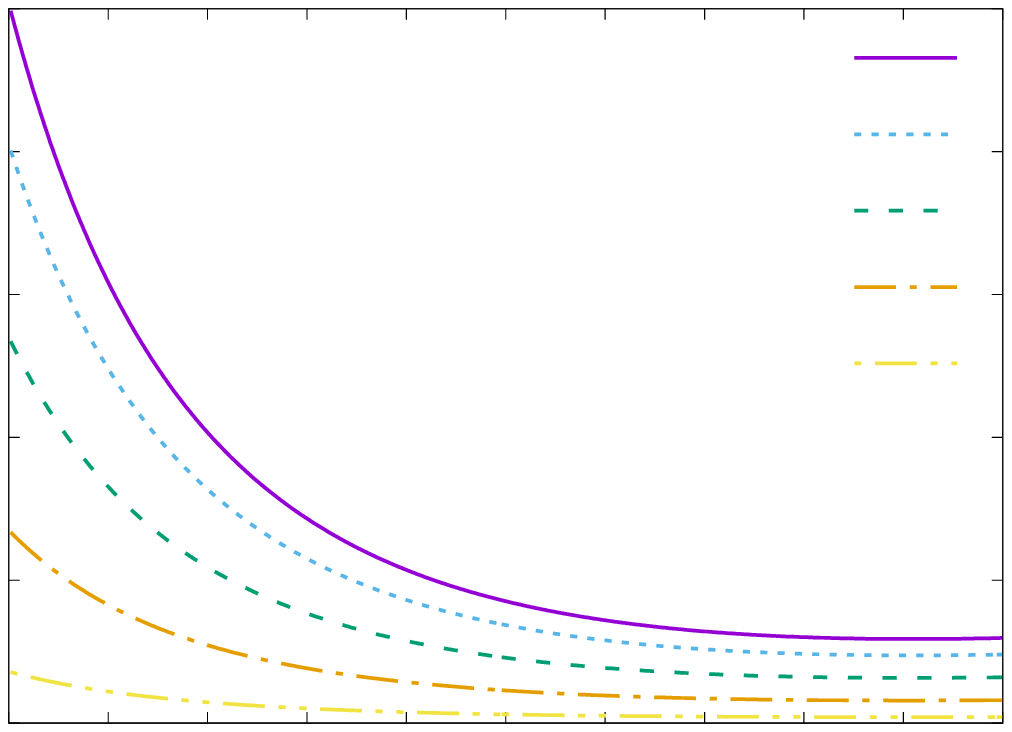}
	\caption{$\hessianDet$ as a function of $|\unodim|$ for networks consisting of uniaxial chains and with $\N = 100$ and $\Std = 0$. 
		The $\polarAvg$ considered are $\pi / 12$, $\pi / 6$, $\pi / 4$, $\pi / 3$, and $5\pi/12$.
		$\hessianDet$ decreases with $|\unodim|$ before leveling off asymptotically to some positive value; and importantly, it does not approach zero.
		Therefore, material instability does not occur.}
	\label{fig:Determinant-vs-kappa_Uni}
\end{figure}

It is reasonable to then ask: does the stability picture change as $|\unodim|$ increases?
\Fref{fig:Determinant-vs-kappa_TI} and \Fref{fig:Determinant-vs-kappa_Uni} show $\hessianDet$ as a function of $|\unodim|$ for architectures consisting of TI and uniaxial chains, respectively.
For each figure, $\N = 100$, $\Std = 0$, and $\polarAvg$ is varied by series: $\pi / 12$, $\pi / 6$, $\pi / 4$, $\pi / 3$, and $5\pi/12$.
\Fref{fig:Determinant-vs-kappa_TI} shows that, for architectures consisting of TI chains, $\hessianDet$ decreases slightly before increasing with $|\unodim|$.
Importantly, it does not approach zero in any case.
For architectures consisting of uniaxial chains (\Fref{fig:Determinant-vs-kappa_Uni}), $\hessianDet$ decreases with $|\unodim|$ before leveling off asymptotically to some number greater than zero.
Therefore, we see, these anisotropic DEs do not become unstable when loaded by a constant electric field.

We conclude this section by explaining why we have not seen instability in these materials.
The reason becomes apparent when we consider how $\ElasticPara$, $\crossModulusPara$ and $\susceptibilityPara$ scale.
Specifically:
\begin{align*}
	\ElasticPara &= \bigoof{\chaindensity \kB \T} + \bigoof{\chaindensity \kB \T \times \unodim} = \bigoof{\chaindensity \kB \T} + \bigoof{\chaindensity \ezeromag^2 \sussymbol_M}, \\
	\susceptibilityPara &= \bigoof{\chaindensity \N \sussymbol_M}, \\
	\crossModulusPara &= \bigoof{\ezeromag \N^{-1}}.
\end{align*}
And, consequently,
\begin{equation*}
\hessianDet = \bigoof{\N^{-1}} \left[\bigoof{\sussymbol_{M}^{-1} \kB \T} + \bigoof{\ezeromag^2}\right] - \bigoof{\ezeromag^2 \N^{-2}}.
\end{equation*}
Since in many cases $\N \sim 100$--$10000$, the strictly nonpositive contributions to $\hessianDet$ are negligible compared to the strictly nonnegative.
The main reason why this scaling occurs is because $\susceptibilityPara$ scales with $\chaindensity \N$ but $\partial \susceptibilityPara / \partial \pstchu$ only scales as $\chaindensity$.

In the next section, we explore how the performance of a dielectric elastomer actuator changes with the network architecture.
The significance of the results developed in this section will be two-fold.
First, as mentioned in \Fref{sec:electrostrictive-stress}, the electrostrictive stress that develops in the DE film due to its polarization response will effect the performance of the DE as an actuator; and, importantly, the electrostrictive stress is not optimal for the same architectures in which the polarization susceptibility is optimal.
Secondly, we will also consider the failure of DEAs due to instability.
When we do, we will be justified in dropping the Hessian-based approach and instead only consider the sign of $\partial^2 \Gibbs / \partial \pstchu^2$--as the sign of $\partial^2 \Gibbs / \partial \pstchu^2$ captures the instability associated with the Coulomb attraction overcoming the stiffness of the film.

\section{Application to Dielectric Elastomer Actuators} \label{sec:deas}

In this section, we explore the effect of our design parameters on the deformation and usable work obtained from a dielectric elastomer actuator (DEA) as a function of its electrical input.
There are two main goals associated with this design: (1.) we would like to maximize the deformation and/or usable work that results from a fixed electrical input; and (2.) we would typically like to maximize the deformation and/or usable work that the DEA can produce before its failure.

\begin{figure} 
	\centering
	\includegraphics[width=\linewidth]{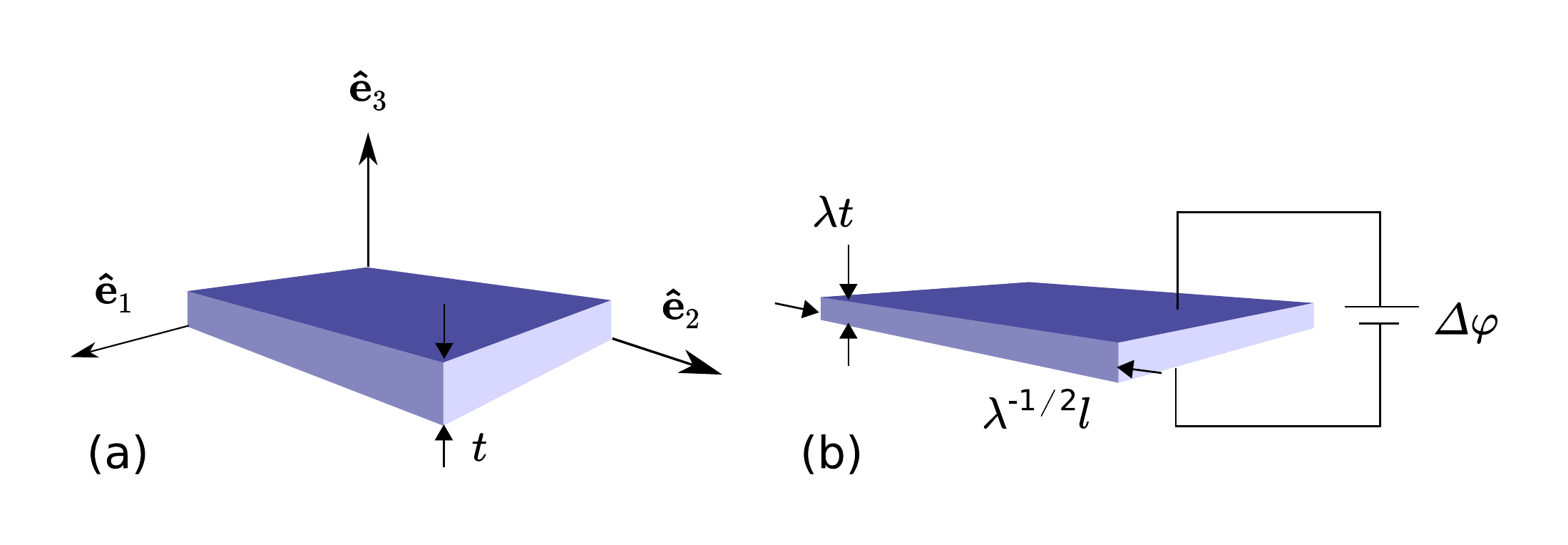}
	\caption{Dielectric elastomer actuator in the reference configuration (a) and deformed state (b).
		A voltage difference is applied across the top and bottom surfaces, which are then attracted to each other in accordance with Coulomb's law.
		The attraction between the electrodes compresses the film across its thickness.
		Additional stress develops as a result of the electromechanical coupling of the material itself~\cite{cohen2016electromechanical,mcmeeking2005electrostatic,zhao2008electrostriction,suo2010theory}.}
	\label{fig:dea-schematic}
\end{figure}

Let the dimensions of the DEA in the reference configuration be $\Lw \times \Lw \times \Le$, where $\Le \ll \Lw$ is the thickness of the DE film.
The top and bottom surfaces of the DE film are assumed to be covered with compliant electrodes with negligible stiffness.
When operating the DEA, a fixed voltage difference, $\epotDiff$, is applied across the top and bottom surfaces.
Consequently, equal and opposite net charges, $\pm Q$, accumulate on the top and bottom surfaces, which are then attracted to each other in accordance with Coulomb's law.
The attraction between the electrodes compresses the film across its thickness.
And an additional stress develops as a result of the electromechanical coupling of the material itself~\cite{cohen2016electromechanical,mcmeeking2005electrostatic,zhao2008electrostriction,suo2010theory} (see also \Fref{sec:electrostrictive-stress}).
The DEA setup is shown in \Fref{fig:dea-schematic}.

Following~\cite{yang2017revisiting} and~\cite{zhao2007method}, the Gibbs free energy of a DEA undergoing homogeneous deformation, $\F = \diag\left(\pstchu^{-1/2},\pstchu^{-1/2},\pstchu\right)$, is:
\begin{equation} \label{eq:DEA-gibbs}
    \gibbs = \Lw^2 \Le \left(\W + \frac{1}{2}\emag^2\right) - \left(\epotDiff\right) Q
\end{equation}
where the first term in the parentheses is the free energy density of the DE film and the second term in the parentheses is the energy density of the electric field; and the last term in \eqref{eq:DEA-gibbs} is the work of the battery.
However, \eqref{eq:DEA-gibbs} is in terms of $\W$ instead of $\Wstar$.
Recalling that $\W = \Wstar + \polarization \cdot \efield$ and using Gauss's law:
\begin{equation} \label{eq:DEA-gibbs2}
    \gibbs = \Lw^2 \Le \left(\Wstar - \frac{1}{2}\emag^2\right).
\end{equation}
In deriving \eqref{eq:DEA-gibbs2}, we have assumed that $\emag = -\left(\epotDiff\right) / \pstchu \Le$ inside the DE film, which is a good approximation when $\Le \ll \Lw$ and the DE film has homogeneous material properties.

As mentioned previously, there is an initial relaxation, in general, after manufacturing an anisotropic dielectric elastomer.
To account for this, we propose the following process:
\begin{enumerate}
	\item The DE film is manufactured; it relaxes to some $\pstchu = \rstate{\pstchu}$.
	In anticipation of this relaxation, its initial, pre-relaxation thickness is $\Le / \rstate{\pstchu}$.
	\item A voltage difference, $\epotDiff$, is applied across the electrodes.
	\item The DE film deforms further by some amount $\pstchpr$ such that $\pstchu = \pstchpr \rstate{\pstchu}$.
\end{enumerate}
To model this process, we reformulate \eqref{eq:DEA-gibbs2} as:
\begin{equation}
    \gibbs\left(\pstchpr\right) = \Lw^2 \Le \left[\Wstar\left(\rstate{\pstchu}\pstchpr, \Emag {\pstchpr}^{-1}\right) - \frac{1}{2} \Emag^2 {\pstchpr}^{-2}\right]
\end{equation}
where $\Emag = \epotDiff / \Le$.
The stable equilibrium states of the DEA are given by $\frac{\dm \gibbs}{\dm \pstchpr} = 0, \frac{\dm^2 \gibbs}{\dm {\pstchpr}^2} > 0$; the DEA becomes unstable when $\frac{\dm \gibbs}{\dm \pstchpr} = 0, \frac{\dm^2 \gibbs}{\dm {\pstchpr}^2} \leq 0$ (this simplified choice of stability criteria was justified in \Fref{sec:material-stability}).
The equation $\frac{\dm \gibbs}{\dm \pstchpr} = 0$, however, is a highly nonlinear function of $\pstchpr$.
Even if one uses a Taylor expansion (about $\Emag=0$) of $\auxf\left(\Emag^2 {\pstchpr}^{-2} \left(\dsus\right) / 2 \kB \T\right)$ terms (see \eqref{eq:free-polarization-Astar}) and truncates higher order terms, the equilibrium equation still requires solving for the roots of a $5$th order polynomial.
Instead, moving forward, we use Newton's method with an initial guess of $\pstchpr = 1$.


\subsection{Deformation}

Let $\Enodim \coloneqq \Emag \sqrt{\N \susceptibilityHotFree / \kB \T}$.
This is a nondimensional measure of the electric field, similar to the dimensionless electric field used in~\cite{zhao2007electromechanical,zurlo2017catastrophic}.
Because we would like to emphasize optimization here, we will focus on three types of networks: 
(1) a uniform (isotropic) network as a baseline, 
(2) network designs in the limit of $\sigma = 0$, and 
(3) the aforeproposed hybrid network.
Similarly, because (for unitype networks) elastomers that consist of TI monomers will have an greater electromechanically induced susceptibility than elastomers consisting of uniaxial monomers, we will consider elastomers such that $\dsus = \sus{2}$ for the uniform and $\sigma = 0$ networks.

\begin{figure} 
	\centering
	\input{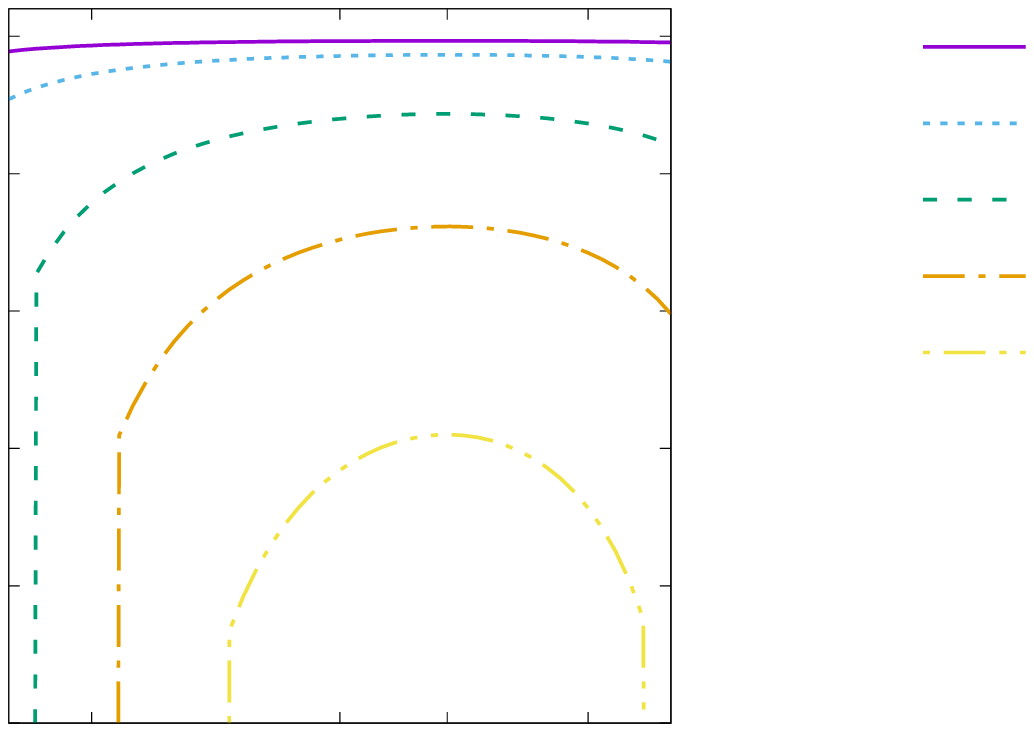}
	\caption{$\pstchpr$ vs $\polarAvg$ contours of constant $\Enodim$ for networks of type (1) and (2).
The end of each data series (vertical line to the x-axis) represents failure of the DEA for the corresponding network architecture and applied electric field.
		For each of the contours, $\polarAvg = \polarAB$ has the least amount of deformation and the contours are symmetric about $\polarAB$.
		However, while the deformation increases as $\polarAvg$ has a larger deviation from $\polarAB$, the DEA also fails at smaller $\Enodim$.
		Note that $\polarAvg = \polarAB$ has an equivalent electromechanical response to the uniform network--so $\polarAvg = \polarAB$ represents our isotropic baseline.}
	\label{fig:VCtrl_lambda-vs-theta0_TI_I0}
\end{figure}

First, \Fref{fig:VCtrl_lambda-vs-theta0_TI_I0} shows $\pstchpr$ vs $\polarAvg$ contours of constant $\Enodim$ for networks of type (1) and (2).
Note that the end of each data series (vertical line to the x-axis) represents failure of the DEA for the corresponding network architecture and applied electric field.
It can be seen in \Fref{fig:VCtrl_lambda-vs-theta0_TI_I0} that, for each of the contours, $\polarAvg = \polarAB$ has the least amount of deformation.
In fact, the contours are symmetric about $\polarAB$.
However, while the deformation increases as $\polarAvg$ has a larger deviation from $\polarAB$, the DEA also fails at smaller $\Enodim$.
Thus, there is a trade-off between the maximizing the deformation for a fixed $\Enodim$ and maximizing the operating $\Enodim$ before failure--and, consequently, the maximum deformation possible.
Put differently, there is a trade-off between electromechanical efficiency and stability.
Note that it can be seen from numerical examples that $\polarAvg = \polarAB$ has an equivalent electromechanical response to the uniform network.
Thus, $\polarAvg = \polarAB$ represents our isotropic baseline.

\begin{figure}
	\centering
	\input{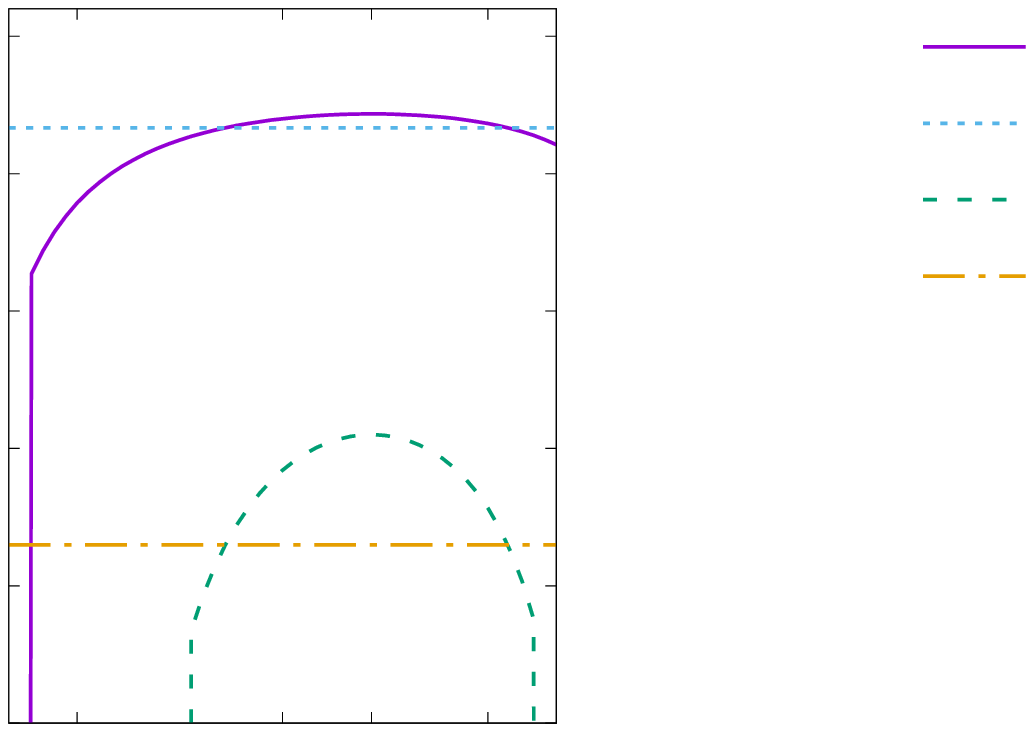}
	\caption{Comparison of the DEA deformation of the hybrid network to the unitype networks with $\Std = 0$.
		The unitype networks have varying $\polarAvg$.
		Notice: for a given $\Enodim$, the hybrid network deforms more than the unitype isotropic network (i.e. $\polarAvg = \polarAB$).
		However, for $\Enodim$ low enough, there is always some $\polarAvg, \Std = 0$ for which the unitype network has a large induced deformation than the hybrid network.
	}
	\label{fig:VCtrl_lambda-vs-theta0_I0-vs-Hybrid}
\end{figure}

Next, we consider the hybrid network within this context.
\Fref{fig:VCtrl_lambda-vs-theta0_I0-vs-Hybrid} compares the DEA deformation of the hybrid network to the unitype networks with $\Std = 0$.
For a given $\Enodim$, the hybrid network deforms more than the unitype isotropic network (i.e. $\polarAvg = \polarAB$).
However, for $\Enodim$ low enough, there is always some $\polarAvg, \Std = 0$ for which the unitype network has a larger induced deformation than the hybrid network.

\subsection{Usable Work}

\begin{figure}
	\centering
	\input{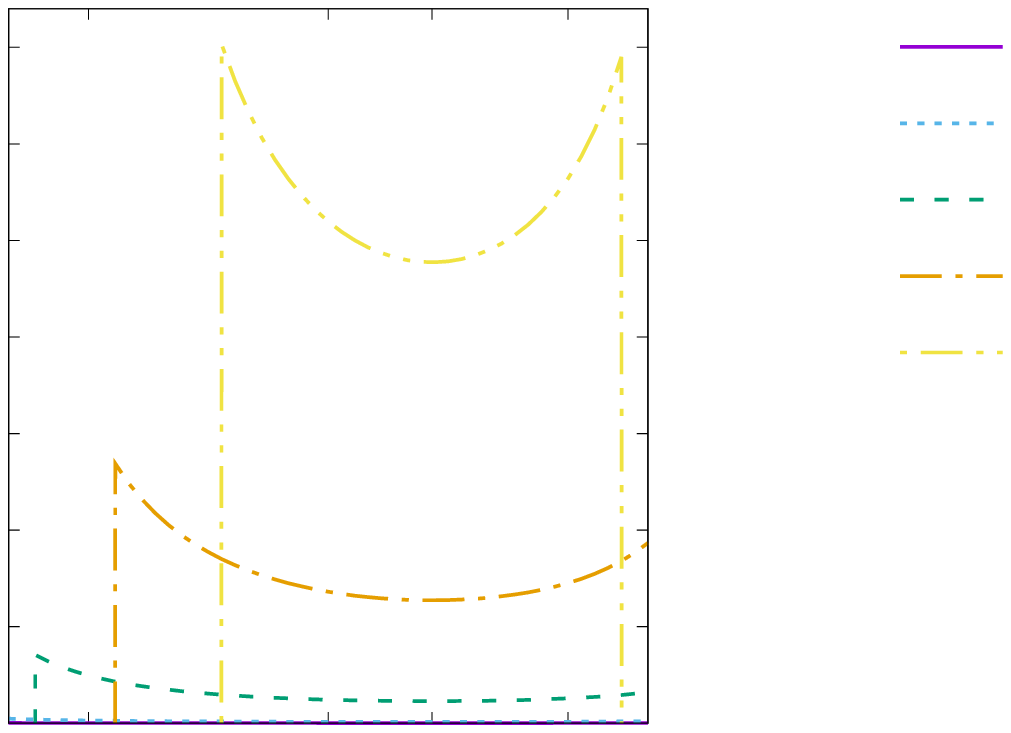}
	\caption{$-\Delta \gibbs / \chaindensityref \kB \T$ vs $\polarAvg$ contours of constant $\Enodim$ for networks of type (1) and (2).
		For each of the contours, $\polarAvg = \polarAB$ has the least amount of usable work and the contours are symmetric about $\polarAB$.
		However, while the deformation increases as $\polarAvg$ has a larger deviation from $\polarAB$, the DEA also fails at smaller $\Enodim$.
		Note that $\polarAvg = \polarAB$ has an equivalent electromechanical response to the uniform network--so $\polarAvg = \polarAB$ represents our isotropic baseline.}
	\label{fig:VCtrl_DeltaPsi-vs-theta0_TI_I0}
\end{figure}

Another performance metric for our DEAs is the amount of usable work that can be derived from the system.
We will use $-\Delta \gibbs$ as a measure of the usable work, where
\begin{equation*}
\Delta \gibbs = \gibbs\left(\rstate{\pstchu}\pstchpr, \Enodim\right) - \gibbs\left(\rstate{\pstchu}, \Enodim\right).
\end{equation*}
\Fref{fig:VCtrl_DeltaPsi-vs-theta0_TI_I0} shows $-\Delta \gibbs / \chaindensityref \kB \T$ vs $\polarAvg$ contours of constant $\Enodim$ for networks of type (1) and (2).
Similar to the case of deformation, it can be seen that, for a given $\Enodim$, the usable work can be maximized by picking the $\polarAvg$ such that $|\polarAB - \polarAvg|$ is maximized without an instability occurring.

\begin{figure}
	\centering
	\input{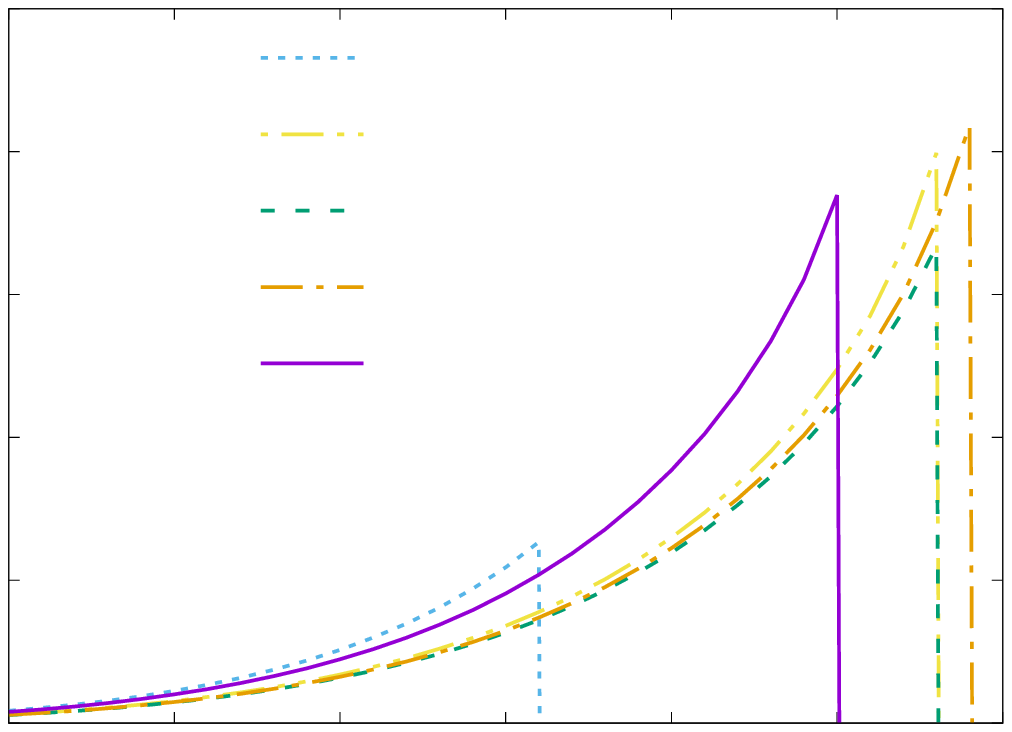}
	\caption{Comparison of the DEA usable work of the hybrid network to the unitype networks with $\Std = 0$.
		While the most usable work can be derived from a unitype, $\Std = 0$ network with properly chosen $\polarAvg$ for a given $\Enodim$, the hybrid network combines both stability and an enhanced usable work.
		In fact, the hybrid network performs $\approx 75\%$ better than the isotropic network.
	}
	\label{fig:VCtrl_DeltaPsi-vs-EL_I0-vs-Hybrid}
\end{figure}

Next, consider the usable work of the hybrid network compared to the unitype $\Std = 0$ networks.
\Fref{fig:VCtrl_DeltaPsi-vs-EL_I0-vs-Hybrid} shows $-\Delta \gibbs / \chaindensityref \kB \T$ as a function of $\Enodim$ for $\Std = 0; \polarAvg = \pi / 6, \pi / 4, \polarAB, \pi / 3$ and the hybrid network.
Again, it can be seen that for a fixed $\Enodim$, a unitype network with properly chosen $\polarAvg$ can outperform the hybrid network.
However, the hybrid network combines both a higher electromechanical coupling (than the isotropic network) and maintains its stability at larger $\Enodim$.
In particular, \emph{the hybrid network shows $\approx 75\%$ increase in usable work over the isotropic network for general $\Enodim$}.
While $\Std = 0, \polarAvg = \polarAB$ can endure a larger $\Enodim$ before failure, the maximum usable work before failure of the hybrid network is still greater than that of the isotropic network.
In summary the unitype, $\Std = 0$ networks can be optimized for a specific electrical load more so than the hybrid network.
However, the hybrid network is more stable and would be much more preferable than the unitype network when used in an application with a wider range of operating $\Enodim$.

\section{Summary of Findings}
In this work, we utilized a multiscale modelling approach to investigate how the microstructure of a dielectric elastomer affects its material properties and performance as an actuator.
The multiscale model lead to important insights such as the phenomenon of electrostatic chain torque, which not only contributes to the electromechanical coupling of the material but could be used as a possible mechanism for aligning chains during the DE manufacturing process.
Towards material design, we proposed the following design parameters: the distribution of chain end-to-end vectors in the network, $\chainpdf$; the mass density (i.e. $\chaindensityref \N$); the density of cross-links; and the fraction of loose-end monomers, $\lefrac$.
Further, we restricted our attention to forms of $\chainpdf$ that would result in a transversely isotropic material.
It was found that substantial gains in the deformation and usable work, for fixed electrical input, can be obtained by designing and manufacturing an anisotropic DE material.

Our key findings include the following:
\begin{enumerate}
    \item To a good approximation, the electromechanical coupling of the network was invariant with respect to $\lefrac$.
    \item The tangent stiffness modulus in the direction of the axis of symmetry, $\rstate{\ElasticPara}$, is a maximum for $\left(\polarAvg, \Std\right) = \left(0, 0\right)$ ($\rstate{\ElasticPara} = 9 \chaindensity \kB \T$ and vanishes as $\polarAvg \rightarrow \pi / 2$, $\Std \rightarrow 0$.
    Similarly, $\rstate{\ElasticPerp}$, is a maximum for $\left(\polarAvg, \Std\right) = \left(\pi / 2, 0\right)$ ($\rstate{\ElasticPerp} = 9 / 4 \chaindensity \kB \T$ and vanishes as $\polarAvg \rightarrow 0$, $\Std \rightarrow 0$.
    These results are intuitive as they suggest that the elastomer is more stiff in directions in which more chains are aligned.
    In general, the stiffness, at fixed density, is proportional to the density of cross-links in the network.
    \item The total susceptibility scales as $\bigoof{\N \chaindensity}$ while (to a good approximation) the influence of the network architecture on the susceptibility only scales as $\bigoof{\chaindensity}$.
    This means that unless the cross-linking density is high--which would reduce compliance--the potential increase in susceptibility from tuning the network architecture would likely only be modest.
    \item Surprisingly, the true electrostriction of a transversely isotropic DE is invariant with respect to its network architecture.
    \item Despite the gains in susceptibility that result from adjusting the network architecture being somewhat modest, its influence (as well as the influence of some of the other aforementioned material properties) on the deformation and usable work of a dielectric elastomer actuator is nonlinear.
    In fact, the performance of a DEA can be significantly increased by properly designing the elastomer network.
    \item For single chain type networks, there is a trade-off between the maximum amount of deformation and usable work that may be extracted from a DEA before failure and the efficiency of the DEA (i.e. mechanical output per electrical input).
    \item A so-called ``hybrid network'' was proposed which consisted of uniaxial chains oriented with $\edir$ and $-\edir$ and TI chains oriented orthogonal to $\edir$.
    The hybrid network preserved the isotropic stiffness and electromechanical stability of the DE, while increasing its usable work output by $\approx 75\%$. 
\end{enumerate}

\paragraph*{Outlook.}

Here we have chosen to focus on optimizing the microstructure of dielectric elastomers for a general type of actuator; for simplicity, we chose an ansatz for $\chainpdf$ with only two tunable parameters and assumed the material was homogeneous,
However, one could imagine relaxing this and using homogenization, topology optimization, machine learning, and other tools to design a more complex, possibility heterogeneous, microstructure to achieve novel behavior such as bending actuation and shape morphing, or for optimizing the microstructure for specific applications in wearable electronics, energy harvesting, soft robotics, prosthetics, etc.
In general, going beyond isotropic materials and exploiting emerging manufacturing techniques to introduce ordering into the material can lead to significant performance increases, following broadly along the lines of work in nonlinear homogenization~\cite{castaneda2011homogenization,galipeau2013finite,siboni2014fiber}.

\section*{Acknowledgments}

This paper draws from the doctoral dissertation of Matthew Grasinger at Carnegie Mellon University. 
We thank Michael Bockstaller, Timothy Breitzman, Gal deBotton, Richard James, Carmel Majidi, Pedro Ponte Casta\~{n}eda, Prashant Purohit, M. Ravi Shankar, and Pradeep Sharma for useful discussions.
We acknowledge financial support from NSF (1635407), ARO (W911NF-17-1-0084), ONR (N00014-18-1-2528, N00014-18-1-2856), BSF (2018183), and AFOSR (MURI FA9550-18-1-0095).
We acknowledge NSF for computing resources provided by Pittsburgh Supercomputing Center.

\appendix
\section{Polarization as Energy Conjugate to Electric Field}
\label{app:polarization}

We show that a consequence of assuming \eqref{eq:density} as the form of the monomer density function and enforcing the constraints given in \eqref{eq:ccn}, we arrive at the relationship: $\chainpolar = -\takepartial{\A}{\efield}$.
\hl{This was first shown in}~\cite{grasinger2020statistical} \hl{and is provided here again for completeness.}

Taking derivatives of both sides of \eqref{eq:ccn} with respect to $\efield$, we obtain:
\begin{equation} \label{eq:dconstraints}
	\begin{split}
		\takepartial{}{\efield} \intoverSns{\density\left(\nvec\right)} = \intoverSns{\takepartial{\density}{\efield}} &= \takepartial{\N}{\efield} = 0 
		\\
		\takepartial{}{\efield} \intoverSns{\density\left(\nvec\right) \nvec} = \intoverSns{\takepartial{\density}{\efield} \nvec} &= \takepartial{\rvec / \mlen}{\efield} = 0
	\end{split}
\end{equation}
We are able to interchange the operations of derivation and integration because of the smoothness of the integrands; and in the last equalities we use the fact that neither the number of the monomers in the chain nor the end-to-end vector constraint depend on $\efield$.
	
Now, we obtain the desired result by taking derivatives of both sides of \eqref{eq:A-approx}:
\begin{align*}
		-\takepartial{\A}{\efield} &= -\takepartial{}{\efield} \intoverSns{\left(\density \um + \kB \T \density \ln \density\right)} 
		\\
		& = -\intoverSns{\left[\takepartial{\density}{\efield} \um + \density \takepartial{\um}{\efield} + \kB \T \takepartial{\density}{\efield} \ln \density + \density \left(\takepartial{\density}{\efield} / \density\right)\right]} 
		\\
		&= -\intoverSns{\left[\takepartial{\density}{\efield} \um + \density \takepartial{\um}{\efield} + \kB \T \takepartial{\density}{\efield} \left(\ln \C -\um / \kB \T + \mults \cdot \nvec\right) + \takepartial{\density}{\efield}\right]} 
		\\
		&= -\intoverSns{\left[\density \takepartial{\um}{\efield} + \kB \T \takepartial{\density}{\efield} \left(\mults \cdot \nvec\right) + \left(\kB \T \ln\C + 1\right)\takepartial{\density}{\efield}\right]} 
		\\
		&= -\intoverSns{\density \takepartial{\um}{\efield}} - \kB \T \mults \cdot \left(\intoverSns{\takepartial{\density}{\efield} \nvec}\right) - \left(\kB \T \ln\C + 1\right)\intoverSns{\takepartial{\density}{\efield}}
\end{align*}
By \eqref{eq:dconstraints}, the last two terms vanish.
Thus, recalling \eqref{eq:dipole-response} and \eqref{eq:monomer-energy}:
\begin{equation*}
		-\takepartial{\A}{\efield} = -\intoverSns{\density \takepartial{\um}{\efield}} 
		= \intoverSns{\density \takepartial{}{\efield} \left(\frac{1}{2} \efield \cdot \sustens \efield\right)} 
		= \intoverSns{\density \frac{1}{2} \left(\sustens \efield + \sustens^T \efield \right)} 
		= \intoverSns{\density \dipole}
	    = \chainpolar
\end{equation*}
as desired.

It is obvious then that this is not true of all chains with polarizable monomers such that $\um$ includes a dipole-electric field interaction term (i.e. $-\dipole \cdot \efield$).
Instead, this follows from the fact that the dipole susceptibility tensor, $\sustens$, is symmetric since it derives from a potential; and also from the fact that \emph{the energy to separate bound charges in the monomer is harmonic} such that $\um_{\text{bond}} = \frac{1}{2} \dipole \cdot \sustens^{-1} \dipole$. 

As a corollary, we show that
\begin{equation} \label{eq:polarization-constitutive-response}
	-\takepartial{\Wstar}{\efield} = \polarization
\end{equation}
This follows as a consequence of $-\takepartial{\A}{\efield} = \chainpolar$ and the assumption that chains in the network are in negligible interaction (see \Fref{sec:weakly}).
By the negligibly interacting assumption, $\Wstar = \chaindensity \avg{\A}$, where by $\avg{\generic}$ we mean the average over the chain pdf.
Hence, taking derivatives of both sides:
\begin{equation*}
	-\takepartial{\Wstar}{\efield} = -\takepartial{}{\efield}\chaindensity \avg{\A} = \chaindensity \avg{-\takepartial{\A}{\efield}} = \chaindensity \avg{\chainpolar} = \J^{-1} \Polar.
\end{equation*}
(Recall: incompressibility implies that $\J = 1$.)
This result, along with the statistical mechanical derivation of $\Wstar$, establishes $\Wstar\left(\F, \efield\right)$ as the Legendre transform of $\W\left(\F, \Polar\right)$, the Helmholtz free energy density.


\section{Derivatives of $\W$ and $\Wstar$} \label{app:legendre-transform-stuff}

Since we consider the physical regime in which monomer-monomer interactions are negligible (and thereby the nonlocal $\efield$-$\polarization$ relationship simplifies into a local one), and since we assume the material is incompressible, i.e. $\J = 1$, the following hold:
\begin{enumerate}
 \item $\W\left(\F, \Polar\right) = \Wstar\left(\F, \efield\left(\Polar\right)\right) + \Polar \cdot \efield\left(\Polar\right)$,
 \item $\takepartial{\W}{\Polar} = \efield$,
 \item $\takepartial{\Wstar}{\efield} = -\Polar$.
\end{enumerate}
Assume $\efield\left(\Polar\right)$ invertible such that $\Polar = \Polar\left(\efield\right)$, $\takepartial{\Polar}{\efield}$ is invertible, and $\W$, $\Wstar$, $\efield\left(\Polar\right)$ and $\Polar\left(\efield\right)$ are all smooth functions.

By assumption, $\takepartial{\W}{\Polar} = \efield\left(\Polar\right)$. Taking derivatives of both sides:
\begin{equation*}
	\takecrosspartial{\W}{\Polar}{\Polar} = \takepartial{\efield}{\Polar}.
\end{equation*}
Similarly, taking derivatives of both sides of $\takepartial{\Wstar}{\efield} = -\Polar$:
\begin{equation*}
	\takecrosspartial{\Wstar}{\efield}{\efield} = -\takepartial{\Polar}{\efield}.
\end{equation*}
From the two above equations we can conclude that
\begin{equation*}
	-\takecrosspartial{\Wstar}{\efield}{\efield} = \left(\takecrosspartial{\W}{\Polar}{\Polar}\right)^{-1}
\end{equation*}
The physical implication of the above result is that: $-\takecrosspartial{\Wstar}{\efield}{\efield} = \susceptibilityTensor$.

\section{Integration of Gaussian-like Functions}
\label{app:GI}

The integration in \eqref{eq:energy-density} typically has the form:
\begin{equation*}
    \GaussSineInt{k}{\mu}{\sigma} \coloneqq \int_{0}^{\pi} \dx{x} \left(\GaussExpr{x}{\mu}{\sigma} + \GaussExpr{\pi - x}{\mu}{\sigma}\right) \sin \left(k x\right)
\end{equation*}
where $k \in \mathbb{N}$, $\mu \in \left[0, \frac{\pi}{2}\right]$, and $\sigma \in \left[0, \infty\right)$.

In the limit of $\sigma \ll 1$, we have that
\begin{equation} \label{eq:Izero}
    \GaussSineInt{k}{\mu}{\sigma} \approx \GaussSineIntZero{k}{\mu}{\sigma} = 2 \sqrt{2 \pi} \sigma \exp\left[-\frac{k^2 \sigma^2}{2}\right] \sin \left(k \mu\right)
\end{equation}
This is obtained by bringing the sine term into the exponential (by using $\sin\left(k x\right) = \ImPart\left(\euler^{\im x}\right)$), and using the fast decay of the Gaussian when $\sigma \ll 1$ to approximate the domain of integration by the entire real line.

In the limit of $\sigma \gg 1$, we have that
\newcommand{\z}{\frac{k \pi}{2}}
\begin{equation} \label{eq:Iinf}
    \GaussSineInt{k}{\mu}{\sigma} \approx \GaussSineIntInf{k}{\mu}{\sigma} = \frac{\sin\left(\z\right)}{k^3 \sigma^2} \left( 2 k \pi \cos\left(\z\right) + \left(k^2 \left(\pi^2 - 2\pi\mu + 2\mu^2 - 4\sigma^2\right)\right) \sin\left(\z\right)\right)
\end{equation}
which is derived by using a Taylor expansion of the exponential to linear order in its argument.

\section{List of Symbols}
\label{sec:symbols}

\begin{tabular}{p{2.0cm} l}
$\rvec$ & Chain end-to-end vector (current configuration) \\
$\Rvec$ & Chain end-to-end vector (reference configuration) \\
$\N$ & Number of monomers per chain \\
$\mlen$ & Length of a single monomer \\
$\stch$ & Chain stretch, $|\rvec| / \N \mlen$ \\
$\chaindensity$ & Density of chains per unit volume (current configuration) \\
$\chaindensityref$ & Density of chains per unit volume (reference configuration), $\chaindensityref = \J \chaindensity$ \\
$\nb$ & Number of backbone monomers \\
$\nl$ & Number of loose-end monomers \\
$\lefrac$ & Fraction of loose-end monomers, $\nl / \N$ \\
$\chainpdf$ & Probability density function which describes the fraction of chains with reference end-to-end vector $\Rvec$ \\
$\polarAvg, \Std$ & Mean of pdf Gaussian in the upper half of the sphere, standard deviation of the Gaussian \\
$\efield$ & Electric field \\
$\sus{1}$ & Monomer dipole susceptibility along $\nvec$ \\
$\sus{2}$ & Monomer dipole susceptibility orthogonal to $\nvec$ \\
$\dsus$ & $\sus{2}-\sus{1}$ \\
$\chainpolar$ & Net chain dipole \\
$\unodim$ & Ratio of electrical energy per monomer to thermal energy per monomer, $\emag^2 \dsus / 2 \kB \T$ \\
$\A$ & Closed-dielectric free energy of a single chain \\
$\AHelm$ & Helmholtz free energy of a single chain \\
$\Wstar$ & Closed-dielectric free energy density of the polymer network \\
$\W$ & Helmholtz free energy density \\
$\F$ & Deformation gradient \\
$\J$ & $\det \F$ \\
$\polarization$ & Continuum-scale polarization \\
$\Polar$ & Pullback of the continuum-scale polarization \\
$\Elastic$ & Tangent elastic modulus \\
$\susceptibilityTensor$ & Tangent polarization susceptibility \\
$\susceptibilityTensorLab$ & Secant polarization susceptibility \\
$\crossModulusTensor$ & Cross modulus \\
$\rstate{\generic}$ & Material property $\generic$ with respect to the relaxed state \\
$\generic_{\parallel}$ & Material property $\generic$ along the axis of symmetry \\
$\generic_{\perp}$ & Material property $\generic$ perpendicular to the axis of symmetry \\
\end{tabular}

\bibliographystyle{alpha}
\bibliography{master}

\newcommand{\etalchar}[1]{$^{#1}$}
\begin{thebibliography}{HLCF{\etalchar{+}}12}

\bibitem[ABB{\etalchar{+}}17]{ambulo2017four}
Cedric~P Ambulo, Julia~J Burroughs, Jennifer~M Boothby, Hyun Kim, M~Ravi
  Shankar, and Taylor~H Ware.
\newblock Four-dimensional printing of liquid crystal elastomers.
\newblock {\em ACS applied materials \& interfaces}, 9(42):37332--37339, 2017.

\bibitem[AP12]{argudo2012dependence}
David Argudo and Prashant~K Purohit.
\newblock The dependence of dna supercoiling on solution electrostatics.
\newblock {\em Acta biomaterialia}, 8(6):2133--2143, 2012.

\bibitem[BC04]{bar-cohen2001electroactive}
Yoseph Bar-Cohen.
\newblock {\em Electroactive polymer (EAP) actuators as artificial muscles}.
\newblock SPIE--The International Society for Optical Engineering, 2004.

\bibitem[BDO09]{bustamante2009nonlinear}
Roger Bustamante, Alois Dorfmann, and Ray~W Ogden.
\newblock Nonlinear electroelastostatics: a variational framework.
\newblock {\em Zeitschrift f{\"u}r angewandte Mathematik und Physik},
  60(1):154--177, 2009.

\bibitem[Bea03]{beatty2003average}
Millard~F Beatty.
\newblock An average-stretch full-network model for rubber elasticity.
\newblock {\em Journal of Elasticity}, 70(1-3):65--86, 2003.

\bibitem[BFK{\etalchar{+}}16]{bartlett2016stretchable}
Michael~D Bartlett, Andrew Fassler, Navid Kazem, Eric~J Markvicka, Pratiti
  Mandal, and Carmel Majidi.
\newblock Stretchable, high-k dielectric elastomers through liquid-metal
  inclusions.
\newblock {\em Advanced Materials}, 28(19):3726--3731, 2016.

\bibitem[BJDM17]{babaei2017computing}
Mahnoush Babaei, Isaac~C Jones, Kaushik Dayal, and Meagan~S Mauter.
\newblock Computing the diamagnetic susceptibility and diamagnetic anisotropy
  of membrane proteins from structural subunits.
\newblock {\em Journal of chemical theory and computation}, 13(6):2945--2953,
  2017.

\bibitem[Cas60]{case1960branching}
LC~Case.
\newblock Branching in polymers. i. network defects.
\newblock {\em Journal of Polymer Science}, 45(146):397--404, 1960.

\bibitem[Cd16]{cohen2016electromechanical}
Noy Cohen and Gal deBotton.
\newblock Electromechanical interplay in deformable dielectric elastomer
  networks.
\newblock {\em Physical review letters}, 116(20):208303, 2016.

\bibitem[CDd16]{cohen2016electroelasticity}
Noy Cohen, Kaushik Dayal, and Gal deBotton.
\newblock Electroelasticity of polymer networks.
\newblock {\em Journal of Mechanics Physics of Solids}, 92:105--126, 2016.

\bibitem[CG11]{castaneda2011homogenization}
P~Ponte Casta{\~n}eda and E~Galipeau.
\newblock Homogenization-based constitutive models for magnetorheological
  elastomers at finite strain.
\newblock {\em Journal of the Mechanics and Physics of Solids}, 59(2):194--215,
  2011.

\bibitem[CKSLA11]{carpi2011electroactive}
Federico Carpi, Roy Kornbluh, Peter Sommer-Larsen, and Gursel Alici.
\newblock Electroactive polymer actuators as artificial muscles: are they ready
  for bioinspired applications?
\newblock {\em Bioinspriation \& Biommetics}, 6(4):045006, 2011.

\bibitem[D{\etalchar{+}}72]{daniels1972kuhn}
HE~Daniels et~al.
\newblock Kuhn-gr{\"u}n type approximations for polymer chain distributions.
\newblock In {\em Proceedings of the Sixth Berkeley Symposium on Mathematical
  Statistics and Probability, Volume 3: Probability Theory}. The Regents of the
  University of California, 1972.

\bibitem[DDLS19]{darbaniyan2019designing}
Faezeh Darbaniyan, Kaushik Dayal, Liping Liu, and Pradeep Sharma.
\newblock Designing soft pyroelectric and electrocaloric materials using
  electrets.
\newblock {\em Soft matter}, 15(2):262--277, 2019.

\bibitem[DO06]{dorfmann2006nonlinear}
A~Dorfmann and RW~Ogden.
\newblock Nonlinear electroelastic deformations.
\newblock {\em Journal of Elasticity}, 82(2):99--127, 2006.

\bibitem[DO14]{dorfmann2014nonlinear}
Luis Dorfmann and Ray~W Ogden.
\newblock {\em Nonlinear theory of electroelastic and magnetoelastic
  interactions}.
\newblock Springer, 2014.

\bibitem[EAOvL19]{erol2019microstructure}
Anil Erol, Saad Ahmed, Zoubeida Ounaies, and Paris von Lockette.
\newblock A microstructure-based approach to modeling electrostriction that
  accounts for variability in spatial locations of domains.
\newblock {\em Journal of the Mechanics and Physics of Solids}, 124:35--62,
  2019.

\bibitem[Eri98]{ericksen1998introduction}
Jerald~L Ericksen.
\newblock {\em Introduction to the Thermodynamics of Solids}.
\newblock Springer, 1998.

\bibitem[FAK{\etalchar{+}}19]{ford2019multifunctional}
Michael~J Ford, Cedric~P Ambulo, Teresa~A Kent, Eric~J Markvicka, Chengfeng
  Pan, Jonathan Malen, Taylor~H Ware, and Carmel Majidi.
\newblock A multifunctional shape-morphing elastomer with liquid metal
  inclusions.
\newblock {\em Proceedings of the National Academy of Sciences},
  116(43):21438--21444, 2019.

\bibitem[FG08]{fox2008dynamic}
JW~Fox and NC~Goulbourne.
\newblock On the dynamic electromechanical loading of dielectric elastomer
  membranes.
\newblock {\em Journal of the Mechanics and Physics of Solids},
  56(8):2669--2686, 2008.

\bibitem[Flo44]{flory1944network}
Paul~J Flory.
\newblock Network structure and the elastic properties of vulcanized rubber.
\newblock {\em Chemical reviews}, 35(1):51--75, 1944.

\bibitem[GC13]{galipeau2013finite}
Evan Galipeau and Pedro~Ponte Casta{\~n}eda.
\newblock A finite-strain constitutive model for magnetorheological elastomers:
  magnetic torques and fiber rotations.
\newblock {\em Journal of the Mechanics and Physics of Solids},
  61(4):1065--1090, 2013.

\bibitem[GD20]{grasinger2020statistical}
Matthew Grasinger and Kaushik Dayal.
\newblock Statistical mechanical analysis of the electromechanical coupling in
  an electrically-responsive polymer chain.
\newblock {\em Soft Matter}, 16:6265--6284, 2020.

\bibitem[GMD]{grasinger2020inprep}
Matthew Grasinger, Carmel Majidi, and Kaushik Dayal.
\newblock In preparation.

\bibitem[Gor75]{gordon1975rubber}
Manfred Gordon.
\newblock Rubber elasticity. flaws in the theory of networks.
\newblock {\em Macromolecules}, 8(2):247--248, 1975.

\bibitem[HCB13]{henann2013modeling}
David~L Henann, Shawn~A Chester, and Katia Bertoldi.
\newblock Modeling of dielectric elastomers: Design of actuators and energy
  harvesting devices.
\newblock {\em Journal of the Mechanics and Physics of Solids},
  61(10):2047--2066, 2013.

\bibitem[Hil86]{hill1986statistical}
Terrell~L Hill.
\newblock {\em An introduction to statistical thermodynamics}.
\newblock Dover Publications, 1986.

\bibitem[HLCF{\etalchar{+}}12]{huang2012giant}
Jiangshui Huang, Tiefeng Li, Choon Chiang~Foo, Jian Zhu, David~R Clarke, and
  Zhigang Suo.
\newblock Giant, voltage-actuated deformation of a dielectric elastomer under
  dead load.
\newblock {\em Applied Physics Letters}, 100(4):041911, 2012.

\bibitem[JK90]{james1990frustration}
Richard~D James and David Kinderlehrer.
\newblock Frustration in ferromagnetic materials.
\newblock {\em Continuum Mechanics and Thermodynamics}, 2(3):215--239, 1990.

\bibitem[KG42]{kuhn1942beziehungen}
Werner Kuhn and F~Gr{\"u}n.
\newblock Beziehungen zwischen elastischen konstanten und
  dehnungsdoppelbrechung hochelastischer stoffe.
\newblock {\em Kolloid-Zeitschrift}, 101(3):248--271, 1942.

\bibitem[KLS19]{krichen2019liquid}
Sana Krichen, Liping Liu, and Pradeep Sharma.
\newblock Liquid inclusions in soft materials: Capillary effect, mechanical
  stiffening and enhanced electromechanical response.
\newblock {\em Journal of the Mechanics and Physics of Solids}, 127:332--357,
  2019.

\bibitem[Kof08]{kofod2008static}
Guggi Kofod.
\newblock The static actuation of dielectric elastomer actuators: how does
  pre-stretch improve actuation?
\newblock {\em Journal of Physics D: Applied Physics}, 41(21):215405, 2008.

\bibitem[KT07]{kim2007electroactive}
Kwang~J Kim and Satoshi Tadokoro.
\newblock Electroactive polymers for robotic applications.
\newblock {\em Artificial Muscles and Sensors}, 23:291, 2007.

\bibitem[KZSK12]{kollosche2012complex}
Matthias Kollosche, Jian Zhu, Zhigang Suo, and Guggi Kofod.
\newblock Complex interplay of nonlinear processes in dielectric elastomers.
\newblock {\em Physical Review E}, 85(5):051801, 2012.

\bibitem[Liu13]{liu2013energy}
Liping Liu.
\newblock On energy formulations of electrostatics for continuum media.
\newblock {\em Journal of the Mechanics and Physics of Solids}, 61(4):968--990,
  2013.

\bibitem[Liu14]{liu2014energy}
Liping Liu.
\newblock An energy formulation of continuum magneto-electro-elasticity with
  applications.
\newblock {\em Journal of the Mechanics and Physics of Solids}, 63:451--480,
  2014.

\bibitem[LLS15]{li2015geometrically}
Xiaobao Li, Liping Liu, and Pradeep Sharma.
\newblock Geometrically nonlinear deformation and the emergent behavior of
  polarons in soft matter.
\newblock {\em Soft matter}, 11(41):8042--8047, 2015.

\bibitem[LP74]{langley1974relation}
Neal~R Langley and Keith~E Polmanteer.
\newblock Relation of elastic modulus to crosslink and entanglement
  concentrations in rubber networks.
\newblock {\em Journal of Polymer Science: Polymer Physics Edition},
  12(6):1023--1034, 1974.

\bibitem[LP14]{lopez2014elastic}
Oscar Lopez-Pamies.
\newblock Elastic dielectric composites: Theory and application to
  particle-filled ideal dielectrics.
\newblock {\em Journal of the Mechanics and Physics of Solids}, 64:61--82,
  2014.

\bibitem[LS18]{liu2018emergent}
Liping Liu and Pradeep Sharma.
\newblock Emergent electromechanical coupling of electrets and some exact
  relations—the effective properties of soft materials with embedded external
  charges and dipoles.
\newblock {\em Journal of the Mechanics and Physics of Solids}, 112:1--24,
  2018.

\bibitem[Maj14]{majidi2014soft}
Carmel Majidi.
\newblock Soft robotics: a perspective—current trends and prospects for the
  future.
\newblock {\em Soft Robotics}, 1(1):5--11, 2014.

\bibitem[MD14]{marshall2014atomistic}
Jason Marshall and Kaushik Dayal.
\newblock Atomistic-to-continuum multiscale modeling with long-range
  electrostatic interactions in ionic solids.
\newblock {\em Journal of the Mechanics and Physics of Solids}, 62:137--162,
  2014.

\bibitem[ML05]{mcmeeking2005electrostatic}
Robert~M McMeeking and Chad~M Landis.
\newblock Electrostatic forces and stored energy for deformable dielectric
  materials.
\newblock {\em Journal of Applied Mechanics}, 72(4):581--590, 2005.

\bibitem[PKK00]{pelrine2000high}
Ron Pelrine, Roy Kornbluh, and Guggi Kofod.
\newblock High-strain actuator materials based on dielectric elastomers.
\newblock {\em Advanced Materials}, 12(16):1223--1225, 2000.

\bibitem[RYBS19]{rahmati2019nonlinear}
Amir~Hossein Rahmati, Shengyou Yang, Siegfried Bauer, and Pradeep Sharma.
\newblock Nonlinear bending deformation of soft electrets and prospects for
  engineering flexoelectricity and transverse (d 31) piezoelectricity.
\newblock {\em Soft Matter}, 15(1):127--148, 2019.

\bibitem[SC14]{siboni2014fiber}
Morteza~Hakimi Siboni and Pedro~Ponte Casta{\~n}eda.
\newblock Fiber-constrained, dielectric-elastomer composites: finite-strain
  response and stability analysis.
\newblock {\em Journal of the Mechanics and Physics of Solids}, 68:211--238,
  2014.

\bibitem[Sd13]{shmuel2013axisymmetric}
Gal Shmuel and Gal deBotton.
\newblock Axisymmetric wave propagation in finitely deformed dielectric
  elastomer tubes.
\newblock {\em Proceedings of the Royal Society A: Mathematical, Physical and
  Engineering Sciences}, 469(2155):20130071, 2013.

\bibitem[Suo10]{suo2010theory}
Zhigang Suo.
\newblock Theory of dielectric elastomers.
\newblock {\em Acta Mechanica Solida Sinica}, 23(6):549--578, 2010.

\bibitem[SW17]{shen2017electrostatic}
Kevin Shen and Zhen-Gang Wang.
\newblock Electrostatic correlations and the polyelectrolyte self energy.
\newblock {\em The Journal of chemical physics}, 146(8):084901, 2017.

\bibitem[SZG08]{suo2008nonlinear}
Zhigang Suo, Xuanhe Zhao, and William~H Greene.
\newblock A nonlinear field theory of deformable dielectrics.
\newblock {\em Journal of the Mechanics and Physics of Solids}, 56(2):467--486,
  2008.

\bibitem[TM11]{tadmor2011modeling}
Ellad~B Tadmor and Ronald~E Miller.
\newblock {\em Modeling materials: continuum, atomistic and multiscale
  techniques}.
\newblock Cambridge University Press, 2011.

\bibitem[Tou56]{toupin1956elastic}
R.A. Toupin.
\newblock The elastic dielectric.
\newblock {\em Journal of Rational Mechanics and Analysis}, 5(6):849--915,
  1956.

\bibitem[Tre75]{treloar1975physics}
L~R~G Treloar.
\newblock {\em The physics of rubber elasticity}.
\newblock Oxford University Press, 1975.

\bibitem[Ver15]{verron2015questioning}
Erwan Verron.
\newblock Questioning numerical integration methods for microsphere (and
  microplane) constitutive equations.
\newblock {\em Mechanics of Materials}, 89:216--228, 2015.

\bibitem[WBS{\etalchar{+}}16]{ware2016localized}
Taylor~H Ware, John~S Biggins, Andreas~F Shick, Mark Warner, and Timothy~J
  White.
\newblock Localized soft elasticity in liquid crystal elastomers.
\newblock {\em Nature communications}, 7(1):1--7, 2016.

\bibitem[Wei12]{weiner2012statistical}
Jerome~Harris Weiner.
\newblock {\em Statistical mechanics of elasticity}.
\newblock Courier Corporation, 2012.

\bibitem[WM07]{wissler2007mechanical}
Michael Wissler and Edoardo Mazza.
\newblock Mechanical behavior of an acrylic elastomer used in dielectric
  elastomer actuators.
\newblock {\em Sensors and Actuators A: Physical}, 134(2):494--504, 2007.

\bibitem[WTF04]{wang2004self}
Qiang Wang, Takashi Taniguchi, and Glenn~H Fredrickson.
\newblock Self-consistent field theory of polyelectrolyte systems.
\newblock {\em The Journal of Physical Chemistry B}, 108(21):6733--6744, 2004.

\bibitem[WVDG93]{wu1993improved}
PD~Wu and Erik Van Der~Giessen.
\newblock On improved network models for rubber elasticity and their
  applications to orientation hardening in glassy polymers.
\newblock {\em Journal of the Mechanics and Physics of Solids}, 41(3):427--456,
  1993.

\bibitem[YD11]{yang2011completely}
Lun Yang and Kaushik Dayal.
\newblock A completely iterative method for the infinite domain electrostatic
  problem with nonlinear dielectric media.
\newblock {\em Journal of Computational Physics}, 230(21):7821--7829, 2011.

\bibitem[YM16]{yu2016energy}
Ying-Ju Yu and Alan~JH McGaughey.
\newblock Energy barriers for dipole moment flipping in pvdf-related
  ferroelectric polymers.
\newblock {\em The Journal of chemical physics}, 144(1):014901, 2016.

\bibitem[YZS17]{yang2017revisiting}
Shengyou Yang, Xuanhe Zhao, and Pradeep Sharma.
\newblock Revisiting the instability and bifurcation behavior of soft
  dielectrics.
\newblock {\em Journal of Applied Mechanics}, 84(3):031008, 2017.

\bibitem[ZDDP17]{zurlo2017catastrophic}
Giuseppe Zurlo, Michel Destrade, Domenico DeTommasi, and Giuseppe Puglisi.
\newblock Catastrophic thinning of dielectric elastomers.
\newblock {\em Physical review letters}, 118(7):078001, 2017.

\bibitem[ZHS07]{zhao2007electromechanical}
Xuanhe Zhao, Wei Hong, and Zhigang Suo.
\newblock Electromechanical hysteresis and coexistent states in dielectric
  elastomers.
\newblock {\em Physical review B}, 76(13):134113, 2007.

\bibitem[ZS07]{zhao2007method}
Xuanhe Zhao and Zhigang Suo.
\newblock Method to analyze electromechanical stability of dielectric
  elastomers.
\newblock {\em Applied Physics Letters}, 91(6):061921, 2007.

\bibitem[ZS08]{zhao2008electrostriction}
Xuanhe Zhao and Zhigang Suo.
\newblock Electrostriction in elastic dielectrics undergoing large deformation.
\newblock {\em Journal of Applied Physics}, 104(12):123530, 2008.

\end{thebibliography}

\end{document}